\documentclass[%
 reprint,
 amsmath,amssymb,
 aps,
]{revtex4-2}

\usepackage{graphicx}% Include figure files
\usepackage{dcolumn}% Align table columns on decimal point
\usepackage{bm}% bold math
\usepackage{xcolor}
\usepackage{hyperref}% add hypertext capabilities

\renewcommand{\vec}[1]{\mathbf{#1}}
\usepackage[mathlines]{lineno}% Enable numbering of text and display math
\begin{document}

\preprint{APS/123-QED}

\title{Exact Statistics of Helical Wormlike Chains with Twist-Bend Coupling}

\author{Ashesh Ghosh$^{1,2,3}$}
 \email{ashesh@berkeley.edu}
\author{Kranthi K. Mandadapu$^{2,3}$}%
 \email{kranthi@berkeley.edu}
\author{David T. Limmer$^{1,3,4,5}$}
 \email{dlimmer@berkeley.edu}
\affiliation{
$^1$Department of Chemistry, University of California, Berkeley, CA 94720, \\
$^2$Department of Chemical and Biomolecular Engineering, University of California, Berkeley, CA 94720, \\
$^3$Chemical Sciences Division, Lawrence Berkeley National Laboratory, Berkeley, CA 94720, \\
$^4$Kavli Energy Nanoscience Institute, Berkeley, CA 94720, \\
$^5$Material Science Division Lawrence Berkeley National Laboratory, Berkeley, CA 94720, USA.
}

\date{\today}

\begin{abstract}
We present a solution for the Green's function for the general case of a helical wormlike chain with twist-bend coupling, and demonstrate the applicability of our solution for evaluating general structural and mechanical chain properties. 
We find that twist-bend coupling renormalizes the persistence length and the force-extension curves relative to worm-like chains.  
Analysis of intrinsically twisted polymers shows that incorporation of twist-bend coupling results in the oscillatory behavior in principal tangent correlations that are observed in some studies of synthetic polymers.
The exact nature of our solution provides a framework to evaluate the role of twist-bend coupling on polymer properties and motivates the reinterpretation of existing bio-polymer experimental data. 
%Furthermore, direct relevance to experimental observable, such as force extension and dynamic structure factor are drawn.
%
%Recent work on biopolymers mechanics and conjugated polymers show the role of coupled twist-bend modes in understanding the structure and morphology of such structures.
%At quadratic levels of strain energy around a reference configuration, the elastic energy of a 3d wire is well known.
%However, no analytical solution for the chain propagator exists till date. 
%
%Our analysis is applied to
%his work provides the framework to understand and interpret data for bio-polymer (or, beyond) experiments where twist-bend coupling could play a role.
\end{abstract}

%Use showkeys class option if keyword
\maketitle

%\tableofcontents

%\noindent
%Geometrical structures with internal separation of length scales are ubiquitous in nature, most profoundly in polymers with structural anisotropy.
%Polymers are often very large along the body contour  as compared to the dimensions of their cross-section.
%This internal separation of length scale creates an inherent anisotropy in mechanical and physical properties of such structures.
%
The structural and mechanical properties of polymers are largely understood within the context of simple models. This is due in part to the large separation of length scales between their body contour length and dimensions of their cross-section that allow for the abstraction of molecular details. 
For example, the mechanical properties of double stranded DNA (dsDNA) are well described by simple elastic polymer models.
%
%are critical for both its structure and function within the cell nucleus. 
A wormlike chain (WLC) model polymer described with just a bending rigidity~\cite{doi1988theory} correctly reproduces the force extension behavior of dsDNA~\cite{smith1992direct,bustamante1994entropic,bustamante2003ten}.
However, capturing the twist and torsional response of dsDNA~\cite{marko1994fluctuations} require a more sophisticated elastic polymer model. 
To that end, early attempts to include some features of the internal structure of dsDNA apply the helical WLC (HWLC) model, characterized by an additional twist rigidity ~\cite{nomidis2019twist, segers2022mechanical,liu2011statistical,marko1997stretching,bouchiat1998elasticity}.
Quantitative comparisons suggest discrepancies between the predictions of the helical WLC and combined force plus torque induced dsDNA response~\cite{lipfert2010magnetic,lipfert2011freely,nomidis2017twist}.
A promising alternative model was developed by Marko and Siggia~\cite{marko1994bending} in which the helical WLC includes an additional direct coupling between twist and bend degrees of freedom~\cite{nomidis2017twist,skoruppa2018bend,zhang2024influence,qiang2022multivalent,nomidis2019twist,gao2021torsional,nomidis2019twist2,fosado2021twist}.
Such coupled twist and bend motions are not unique to dsDNA, rather most semiflexible filaments such as actin~\cite{enrique2010origin}, conjugated polymers\cite{root2017modelling}, or developable ribbons \cite{giomi2010statistical,yong2022statistics}  display large scale conformations with coupled modes.
Here, we consider this general class of twist-bend coupled helical polymers and solve exactly for the two-point correlation function, or Green's function. From this solution, we extract a number of structural and mechanical properties and compare them to other limiting models.

Despite the simplicity presented by helical WLC as an elastic wire model of a polymer chain, very few analytical results exist to directly compare theoretical predictions with experiments.
Leveraging an isomorphism between the twist-bend coupled helical wormlike chain (TBcHWLC) and the rotational motion of an asymmetric top~\cite{whittaker1964treatise}, we obtain the exact analytical results for the full Green's function of polymer chain statistics.
The orientation only Green's function is shown to capture tangent correlations and the effect of twist-bend coupling on the mean squared end-to-end distance of the polymer.
Making use of the Stone-Fence diagrammatic methods~\cite{chandrasekhar1943stochastic,yamakawa2016helical}, we further compute the free energy as a function of tangential force on the polymer chain, as well as its ring-closure probability~\cite{jacobson1950intramolecular}.

\begin{figure}[b!]
    \centering
    \includegraphics[scale=0.30]{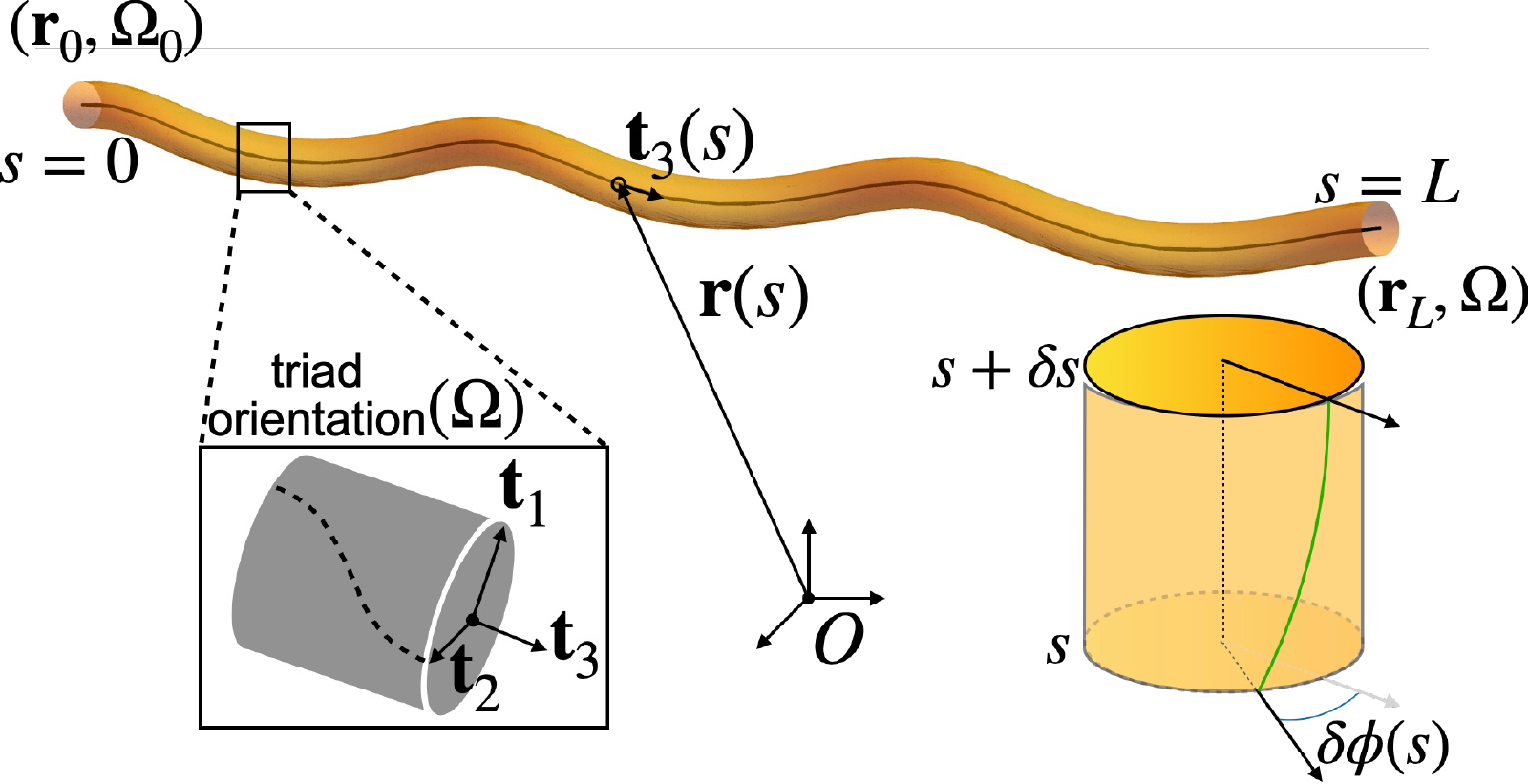}
    \caption{Schematic of the twisted wormlike chain model, parametrized by the arc length variable $s\in [0,L]$. The center-line of the chain is shown in a solid black line. The orthonormal triad orientations $\{\vec{t}_i\}$ are shown in the inset (bottom left). The twisting of the chain is schematically considered as infinitesimal rotation of the cylinder $\delta\phi(s)$ in going from $s\!\to \!s+\delta s$ (bottom right). An integral over the rotations result in the overall twist of the polymer conformation.
    }
    \label{fig:schematic_fig1}
\end{figure}
We model the polymer chain as a continuous curve $\vec{r}(s)$ with origin at $O$ (Fig.~\ref{fig:schematic_fig1}), where $s\in [0,L]$ is an arc length variable that runs along the body of the chain. 
We define an orthonormal triad at every position $\{\vec{t}_i(s)\}$ such that $\vec{t}_i\cdot \vec{t}_j=\delta_{ij}$ and $\vec{t}_i=\epsilon_{ijk}\vec{t}_j\times \vec{t}_k$.
The body tangent is aligned with $\vec{t}_3$, hence, $\vec{t}_3(s)=\partial_s \vec{r}(s)$ and assume  perfect inextensibility, $|\partial_s \vec{r}(s)|=1$.
As a set of orthonormal vectors, the triad fully defines the material orientation of the chain~\cite{o1997computation,green1974theory}, and can be equivalently represented by a rotation vector $\vec{\Omega}(s)\equiv\{\Omega_1(s),\Omega_2(s),\Omega_3(s)\}$, with $\partial_s \vec{t}_i=(\bar{\Omega}\vec{t}_3+\vec{\Omega})\times\vec{t}_i$ (SM Section 2).
Here $\bar{\Omega}$ represents an intrinsic rotation of the triad orientation along the polymer chain and corresponds to an equilibrium twist of the polymer. The two components, $\Omega_1$ and $\Omega_2$ are related to bending modes~\cite{skoruppa2018bend} \footnote{Known as ``tilt'' and ``roll'' deformations in the context of dsDNA~\cite{skoruppa2018bend}} and $\Omega_3$ gives twist modes of deformation away from  $\bar{\Omega}$~\cite{spakowitz2006wormlike}. Further, any orientation of the triad can be represented in terms of Euler angles~\cite{whittaker1964treatise,edmonds1996angular} whose explicit expressions are given in the SM (Section 3). 

We study elastic deformations of the polymer by considering the energy that is quadratic in the strain integrated along the chain~\cite{landau2012theory,love2013treatise},
\begin{align}
\label{eq:Total_energy}
    \beta \mathcal{E} & = \frac{1}{2}\int_0^L ds \, \varepsilon(s) \\
    \varepsilon &=A_1\Omega_1^2+A_2\Omega_2^2+A_3 \left (\Omega_3-\bar{\Omega} \right )^2+2A_4\Omega_2\Omega_3 \nonumber
\end{align}
where $\beta$ is inverse temperature times Boltzmann's constant, $A_1$ and $A_2$ are  anisotropic bending moduli, $A_3$ is the modulus for twist, and $A_4$ the twist-bend coupling. Without loss of generality, we assume the polymer is symmetric with respect to inversion of its ends, so that the energy is invariant under $\Omega_1\to -\Omega_1$\cite{marko1994bending} and there is no coupling between $\Omega_1$ and $\Omega_3$ or, $\Omega_1$ and $\Omega_2$. 

Because the energy is harmonic, the statistical behavior of the polymer chain can be expressed in terms of a two point function, or Green's function.
The Green's function can be evaluated by summing over all polymer configurations that start and end at some fixed positions and orientations,  weighted by their Boltzmann weights~\cite{yamakawa2016helical}.
For our purposes, it is most convenient to evaluate this function in Fourier space, $\mathcal{G}(k, {\Omega}|{\Omega}_0; L)$, 
\begin{align}
    \label{eq:full_G_formal}
    \mathcal{G}(k, {\Omega}|{\Omega}_0; L) & =\int d\vec{R}  \int_{{{\Omega}_0}}^{{\Omega}} \mathcal{D}[{\Omega}(s)] e^{-\beta  \mathcal{E}[{\Omega}(s)]} e^{- i \vec{k} \cdot \vec{R} }
\end{align}
which defines a characteristic function for the probability that a chain with orientation ${\Omega}_0$ at $s=0$ will have an orientation ${\Omega}$ at $s = L$ with end-to-end vector $\vec{R}=\int_0^L ds \, \vec{t}_3(s)$ (\emph{cf.} Fig.~\ref{fig:schematic_fig1}).
The Fourier variable $k$ is oriented along the $z-$axis of the chain.
From the Feynman-Kac theorem \cite{moral2004feynman}, the corresponding diffusion, or Schrödinger equation for $\mathcal{G}(k, {\Omega}|{\Omega}_0; L)$ is given by, 
\begin{equation}\label{eq:GEQ}
\left(\partial_L - \varepsilon+i\vec{k}\cdot\vec{t}_3\right){{\mathcal{G}}}(\Vec{k}, {\Omega}|{\Omega}_0; L) = \delta(L)\delta({\Omega}-{\Omega}_0)
\end{equation}
subject to the normalization condition
\begin{equation}
\int d\, {\Omega} \, \mathcal{G}(0, {\Omega}|{\Omega}_0; L) =1
\end{equation}
when evaluated at $k=0$, as expected for a characteristic function. 

Equation \ref{eq:GEQ} is exactly solvable as an eigenfunction expansion in terms of the Wigner-$D$ functions~\cite{rose1995elementary,leader2001spin} $\mathcal{D}_l^{mj}({\Omega})$ with integer angular momentum variables, $l,m$ and $j$. 
This is most compactly represented by Laplace transforming the characteristic function with respect to $L$,
\begin{equation}
\tilde{\mathcal{G}}(k, {\Omega}|{\Omega}_0; p)= \int_0^\infty dL \, e^{-p L} \mathcal{G}(k, {\Omega}|{\Omega}_0; L)
\label{eq:laplace_green}.
\end{equation}
Using the transformation of Eq.~\ref{eq:laplace_green}, the solution to Eq.~\ref{eq:GEQ} is given by
\begin{equation}
    \tilde{{\mathcal{G}}}(\Vec{k}, {\Omega}|{\Omega}_0; p) = 
    \sum_{l,l',m,j}
    %\sum_{l_0, l_f = 0}^{\infty} \sum_{m,j=-\text{min}(l_0,l_f)}^{\text{min}(l_0,l_f)} 
    \mathcal{D}_{l}^{mj} ({\Omega})
    \mathcal{D}_{l'}^{m j^{*}} ({\Omega}_{0}) \mathfrak{G}_{l l'}^{mj}(k,p)
    \label{eq:G_full_final} 
\end{equation}
with expansion coefficients $\mathfrak{G}_{l l'}^{mj}$ derived from Stone-Fence diagrammatic methods~\cite{chandrasekhar1943stochastic,yamakawa2016helical}
and the sums have limits $\{l,l'\} \in[0,\infty)$ and $ \{m,j\}\in [-\mathrm{min}(l,l'), \mathrm{min}(l,l')]$. 
The Wigner functions and explicit expressions for $\mathfrak{G}_{l l'}^{mj}$ are in the SM (Section 4). 

\begin{figure}[t]
     \centering
     \includegraphics[width=7.5cm]{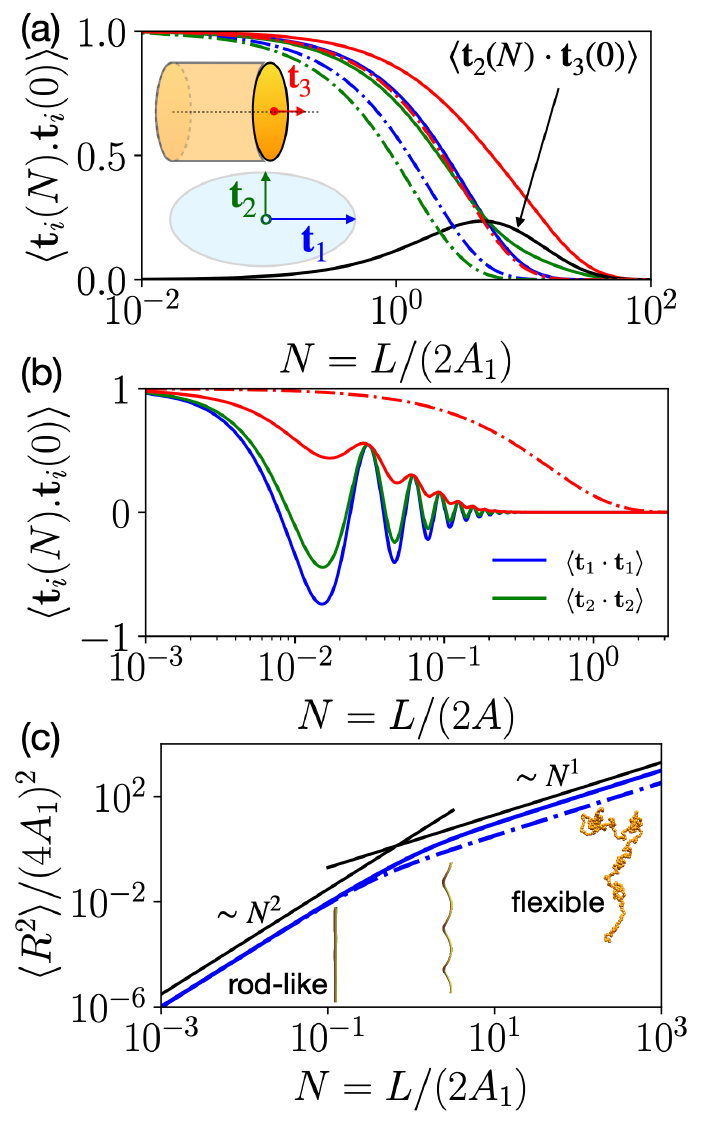}
     \caption{
     %\textbf{Structural Properties of Coupled HWLC:} 
     (a) Tangent-tangent correlations $\langle \vec{t}_i(N).\vec{t}_i(0)$ for $i=1 \text{ (orange)}, 2 \text{ (blue) 
     and}, 3 \text{ (green)}$. Inset shows the three tangent directions considering an infinitesimal polymer as an elliptic cylinder. (b) Tangent correlations for a polymer with energy functional given as, Eq.~\ref{eq:Total_energy} with $A_1\!\!=\!\!A_2\!=50\text{nm}, A_3\!\!=\!\!100\text{nm}$ and  $A_4\!=\!25\text{nm}$ with $2A_1\bar{\Omega}=\pi/2$. Dash-dotted curve shows the $\vec{t}_3$-tangent correlations for the exact same parameters with $A_4\!\!=\!0$. (c) Mean squared end-to-end distance is shown as a function of chain length. Dash-dotted line represents $A_4\!\!\to\! 0$. Schematic of rod-like ($\alpha=2$), intermediate ($1\!<\!\alpha\!<\!2$) and flexible ($\alpha=1$) configurations are shown that follow different scaling laws ($\langle R^2 \rangle\!\!\propto \!N^\alpha$). Plot (a) and (c) are for the parameters $A_1=10, A_2=2, A_3=1, \text{and }A_4=0.7$.}
     \label{fig:fig2}
     \vspace{-10 pt}
 \end{figure}

The full Green's function %cannot be simplified further, but its reduced representations can, and these encode 
can be used to evaluate the structural and mechanical properties from its reduced representations.
To that end, the orientation only Green's function, defined by integrating over $R$,  is given by evaluating ${\tilde{\mathcal{G}}}(\Vec{k}, {\Omega}|{\Omega}_0; p)$ at $k=0$ and performing an inverse Laplace transform ~\cite{rose1995elementary,leader2001spin}. 
%It has a expansion,
%\begin{align}
%    \label{eq: G_orient}
%    \mathcal{G}_\mathrm{O}({\Omega}|{\Omega}_{0}; L) \!=\! \sum_{l=0}^\infty \! \sum_{m,j,j'=-l}^{l}
%    \!\!\!\!\mathcal{D}_{l}^{mj} ({\Omega})
%    \mathcal{D}_{l}^{m j'^{*}} ({\Omega}_{0})
%    g_{l}^{jj'}(L)
%\end{align}
%where, $g_{l}^{jj'}(L)$ are  expansion coefficients.
The orientation only Green's function allows us to calculate tangent correlation functions (SM Section 4).
The decay of tangent correlations as a function of the distance between two points along the polymer backbone are determined by the chain's rigidity. For the twist-bend coupled helical wormlike chain, for $\bar{\Omega}=0$, the principle tangent correlation averaged at fixed temperature is
\begin{align}
\langle \vec{t}_{3}(L)\cdot \vec{t}_{3}(0) \rangle = & \exp{\left(\frac{1}{4 A_1} \left(\Delta _{12}+2 \Delta _{13}-4\right) L\right)} \\
& \left[\cosh \left(\frac{\Delta  L}{4 A_1}\right)-\frac{\Delta _{12}}{\Delta} \sinh \left(\frac{\Delta L}{4 A_1}\right)\right] \nonumber
\end{align}
where 
%$\Delta_{1(2,3)}\text{and } \Delta$ are functions of $\{A_i\}$'s and are explicitly provided in SM Section 4. 
$\Delta_{1(2,3)} = 1+A_1A_{(2,3)} (2C_-+C_+)/C_-^2, \Delta_{14} = 4 A_1A_4^3/C_-^2, \Delta=\sqrt{\Delta _{14}^2+\Delta_{12}^2}$ and $C_\pm  = A_4^2 \pm A_2 A_3 $. 
 Figure~\ref{fig:fig2}(a) shows tangent correlations for $\langle \vec{t}_i(L) \cdot \vec{t}_i(0)\rangle$ for $i\!=\!1,2,3$ as a function of dimensionless chain length for $\{A_1,A_2,A_3,A_4\}\!\equiv\!\{10,2,1,0.7\}$.
As can be seen in Fig.~\ref{fig:fig2}(a), the slowest decaying correlation corresponds to the body tangent $\vec{t}_3$.
%The primary correlation lengthscale can be identified as the slowest decaying correlation and corresponds to the $\vec{t}_3$ direction. 
In the presence of twist-bend coupling, the characteristic correlation length of the chain is longer than an equivalent chain with $A_4=0$. For the model we consider, only $\langle \vec{t}_2(L)\!\cdot\!\vec{t}_3(0)\rangle\ne 0$ among all possible cross correlations.
This physically stems from $A_4\ne 0$ that couples $\Omega_2$ with $\Omega_3$.
For this model we find the correlation to be symmetric such that, $\langle \vec{t}_2(L)\!\cdot\!\vec{t}_3(0)\rangle \!=\! \langle \vec{t}_3(L)\!\cdot\!\vec{t}_2(0)\rangle$.
As shown in Fig.~\ref{fig:fig2}(a),  for $A_4=0$ these cross correlations vanish, and as $L\to 0$, the cross correlation goes to zero. 
This is because, for a purely circular disc bend and twist cannot be coupled to one another.

 In the absence of twist-bend coupling  with $A_1=A_2=A$ (\emph{i.e.} symmetric bending rigidity), $A_4=0$, but with finite intrinsic twist, $\bar{\Omega}>0$, the correlations have compact forms. For example $\langle \vec{t_3}(L)\cdot\vec{t}_3(0)\rangle  =\exp{(-L/A)}$, the same as the WLC.
 In this limit of a symmetric twisted chain, the other tangent correlations are oscillatory, for $i=1,2$, \emph{i.e.},
  \begin{align}\label{corr12}
 \langle \vec{t_i}(L)\cdot\vec{t}_i(0)\rangle  =e^{ -L(A_3+A)/2 A A_3} \cos \left ( \bar{\Omega} L\right )
\end{align}
where the period of oscillation is $\bar{\Omega}$.
These results implies that in the absence of twist-bend coupling, the principal tangent correlation decays as a simple exponential, oscillations are expected for $\vec{t}_1$ and $\vec{t}_2$ with the period given by the natural twist of the polymer chain.
This behavior is illustrated in Fig.~\ref{fig:fig2}(b) as a function of the dimensionless chain length.
Also shown is the case of $A_4>0, \bar{\Omega}>0$, which is computed numerically, and for which the principle tangent correlation function is also oscillatory.  
It is interesting to note that simulation studies of model coupled developable ribbons~\cite{giomi2010statistical,yong2022statistics,freddi2016corrected} and conjugated polymers~\cite{root2017modelling} show this qualitative behavior in the principle tangent correlation function, although in these studies the origin of the oscillations are mis-attributed to intrinsic twist in the absence of twist-bend coupling. An origin that we now rule our with our exact expression.

Using the exact solution for the Greens' function, we may also calculate the end-to-end distance of the chain, which can be obtained as an integral of the principle tangent correlation function, 
\begin{align}
    \label{end_to_end}
    \langle {R}^2 (N)\rangle & =  \int_0^L ds_2 \int_0^L ds_1 \langle \vec{t}_3(s_2) \cdot \vec{t}_3(s_1)\rangle \, .
\end{align}
 For the case of $\bar{\Omega}=0$, $ \langle {R}^2 (N)\rangle $ can be evaluated exactly% though its expression is complex 
 (SM Section 4), and is plotted in Fig.~\ref{fig:fig2}(c).  
Asymptotic forms %$L/2 A_1 \ll 1$, 
for the end-to-end distance are given by, %the expansion
\begin{align}
\langle {R}^2 (L)\rangle \approx \begin{cases} 
      \displaystyle L^2+\frac{1}{3} \left(\Delta _{13}-2\right) L^3 & \!\!{L}/{2A_1} \!\ll\! 1 \\
      \displaystyle \frac{8 \Delta_{123}}{\Delta _{14}^2-4 \left(\Delta _{13}-2\right) \Delta_{123}}L & \!\!{L}/{2A_1} \!\gg\! 1 
   \end{cases}
%&\approx L^2+\frac{1}{3} \left(\Delta _{13}-2\right) L^3
%&+\frac{1}{48} \left(4 \Delta _{13}^2-16 \Delta _{13}+\Delta _{14}^2+16\right) L^4 \nonumber
\end{align}
% while asymptotically for $L/2 A_1 \gg 1$ 
% \begin{align}
% \langle \vec{R}^2 (L)\rangle  \sim \frac{8 \left(2-\Delta _{12}-\Delta _{13}\right)}{4 \left(\Delta _{13}-2\right) \left(\Delta _{12}+\Delta _{13}-2\right)-\Delta _{14}^2}L
% \end{align}
where, $\Delta_{123}=\left(\Delta _{12}+\Delta _{13}-2\right)$.
The results for the end-to-end distance are identical for short chain lengths (\emph{i.e.} $L/2A_1\ll 1$) with or without twist-bend coupling term and both behave as rigid rods.
% \begin{align}
%     \langle R^2\rangle & =N^2+\frac{N^3}{3} \left(\Delta _{13}-2\right) \nonumber \\ 
%     & \hspace{0.4 in}+\frac{N^4}{48} \big(4 \Delta _{13}^2-16 \Delta _{13}  +\Delta _{14}^2+16\big)+\mathcal{O}\left(N^5\right).
% \end{align}
%where, $\Delta_{13}=\left({A_1} {A_3} (3 {A_4}^2-{A_2} {A_3})/\left({A_4}^2-{A_2} {A_3}\right)^2+1\right)$ 
%(hence, $\displaystyle\lim_{A_4\to 0}\Delta_{13}=1-A_1/A_2$) 
%and $\Delta_{14}=4 {A_1} {A_4}^3/\left({A_4}^2-{A_2} {A_3}\right)^2$. 
%(hence, $\displaystyle\lim_{A_4\to 0}\Delta_{14}=0$).
Deviations due to $A_4\ne 0$ become apparent at cubic order in $L$ through intrinsic dependence of $\Delta_{13}$ on $A_4$, increasing the end-to-end distances like $A_4^2$ for small $A_4$.
At large chain lengths, all tangent correlations of the polymer de-correlate from one another and we obtain random walk like behavior, \emph{i.e.} $\langle R^2\rangle\!\sim\!L$, albeit with a higher amplitude for $A_4\ne 0$ compared to $A_4=0$. 
The cross-over between the two scaling regimes occurs at larger $L$ for the twist-bend coupled chain compared to the WLC, due to the larger correlation lengths in the former.

\begin{figure}[t]
     \centering
     \includegraphics[width=8.0cm]{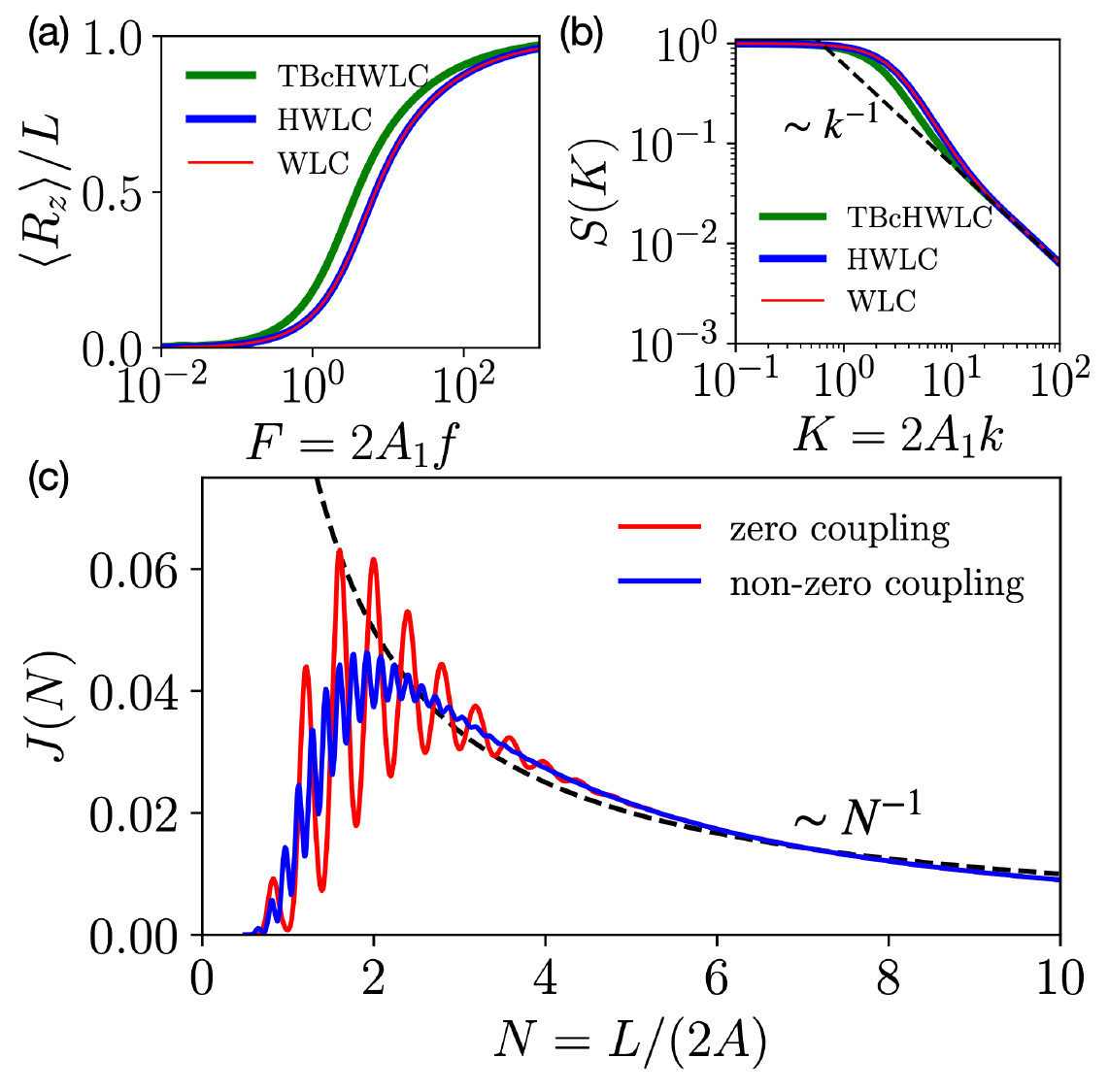}
     \caption{(a) Average extension along the applied force direction, $\langle R_z/L\rangle$ (Eq.~\ref{eq:extension}) is plotted as a function for applied force for the twist-bend coupled chain for the same set of parameters with $N=5$. The limits of HWLC and WLC can be obtained by considering $A_4=0$ and $(A_1=A_2)\wedge(A_3=A_4=0)$. (b) Plot shows the static structure factor of the polymer chain that shows the characteristic $\sim k^{-1}$ scaling behavior of WLC. (c) J-factor(Eq.~\ref{eq:J_factor}) or, polymer cyclization probability as a function of chain length for $A\!\!=\!\!50\text{nm}, \text{ } A_3\!\!=\!\!100\text{nm}$ and $A_4\!=\!30\text{nm}$(blue) or, $A_4\!=\!0\text{nm}$ (red) with $2A\bar{\Omega}=5\pi$.}
     \label{fig:new3}
     \vspace{-10 pt}
 \end{figure}

While the orientational Green's function contains information on tangent correlations, similarly useful information is encoded in the positional Green's function. In Fourier-Laplace space, we can evaluate the positional Green's function, $\tilde{\mathcal{G}}_\mathrm{P}(\Vec{k}; p)$, by integrating over initial and final orientations of the polymer chain, which is equivalent to including only the $l\!\!=\!l'\!=\!\!0$ term in Eq.~\ref{eq:G_full_final},  $\tilde{\mathcal{G}}_\mathrm{P}(\Vec{k}; p)=\mathfrak{G}_{00}^{00}(k,p)$. The partition function of the elastic polymer chain subject to a tensile force $\vec{f}$ acting on the end-to-end distance along the $z$ direction ${R}_z$ is isomorphic to the positional Green's function, subject to the transformation $i\vec{k}\leftrightarrow\beta\vec{f}$.
From this, we can estimate the average extension along the applied tensile force direction as (SM Section 4),
\begin{align} \label{eq:extension}
    \langle R_z\rangle &= %2 A_1
    \frac{\partial  }{\partial \beta f}\ln \left \{ \mathcal{L}^{-1} \left [ {\tilde{\mathcal{G}}}_\mathrm{P}(-i\beta \vec{f};p) \right ] \right \}\\
    &\approx \frac{2 \left(e^{-(2-\Delta_{13}) L}-1 + (2-\Delta_{13}) L\right)}{ 3(2 - \Delta_{13})^2} \beta f +\mathcal{O}(f^2), \nonumber
    % &\approx \frac{16 {A_1} \left(e^{-(2-\Delta_{13}) L}-e^{\Delta_{13} L/4}\right)}{\sqrt{3} (3 \Delta_{13}-8)} \beta f +\mathcal{O}(f^2) \nonumber
\end{align}
where, $\mathcal{L}^{-1}[\cdot]$ indicates the inverse Laplace transform, and in the limit of small forces we define an effective renormalization of the mechanical response due to twist-bend coupling. 
The full force-extension curve is shown in Fig.~\ref{fig:new3}(a), and compared to the WLC. The qualitative dependence of $\langle R_z\rangle$ on the applied force in both cases is similar, but the 
%{\color{red} 
effect of twist-bend coupling is to soften the chain at intermediate range of forces, allowing for large extensions for the same applied force compared to the WLC.
This suggests that the bending constants previously inferred from such mechanical response~\cite{bustamante2003ten,bustamante1994entropic} may need to be reevaluated using models with explicit twist-bend coupling.
%}

The positional Green's function also determines the static structure factor for the polymer chain, defined as
\begin{align}
    S(k) & =\frac{1}{L^2}\int_0^L \!ds_1 \int_0^L \!ds_2 \langle e^{i\vec{k}\cdot[\vec{r}(s_1)-\vec{r}(s_2)]}\rangle \nonumber \\
    & = \frac{8 A_1^2}{L^2}\mathcal{L}^{-1}\left[\frac{\tilde{\mathcal{G}}_\mathrm{P}(\vec{k};p)}{p^2}\right]
\end{align}
and shown in Fig.~\ref{fig:new3}(b).
For large wavevectors, $k\!\gg\!1$ and near $k=0$, the twist-bend coupled chain is identical to the WLC structure factor.
However, at intermediate lengthscales, they deviate from each other, an indication of the additional correlation lengths induced by twist-bend coupling, which may be discernible from small angle x-ray scattering experiments. 

%The qualitative behavior the force extension curves remain the same for the various models considered, however they differ quantitatively.
%Moreover, intuitively we obtain that quantitatively helical WLC and WLC behave the same since pure twist deformations do not store a length.
%Twist-bend coupling seems to alleviate some of the twist induced tension of the polymer chain that is responsible for higher extension along the applied force direction than in absence of coupling. Similar observations remain true for static structure factor of the polymer chain shown in Fig.~\ref{fig:new2}(d), where structure of the polymer remains unaltered doe to twisting (HWLC) but qualitative differences emerge in presence of coupling.

Finally, the full Green's function allows us to consider the problem of cyclization or looping probability, which in the context of Jacobson-Stockmayer theory~\cite{jacobson1950intramolecular,yamakawa2016helical} is known as the $J$-factor and calculated as~\cite{yamakawa2016helical,spakowitz2006wormlike,shimada1984ring},
\begin{align}
    \vspace{-10 pt}
    J(L) & = %\! 8\pi^2 \mathcal{G}(\Vec{R}=\vec{0}, \hat{\Omega}_N=\hat{\Omega}_0|\hat{\Omega}_0; N) \nonumber \\
    %& = 
    \frac{1}{2\pi^2}\int_0^\infty dk k^2 \left[\sum_{l=0}^\infty\sum_{m,j=-l}^{l}\mathcal{L}^{-1}\left\{ \mathfrak{G}_{ll}^{mj}(k,p)\right\}\right]
    \label{eq:J_factor}
    \vspace{-10 pt}
\end{align}
Conceptually, the $J$-factor is the conditional probability that the chain ends have the same orientation and a zero distance between them, so that the inverse Fourier transform of Eq. \ref{eq:full_G_formal} is evaluated at $R=0$. 
Note that while $\tilde{{\mathcal{G}}}_\mathrm{P}(\Vec{k}; p)=\mathfrak{G}_{00}^{00}(k,p)$, the sum over all indices in Eq.~\ref{eq:J_factor} conceptually means contributions arising from various topological isomers of a circular structure with different linking numbers. 
In Fig.~\ref{fig:new3}(c), we plot the cyclization probability as a function of chain length.
In both short and long chain limits the cyclization probability approaches zero due to energetic cost of bending a short chain and entropic cost of bringing two distant points with correct orientations together.
In absence of twist-bend coupling we observe the chains that contain an integer number of helical pitches (or, turns) tend to cyclize with a higher probability.
Twist-bend coupling alleviates some of this tension by changing both the required pitch and the amplitude of cyclization probability.
For a symmetric chain, $A_1\!=\!A_2\!=\!A$, the period of oscillation for $A_4=0$ is $\bar{\Omega}$, as found for the frequency in the correlation functions, Eq.~\ref{corr12}, while this is renormalized by finite $A_4$ to $(A A_3\bar{\Omega})/(A A_3 - A_4^2)$.  For larger chains, $N\!>\!5$ cyclization probability decreases roughly as $\propto N^{-1}$ indicating a search probability of one bead out of $N$ beads in a discrete polymer. The analysis shows, twist-bend coupling modifies the cyclization probability for small as well as moderate chain lengths.

%\emph{\label{sec:Conclusion}Conclusion.-} 
In summary, we have developed an exact analytical solution for the Green's function  of the twist-bend coupled helical wormlike chain model, which can be considered as the most generalizable elastic wire model with arbitrary cross couplings.
The utility of the full solution is analyzed in the context of experimentally realizable force extension curves, static structure factors and the looping probabilities. %while direct comparisons of correlation functions can be made with more detailed models computationally.
The study can be applied to compare and understand magnitude of anisotropy in dsDNA due to presence of major and minor grooves by comparing with force extension data and twist, bend and tension mediated coupling along parts of dsDNA.
Furthermore, the methods in this study would be useful in obtaining the exact solution for the twist waves for circular DNA chains~\cite{skoruppa2018bend} as well as to elucidate the importance of coupling for both twist and force induced response of biopolymers.
Further analysis would be required to decipher the nature of twist-bend coupling in determining the structure and properties of highly bent structures such as dsDNA wrapping around histone.

\begin{acknowledgments}

{\noindent A.G. and D.T.L. acknowledge
the support of the U.S. Department of Energy, Office of Science, Office of Advanced Scientific Computing
Research, and Office of Basic Energy Sciences, via the
Scientific Discovery through Advanced Computing (SciDAC) program.
K.K.M is supported
by Director, Office of Science, Office of Basic Energy Sciences, of the U.S. Department of Energy under contract No. DEAC02-05CH11231.
A.G. thanks Andy Spakowitz for conversations regarding aspects of the wormlike chain model.}
\end{acknowledgments}

% \appendix

% \section{Appendixes}

%\clearpage

% \section{A little more on appendixes}

\bibliographystyle{apsrev4-2}
%\nocite{*}
\bibliography{apssamp}% Produces the bibliography via BibTeX.

\end{document}

\maketitle
\begin{sloppypar}
\begin{center}
    \tableofcontents
\end{center}

\vspace{.25in}

\section{Overview of the Supplementary Material}
This Supplemental Material (SM) is an accompaniment to the main text, detailing all derivation steps for the theory of twisted wormlike chain model with twist-bend coupling proposed in this work.
The SM is written in a self contained manner, and as such, there are sections containing review material that are useful for derivation steps.
Here, we provide a brief overview of each section to allow readers to decide which section may be skipped or not:

% $\bullet$  In Section~\ref{Section1}, we provide an overview of the basics of wormlike chain models~\cite{yamakawa1997helical}. 
% We briefly review the formalism in terms of the Green function.
% The section is supposed to set the statistical mechanics language that we will use throughout this document and procedurally involves similar steps we shall use to solve for the twist-bend coupled helical wormlike chain.
% Readers who are already familiar with the WLC solutions can skip this section completely.

$\bullet$  In \textbf{Section~\ref{Section2}}, we provide a brief account of the differential geometry of space curves.
While the wormlike chain is completely defined in terms of the bending energy along it's body contour, defining a general class of elastic wire models require reference frames that go beyond the Frenet-Serret description.
We describe the generalized Frenet-Serret or, often called the Darboux frame that provides the natural frame to understand behavior of elastic wires.
The proper choice of frame lets us write the elastic energy in terms of strains away from the reference configuration of an elastic rod.
Eventually, the elastic rod energy is given in terms of rotations of the orthogonal reference frame that sets the initial configuration of the wire.

$\bullet$  The mathematical preliminaries needed for the solution of Green function of twist-bend coupled polymer elastic chain in summarized in \textbf{Section~\ref{Section3}}.
Specifically, a brief discussion on Euler angles, Wigner-D functions are given.
Utility of both Euler angles and Wigner functions are shown by considering to express the tangent vectors in terms of Wigner functions involving Euler angles.
Readers already familiar with the Euler angles and Wigner functions can directly skip to Subsection~\ref{subsec:4_3}.

$\bullet$ \textbf{Section~\ref{section:Green}} introduces the problem of twist-bend coupled helical wormlike chains and the solution methodology. 
%that uses tools and techniques discussed in the previous sections of the SM so far.
We systematically go from the Hamiltonian formalism of the energy of the system to writing down the path integral formulation and corresponding diffusion equation for the polymer chain propagator.
The limit of no `external potential' gives the free `particle' solution in analogy to quantum mechanical systems.
The utility of free particle solution is explained in terms of a few chain properties before we describe the method for solving for the full Green function with both spatial and angular dependence.
Finally, the utility of the full solution to the propagator is shown in terms of obtaining force extension behavior of the polymer, static structure factor as well as ring closure probability.
These considerations close our discussion of the model and contains the derivations to the central results of the article.

\begin{tcolorbox}[colback=green!2!white,colframe=green!75!black]
$\bullet$ In obtaining the solution to the coupled elastic chain, we realize a systematic procedure for the entire derivation can be written almost as an algorithm. 
These step by step procedures are summarized in the colored box as a $8$-step process that starts from the elastic energy of the system to the final steps of using the full propagator to understand static structure factor or, ring closure probability.
Interested readers who want to obtain the full solution to the Green function of any arbitrarily defined elastic energy could use these check points as the necessary steps to solve for the statistical mechanics of their custom defined elastic polymer chain (some unsolved cases so far could be, arbitrary reference curvature along the polymer backbone, existence of stretching and other coupling constants etc.).
\end{tcolorbox}

$\bullet$ \textbf{Section~\ref{sec:naturaA_3wist}} utilizes the procedure mentioned in the step-by-step guide to rewrite the Hamiltonian in appropriate form.
The specific energy function considered for this case is strictly relevant for double stranded DNA (dsDNA).
We assume isotropic bending resistance for dsDNA, a twist persistence length, corresponding natural twist for dsDNA and a twist-bend coupling to analyze our results.

$\bullet$ The final section, \textbf{Section~\ref{sec:general}} of this supplementary material generalizes the elastic polymer model to contain all possible coupling constants between bending and twist degrees of freedom.
Again, only the Hamiltonian is derived and written in terms of the angular momentum operators.
The rest of the steps are left as an open problem for curious and interested readers.

\newpage
\renewcommand{\theequation}{S2.\arabic{equation}}

\section{Formulation of the Helical Wormlike Chain \label{Section2}}
In this section, we describe the formulation of the model, discussing the choice of particular reference frame and how to properly formulate the deformation modes of an elastic wire.
We briefly overview the differential geometry of space curves, first in terms of the Frenet-Serret reference frame.
The discussion leads to the requirement of defining independent tangents not offered by Frenet-Serret frame.
Hence, the introduction of Darboux frame, or the generalized Frenet-Serret frame serves as the particular choice for studying biopolymers with helical structures.
\subsection{Differential Geometry of Space Curves: Frenet-Serret Frame}
\begin{figure}[h!]
    \centering
    \includegraphics[scale=0.40]{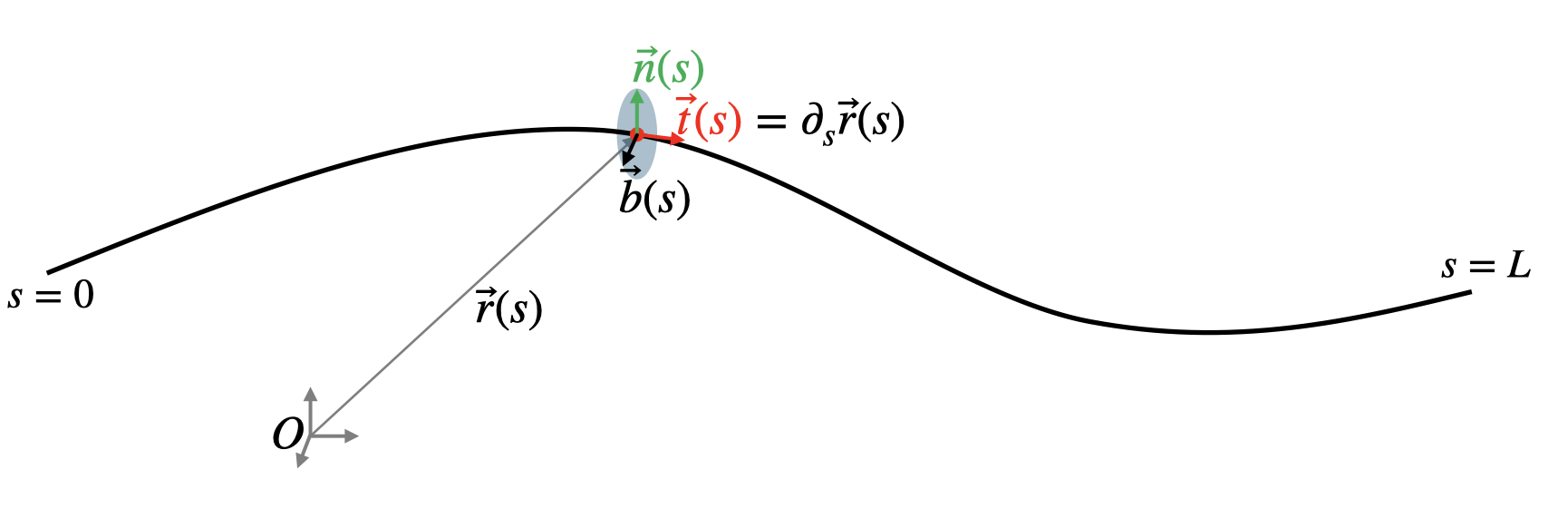}
    \caption{A continuous space curve with arc length parametrization $s\in [0,L]$. 
    The body tangent is defined as, $\vec{t}(s)=\partial_s \vec{r}(s)$.}
    \label{fig:rf_frenet}
\end{figure}

\noindent Let us consider a continuous space curve as shown in Fig.~\ref{fig:rf_frenet} parametrized by the arc length variable $s$ that runs from $s=0$ at one end, to, $s=L$ at the opposite end.
The principle tangent can be defined by the body tangent as, $\displaystyle \vec{t}(s)=\frac{\partial \vec{r}(s)}{\partial s}$ (same as that of $\vec{u}(s)$ in the context of WLC), or, more concretely,
\begin{align}
    \vec{t}(s)\equiv \hat{t}(s)=\frac{\partial \vec{r}(s)}{\partial s}=\lim_{\epsilon \to 0} \frac{\vec{r}(s+\epsilon)-\vec{r}(s)}{\epsilon}
\end{align}
Since, $\vec{r}(s+\epsilon)-\vec{r}(s)=\epsilon \vec{t}(s)$ in the limit $\epsilon \to 0$, $\vec{t}(s)$ is also known as simply the tangent at $s$ to the space curve.
The tangent can be constrained to be unit vector at every point along the curve that leads to inextensibility of the space curve.
We note,
\begin{align}
    \frac{\partial \vec{t}(s)}{\partial s}=\frac{\partial^2 \vec{r}(s)}{\partial s^2}
\end{align}
Since, $\vec{t}(s)\cdot \vec{t}(s)=1$, we have $\vec{t}(s)\cdot\partial_s\vec{t}(s)=0$, indicating, $\vec{t}(s)\!\perp\!\partial_s\vec{t}(s)$.
We can write the magnitude of $\partial_s\vec{t}(s)$ as, $|\partial_s\vec{t}(s)|=\kappa(s)$ and the direction of $\partial_s\vec{t}(s)$ as $\vec{n}(s)$, such that,
\begin{align}
    \partial_s\vec{t}(s)=\kappa(s)\vec{n}(s)
    \label{eq:FS_eq1}
\end{align}
The vector $\vec{n}(s)$ is known as the principal normal to the space curve at $s$ and the scalar $\kappa(s)$ is known as the curvature at $s$.
Finally another vector can be defined as the `bi-normal' at $s$ as,
\begin{align}
    \vec{b}(s)=\vec{t}(s)\times\vec{n}(s).
\end{align}
It should be noted that the set of vectors $\{\vec{t},\vec{n},\vec{b}\}$ at any $s$ for the space curve forms a orthonormal triad in the right-handed sense, since, $\vec{b}(s)\cdot (\vec{t}(s)\times\vec{n}(s))=0$.
Usually the set of triads is known as the Frenet-Serret frame of reference.
Eq.~\ref{eq:FS_eq1} is the first of Frenet-Serret set of equations.
Since, $\vec{b}(s)\!\cdot\! \vec{b}(s)=1$, we have ${\partial_s\vec{b}(s)}\cdot\vec{b}(s)=0$, indicating $\partial_s\vec{b}(s)$ has components along $\vec{t}(s)$ and $\vec{n}(s)$.
Starting from $\vec{b}(s)\cdot \vec{t}(s)=0$ and differentiating, we find,
\begin{align}
    & \partial_s \vec{b}(s)\cdot\vec{t}(s)+\partial_s \vec{t}(s)\cdot\vec{b}(s) =0 \\
    \text{or, } & \partial_s \vec{b}(s)\cdot\vec{t}(s)+ \kappa(s) \vec{n}(s)\cdot\vec{b}(s)=0 \nonumber\\
    \text{or, } & \partial_s \vec{b}(s)\cdot\vec{t}(s)=0
\end{align}
Hence, $\partial_s\vec{b}(s)$ has components only along $\vec{n}(s)$, or, $\partial_s\vec{b}(s)\parallel \vec{n}(s)$. 
We can write,
\begin{align}
    \partial_s\vec{b}(s)=-\tau(s) \vec{n}(s)
    \label{eq:FS_eq2}
\end{align}
where, $\tau(s)$ defines the torsion of the space curve at $s$ and the above Eq.~\ref{eq:FS_eq2} gives second of the Frenet-Serret relations.
Finally, we note,
\begin{align}
    \partial_s \vec{n}(s) & =\partial_s(\vec{b}(s)\times \vec{t}(s)) \nonumber \\
    & =  -\tau(s)\vec{n}(s)\times \vec{t}(s) + \vec{b}(s)\times \kappa(s)\vec{n}(s) \nonumber \\
    & = -\kappa(s)\vec{t}(s) + \tau(s)\vec{b}(s) \label{eq:FS_eq3}
\end{align}
The above Eq.~\ref{eq:FS_eq3} gives the third of Frenet-Serret relations.
The above construction is unique since given any set of functions $\kappa(s)$ and $\tau(s)$ for $s\in[0,L]$, the space curve is uniquely defined.
We can define a ``Darboux" vector for the Frenet-Serret frame as~\cite{o2019engineering}, $\vec{\Omega}_{\mathrm{SF}}(s)$ as,
\begin{align}
    \vec{\Omega}_{\mathrm{SF}}(s)=\kappa(s)\vec{b}(s)+\tau(s)\vec{t}(s).
\end{align}
Using the Darboux vector, we can easily write the Frenet-Serret relations as,
\begin{align}
    \partial_s \vec{\mathbf{x}}(s)=\vec{\Omega}_{\mathrm{SF}}(s)\times \vec{\mathbf{x}}(s) 
\end{align}
where, $\vec{\mathbf{x}}(s)\equiv \{\vec{t}(s),\vec{n}(s),\vec{b}(s)\}$.

\noindent The above description of thread-like space curves work really well for (essentially) $1d$ polymer chains and the curve can be defined by knowing the curvature $\kappa(s)$ and torsion, $\tau(s)$.
The reason the above description does not generalize to any arbitrary dimension is due to enslaving of the normal and binormal to tangent orientations at every arc-length parametrization, $s$.
\begin{figure}[h!]
    \centering
    \includegraphics[scale=0.50]{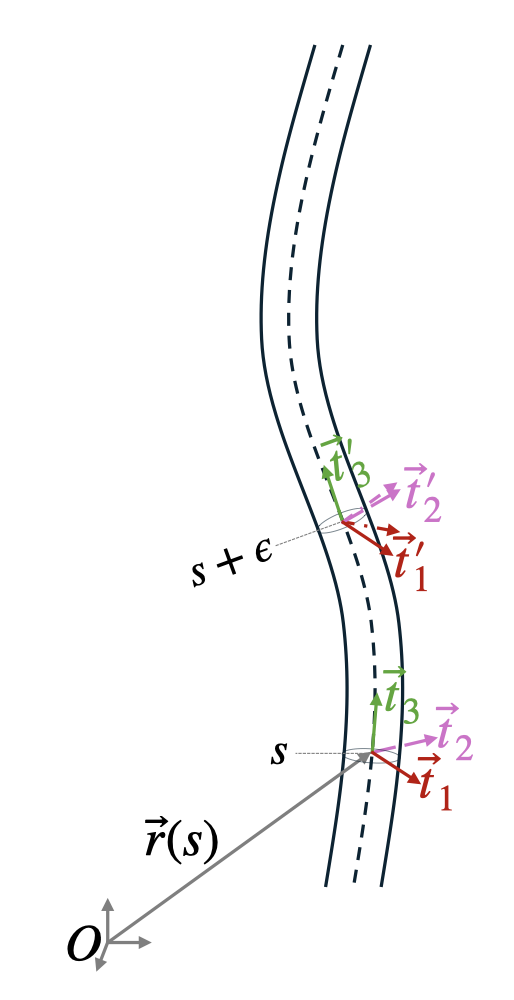}
    \caption{Local coordinate system or, the material frame of the curve $\Gamma(s)\equiv (\vec{r}(s);\{\vec{t}_i(s)\})$ parametrized in terms of the arc length variable $s$.}
    \label{fig:RF_hwlc}
\end{figure}
In essence once the space curve is complete parametrized in $3d$, the curvature and twist are uniquely defined since:
\begin{align}
    \kappa(s) & = \Bigg| \frac{d^2\vec{r}(s)}{ds^2}\Bigg| \\
    \tau(s) & = \frac{d\vec{r}(s)}{ds}\cdot \left(\frac{d^2\vec{r}(s)}{ds^2}\times \frac{d^3\vec{r}(s)}{ds^3}\right) \Bigg| \frac{d^2\vec{r}(s)}{ds^2}\Bigg|^{-2}
\end{align}

\subsection{Generalized Frenet-Serret or the Darboux Frame}
As introduced in the previous subsection, we consider the polymer path and a bend-able twist-able $3d$ cylinder or rod (Fig.~\ref{fig:RF_hwlc}) as an extension to the line element.
Considering a $3d$ cylinder of length $L$ and cross sectional area $a$ (not necessarily circular), we can associate two orthogonal vectors in the cross sectional plane such that they coincide with the principal axes of inertia at $s$.
This description is schematically presented in Fig.~\ref{fig:RF_hwlc} where the localized coordinate system at $s+\epsilon$ can be obtained by an infinitesimal rotation of the coordinate system at $s$ as~\cite{yamakawa1997helical},
\begin{align}
    \vec{\mathbf{t}}_i(s+\epsilon)= \vec{\mathbf{t}}_i(s)+\vec{{\delta \mathbf{\Omega}}}\times \vec{\mathbf{t}}_i(s)
\end{align}
The above description lets us write the angular velocity at which the reference frame rotates as,
\begin{align}
    \vec{\Omega}(s)=\lim_{\epsilon\to 0}\frac{\vec{{\delta \mathbf{\Omega}}}}{\epsilon}
\end{align}
where, the components of $\vec{\Omega}(s)$ are $\{\Omega_1(s),\Omega_2(s),\Omega_3(s)\}$.
%given as, $\Omega_\alpha(s)=\partial_s {\Omega}_\alpha(s)$.
The two equations above lets us write a generalized Frenet-Serret like equation, which is often known as the Darboux frame of reference as,
\begin{align}
    \partial_s\vec{\mathbf{t}}_i(s)&=\vec{\Omega}(s)\times \vec{\mathbf{t}}_i(s) \label{eq:d_frame}\\
    & =\epsilon_{ijk}\Omega_j(s)\vec{\mathbf{t}}_k(s)
\end{align}
where, $\epsilon_{ijk}$ is the Levi-Civita tensor.
The corresponding Darboux vector can be written as, $\displaystyle \vec{\Omega}(s)=\Omega_1(s)\vec{t}_1(s)+\Omega_2(s)\vec{t}_2(s)+\Omega_3(s)\vec{t}_3(s)$.
It is customary to align the $\vec{t}_3(s)$ direction to be along the body tangent to the center-line; hence, $\vec{t}_3(s)\equiv \vec{t}(s)=\partial_s \vec{r}(s)$.
We note,
\begin{align}
    (\partial_s\vec{\mathbf{t}}_i)\cdot(\partial_s\vec{\mathbf{t}}_j) &  = \epsilon_{imn}\Omega_m \epsilon_{jkl}\Omega_k\vec{\mathbf{t}}_n\cdot \vec{\mathbf{t}}_l=\epsilon_{imn}\epsilon_{jkn}\Omega_m\Omega_k \text{ (using $\vec{\mathbf{t}}_n\cdot \vec{\mathbf{t}}_l=\delta_{nl}$)}.
\end{align}
The above form helps us write the elastic energy in the next subsection.

\subsection{Elastic Energy}
\noindent Now, we have the reference frame needed to define our Hamiltonian for the chain, however, so far we have ignored any stretching of the line element. 
For a more general case a stretching scalar at any point along the polymer contour can be calculated as, $\displaystyle \gamma(s)=| \partial_s \vec{t}_3(s)|$.
With this reference frame, we can now define the elastic energy of a wire per unit length as quadratic in the deviations away from the rest energy as\footnote{\noindent A more general expression can be written by the Taylor series expansion given as,

$\displaystyle \beta \mathcal{E}=\int_0^L ds \left(\frac{1}{2}\sum_{kl}A_{kl}\Omega_k\Omega_l+\frac{1}{6}\sum_{klm}A_{klm}\Omega_k\Omega_l\Omega_m+\dots\right)$, where the matrices are symmetric under all possible permutations.}~\cite{yamakawa1997helical,marko1994bending},
\begin{align}
    \beta \mathcal{E} & = \frac{1}{2}\int_0^L ds \left(\mathbf{\Omega}(s) - \mathbf{\bar{\Omega}}(s)\right)^{T} [\mathbb{A}] \left(\mathbf{\Omega}(s) - \mathbf{\bar{\Omega}}(s)\right) \\
    & = \frac{1}{2}\int_0^L ds \sum_{k,l=1}^{4}\left({\Omega}_k(s) - {\bar{\Omega}_k}(s)\right) {A}_{kl} \left({\Omega}_l(s) - {\bar{\Omega}_l}(s)\right)
\end{align}
Here, $\mathbf{\Omega}(s) \equiv (\{ \Omega_i(s)\}, \gamma(s))$ defines the deformation $4-$vector and $\mathbf{\bar{\Omega}}(s)$ is the equilibrium conformation (or, the lowest energy) of the elastic wire.
$\mathbb{A}$ defines the amplitude of each deformation contribution, or, the element ${A}_{mn}$ correspond to the energy cost (or, modulus) associated with the quadratic element $\tilde{\Omega}_m \tilde{\Omega}_n$, where $\tilde{\Omega}\equiv (\Omega-\bar{\Omega})$.
Here, we can note that if the deformation $4$-vector contains only stretching energy, the corresponding model is known as Gaussian chain in the context of Flexible chain models (a continuous limit of Freely Jointed Chain (FJC)) with total energy given as $\displaystyle \beta E_{\text{flexible}}=\frac{k_s}{2}\int_0^L ds \gamma(s)^2$, where $k_s$ is the stretching modulus~\cite{doi1988theory}.

\noindent In the WLC model, we usually assert point-wise inextensibility as, $|\vec{t}_3(s)|=1$ for all $s\in [0,L]$.
Using Eq.~\ref{eq:d_frame} we can find, 
\begin{align}
    \partial_s \vec{t}_3(s) & = \Omega_1(s)\vec{t}_2(s)-\Omega_2(s)\vec{t}_1(s) \\
    \text{or, } (\partial_s \vec{t}_3(s))^2 & = \Omega_1(s)^2+\Omega_2(s)^2
\end{align}
Using these definitions, we could re-write the energy of WLC model as,
\begin{align}
    \beta \mathcal{E}_0=\frac{A}{2}\int_0^L ds \left(\frac{\partial\vec{t}_3(s)}{\partial s}\right)^2=\frac{A}{2}\int_0^L ds \left(\Omega_1(s)^2+\Omega_2(s)^2\right)
    \label{eq:hamiltonian_WLC_reqrite}
\end{align}
which means the deformation $3$-vector for the WLC model is,
\begin{align}
    \mathbb{A} = \begin{pNiceMatrix}[first-row,first-col]
 & \Omega_1 & \Omega_2 & \Omega_3 \\
 \Omega_1 & A/2 & 0 & 0  \\
 \Omega_2 & 0 & A/2 & 0  \\
 \Omega_3 & 0 & 0 & 0 
\end{pNiceMatrix}
\end{align}
where, we have ignored contributions from $\gamma(s)$ since we will not consider stretching in our work (hence, deformation is a $3$-vector).
We can also identify contributions relating to $\Omega_1(s)$ and $\Omega_2(s)$ as bending contributions given WLC contains only bending energy for the polymer with isotropic bending stiffness given by the persistence length of the polymer chain (as we have identified, the persistence length gives the principal tangent de-correlation lengthscale).

\noindent For a more general case of anisotropic polymer chain we will consider here, the modulus tensor can be generally written as,
\begin{align}
    \mathbb{A} = \begin{pNiceMatrix}[first-row,first-col]
 & \Omega_1 & \Omega_2 & \Omega_3 \\
 \Omega_1 & A_1 & A_{12} & A_{13}  \\
 \Omega_2 & A_{12} & A_2 & A_{23}  \\
 \Omega_3 & A_{13} & A_{23} & A_3
\end{pNiceMatrix}
\end{align}
The limit of $A_{ij}=0$ for $i\ne j$ corresponds to the Helical Wormlike Chain (HWLC) model that has no cross coupling between different deformation modes.
In particular, the deformation mode that corresponds to $\Omega_3$ is called the twist deformation.
Hence, in the HWLC, we have two modes of bending and one degree of twisting.
\begin{figure}[h!]
    \centering
    \includegraphics[scale=0.50]{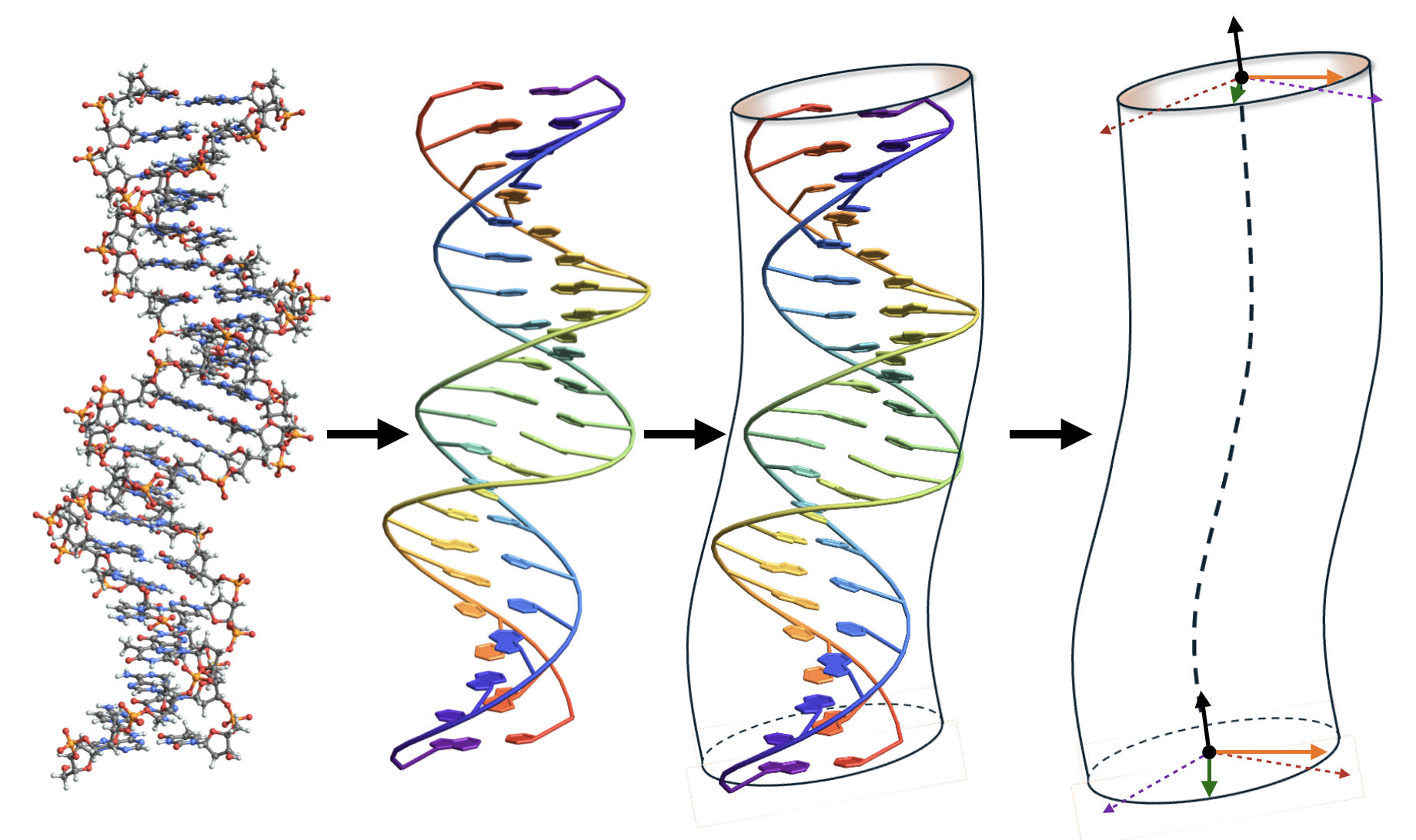}
    \caption{We schematically represent the coarse-graining for a ds-DNA to an effective twist-bent cylinder (or, rod). The tangent vectors (black, orange and green) represent the $\{\vec{t}_i\}$ at two ends of the cylinder. The two dashed arrows represent the location of two strands with respect to the center of the cylinder at fixed arc-length position(s). While the principal tangent follows the center-line curvature the other two tangents co-rotate around $\vec{t}_s$ continuously.}
    \label{fig:DNA_CG}
\end{figure}

% \noindent [{\color{red}Discuss in detail (logic/math) here why do we choose the specific A matrix given below:}]

\noindent For our case, that considers explicit twist-bend coupling on top of HWLC (that we call twist-bend coupled HWLC or, TBcHWLC)
\begin{align}
    \mathbb{A} = \begin{pNiceMatrix}
 A_1 & 0 & 0  \\
 0 & A_2 & A_4  \\
 0 & A_4 & A_3
\end{pNiceMatrix}
\end{align}
While, the specific matrix structure is highly relevant to dsDNA, in section~\ref{sec:general} we consider the case for the most general structure of the deformation matrix with all possible coupling(s) between twist and bend. 
Hence, for further development and to keep the discussion streamlined, we consider the following elastic energy of the wire~\cite{marko1994bending},
\begin{align}
    \beta \mathcal{E} & = \frac{1}{2}\int_0^L ds \left(A_1\Omega_1^2+ A_2\Omega_2^2+A_3\Omega_3^2+A_4(\Omega_2\Omega_3+\Omega_3\Omega_2)\right)
    \label{eq:elastic_energy}
\end{align}
where, $\Omega_\alpha\equiv \Omega_\alpha(s)$.

\newpage
\renewcommand{\theequation}{S3.\arabic{equation}}
\section{Mathematical Reviews \label{Section3}}
In this section we briefly review the necessary mathematical tools required to solve for the exact propagator of TBcHWLC.
Specially, we review on Euler angles~\cite{varshalovich1988quantum} and Wigner-D matrices~\cite{shiraishi2013probing} in the following subsections.
The utility of Wigner-D functions is further shown to represent the material triad $\{\vec{t}_1,\vec{t}_2,\vec{t}_3\}$ in terms of Wigner functions.
\subsection{Euler Angles}
According to Euler's rotation theorem, any rotation may be described using three angles. If the rotations are written in terms of rotation matrices $\mathbb{D}$,  $\mathbb{C}$, and $\mathbb{B}$, then a general rotation $\mathbb{A}$ can be written as
\begin{equation}
    \mathbb{A}= \mathbb{B C D}
\end{equation}
The three angles giving the three rotation matrices are called Euler angles.
\begin{figure}[h!]
    \centering
    \includegraphics[scale=0.50]{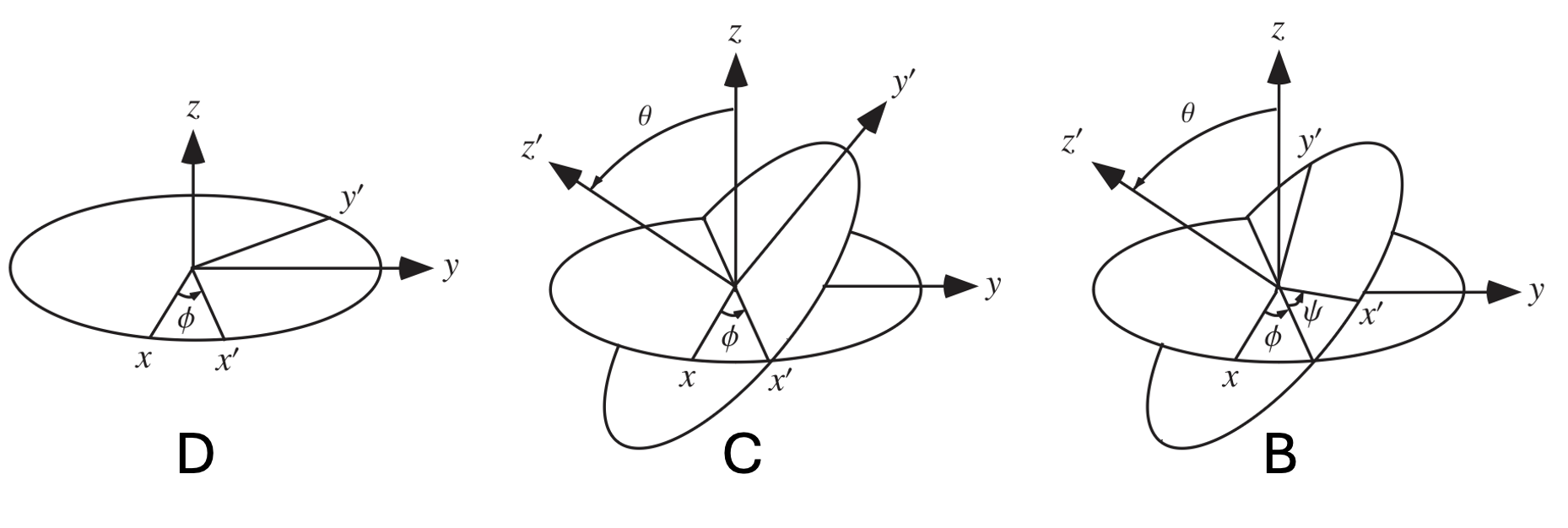}
    \caption{Euler angles in the ZXZ or the so called x-convention. For example, operation D in the figure involves rotating the xy plane around the z-axis by angle $\phi$.}
    \label{fig:enter-label}
\end{figure}
There are several conventions for Euler angles, depending on the axes about which the rotations are carried out.
In the so-called ``ZXZ-convention''
convention, the rotation given by Euler angles $(\phi,\theta,\psi)$, where

1. the first rotation is by an angle $\phi\in[0,2\pi]$ about the z-axis using  $\mathbb{D}$,

2. the second rotation is by an angle $\theta \in [0,\pi]$ about the former $x$-axis (now $x^\prime$) using $\mathbb{C}$, and,

3. the third rotation is by an angle $\psi\in[0,2\pi]$ about the former z-axis (now $z^\prime$) using $\mathbb{B}$.

\noindent We instead use the ``ZYZ-convention'', where the three material triads are given as elements of the $\mathbb{A}$ matrix as\footnote{Follows from equations (36)-(47) from \href{https://mathworld.wolfram.com/EulerAngles.html}{Wolfram Mathworld Euler Angles}},
\begin{align}
    \begin{pmatrix}
        \hat{t}_1\\
        \hat{t}_2\\
        \hat{t}_3
    \end{pmatrix}
    =
    \begin{pmatrix}
        \sin{\psi} & -\cos{\psi} & 0 \\
        \cos{\psi} &  \sin{\psi} & 0 \\
        0 & 0 & 1
    \end{pmatrix}
    \begin{pmatrix}
        1 & 0 & 0 \\
        0 & \cos{\theta} & \sin{\theta} \\
        0 & -\sin{\theta} & \cos{\theta}
    \end{pmatrix}
    \begin{pmatrix}
        -\sin{\phi} & \cos{\phi} & 0 \\
        -\cos{\phi} & -\sin{\phi} & 0 \\
        0 & 0 & 1
    \end{pmatrix}
\end{align}
or,
\begin{align}
    \hat{t}_1 & = (-\sin{\phi}\sin{\psi} + \cos{\theta}\cos{\phi}\cos{\psi}; \cos{\phi}\sin{\psi} + \cos{\theta}\sin{\phi}\cos{\psi};-\sin{\theta}\cos{\psi}) \\
    \hat{t}_2 & = (-\sin{\phi}\cos{\psi} - \cos{\theta}\cos{\phi}\sin{\psi}; \cos{\phi}\cos{\psi} - \cos{\theta}\sin{\phi}\sin{\psi};\sin{\theta}\sin{\psi})\\
    \hat{t}_3 & = (\sin{\theta}\cos{\phi} ; \sin{\theta}\sin{\phi};\cos{\theta})
\end{align}
The above equations also give (from generalized Frenet-Serret frame equations),
\begin{align}
\Omega_1 & = -(\sin{\theta}\cos{\psi})\partial_s{\phi} + \sin{\psi}\partial_s{\theta}\\
\Omega_2 & = (\sin{\theta}\sin{\psi})\partial_s{\phi} + (\cos{\psi})\partial_s{\theta}\\
\Omega_3 & = (\cos{\theta})\partial_s{\phi} + \partial_s{\psi}
\end{align}
An equivalent representation of the above set of equations in terms of discrete infinitesimal rotations are given as,
\begin{align}
    \begin{pmatrix}
        \delta \Omega_1 \\
        \delta \Omega_2 \\
        \delta \Omega_3
    \end{pmatrix}
    =
    \begin{pmatrix}
        -\sin{\theta}\cos{\psi}  & \sin{\psi} & 0 \\
        \sin{\theta}\sin{\psi} & \cos{\psi} & 0 \\
        \cos{\theta} & 0 & 1
    \end{pmatrix}
    \begin{pmatrix}
        \delta \phi \\
        \delta \theta \\
        \delta \psi
    \end{pmatrix}
\end{align}
By matrix inversion we find,
\begin{align}
    \begin{pmatrix}
        \delta \phi \\
        \delta \theta \\
        \delta \psi
    \end{pmatrix}
    =
    \begin{pmatrix}
       \displaystyle  -\frac{\cos{\psi}}{\sin{\theta}}  & \displaystyle \frac{\sin{\psi}}{\sin{\theta}} & 0 \\
        \sin{\psi} & \cos{\psi} & 0 \\
        \cos{\psi}\cot{\theta} & -\sin{\psi}\cot{\theta} & 1
    \end{pmatrix}
    \begin{pmatrix}
        \delta \Omega_1 \\
        \delta \Omega_2 \\
        \delta \Omega_3
    \end{pmatrix}
\end{align}
Using these, we can express the generalized momentum operators in terms of the 3 Euler angles as,
\begin{align}
    \hat{L}_{{i}} = \frac{\partial \phi}{\partial \Omega_i} \frac{\partial }{\partial \phi} + \frac{\partial \theta}{\partial \Omega_i} \frac{\partial }{\partial \theta} + \frac{\partial \psi}{\partial \Omega_i} \frac{\partial }{\partial \psi}
\end{align}
Evaluating the partial derivatives, we find the expressions as,
\begin{align}
    \hat{L}_1 & = -\frac{\cos{\psi}}{\sin{\theta}}\frac{\partial }{\partial \phi} +  \sin{\psi}\frac{\partial }{\partial \theta} + \cos{\psi}\cot{\theta}\frac{\partial }{\partial \psi} \\
    \hat{L}_2 & = \frac{\sin{\psi}}{\sin{\theta}}\frac{\partial }{\partial \phi} +  \cos{\psi}\frac{\partial }{\partial \theta} - \sin{\psi}\cot{\theta}\frac{\partial }{\partial \psi} \\
    \hat{L}_3 & = \frac{\partial }{\partial \psi}
\end{align}

\subsection{Wigner D-matrix}
The normalized Wigner functions used here $\mathcal{D}_l^{m,j}(\Omega)$ are given as~\cite{yamakawa1997helical},
\begin{equation}
    \mathcal{D}_l^{m,j}(\Omega) \equiv \mathcal{D}_l^{m,j}(\theta, \phi, \psi) = c_l e^{-im\phi} e^{-ij\psi} d_l^{m,j}(\theta)
\end{equation}
where, $\displaystyle c_l = \left(\frac{2l+1}{8\pi^2}\right)^{1/2}$,
with,
\begin{align}
     d_l^{m,j}(\theta) = \left[\frac{(l+j)!(l-j)!}{(l+m)!(l-m)!}\right]^{1/2} \left(\cos{\frac{1}{2}\theta}\right)^{j+m} \left(\sin{\frac{1}{2}\theta}\right)^{j-m} P_{l-j}^{j-m,j+m}(\cos{\theta}) \\
     P_{n}^{\alpha,\beta}(x) = \frac{(-1)^n}{2^n n!}(1-x)^{-\alpha} (1+x)^{-\beta} \frac{d^n}{dx^n}\left[(1-x)^{\alpha+n}(1+x)^{\beta+n}\right] \text{   } (\forall |x|\le 1)
\end{align}
Here, $P_{n}^{\alpha,\beta}(x)$ is known as the Jacobi polynomial.
In particular we have the following relation between Wigner D-functions and spherical harmonics as,
\begin{align}
    \mathcal{D}_l^{m,j=0}(\theta, \phi, \psi) = \frac{1}{\sqrt{2\pi}} (-1)^{(m+|m|)/2}Y_l^m(\theta,\phi) \\
    \mathcal{D}_l^{m=0,j}(\theta, \phi, \psi) = \frac{1}{\sqrt{2\pi}} (-1)^{(j-|j|)/2}Y_l^j(\theta,\psi)
\end{align}
We have the property of complex conjugation as~\cite{yamakawa1997helical},
\begin{align}
    \mathcal{D}_{l}^{{m,j}^\star} = (-1)^{m-j} \mathcal{D}_{l}^{-m,-j} \text{ or, } \mathcal{D}_{l}^{{m,j}} = (-1)^{m-j} \mathcal{D}_{l}^{-m,-j^\star}
\end{align}
The orthonormality of Wigner D functions are given as~\cite{yamakawa1997helical},
\begin{align}
    \int \mathcal{D}_{l_1}^{{m_1,j_1}^\star}(\Omega) \mathcal{D}_{l_2}^{m_2,j_2}(\Omega) d\Omega & = \delta_{l_1,l_2} \delta_{m_1,m_2} \delta_{j_1,j_2} \\
    \sum_{l=0}^\infty \sum_{m=-l}^{l} \sum_{j=-l}^{l} \mathcal{D}_{l}^{{m,j}^\star}(\Omega) \mathcal{D}_{l}^{m,j}(\Omega^\prime) & = \frac{1}{\sin{\theta}}\delta(\theta-\theta^\prime) \delta(\phi-\phi^\prime) \delta(\psi-\psi^\prime) \nonumber \\
    & = \delta(\Omega-\Omega^\prime)
\end{align}
Moreover, the product of two Wigner D functions can be written as a sum over a single Winger D function using the Wigner $3j$ symbol as~\cite{yamakawa1997helical},
\begin{equation}
    \mathcal{D}_{l_1}^{m_1,j_1} \mathcal{D}_{l_2}^{m_2,j_2}=8\pi^2\sum_{l_3=|l_1-l_2|}^{l_1+l_2} \sum_{m_3=-l_3}^{l_3} \sum_{j_3=-l_3}^{l_3}c_{l_1} c_{l_2} c_{l_3} \begin{pmatrix}
        l_1 & l_2 & l_3 \\
        m_1 & m_2 & m_3
    \end{pmatrix}
    \begin{pmatrix}
        l_1 & l_2 & l_3 \\
        j_1 & j_2 & j_3
    \end{pmatrix}
    \mathcal{D}_{l_3}^{{m_3,j_3}^\star} 
\end{equation}
The explicit expression can be found from elsewhere~\cite{goldsteinclassical,varshalovich1988quantum,yamakawa1997helical}.
The above expression also lets us write,
\begin{equation}
    \int \mathcal{D}_{l_1}^{m_1,j_1} \mathcal{D}_{l_2}^{m_2,j_2} \mathcal{D}_{l_3}^{m_3,j_3} d\Omega = 8\pi^2 c_{l_1} c_{l_2} c_{l_3} \begin{pmatrix}
        l_1 & l_2 & l_3 \\
        m_1 & m_2 & m_3
    \end{pmatrix}
    \begin{pmatrix}
        l_1 & l_2 & l_3 \\
        j_1 & j_2 & j_3
    \end{pmatrix}
\end{equation}

\subsection{Material Triad in Terms of Wigner D-matrix\label{subsec:4_3}} Here, we write the material triads in terms of the Wigner D-functions.
It is helpful to write the $d_l^{m,j}(\theta)$ functions explicitly for $l=1 \text{ and } 2$. To note the following symmetry holds, $d_l^{m,j}=(-1)^{j-m}d_l^{j,m}=d_l^{-j,-m}$.
\begin{align}
    d_1^{m,j}(\theta)= \begin{cases} 
      \cos{\theta} & m=0, j=0 \\
      \frac{1}{2}\left(1-\cos{\theta}\right) & m=1, j=-1 \\
      -\frac{1}{\sqrt{2}}\sin{\theta} & m=1, j=0 \\
      \frac{1}{2}\left(1+\cos{\theta}\right) & m=1, j=1 
   \end{cases}
\end{align}
\begin{align}
    d_2^{m,j}(\theta)= \begin{cases} 
      \frac{1}{2}\left(3\cos^2{\theta}-1\right) & m=0, j=0 \\
      \frac{1}{2}\left(-2\cos^2{\theta}+\cos{\theta}+1\right) & m=1, j=-1 \\
      -\sqrt{\frac{3}{8}}\sin{2\theta} & m=1, j=0 \\
      \frac{1}{2}\left(2\cos^2{\theta}+\cos{\theta}-1\right) & m=1, j=1 \\
      \frac{1}{4}(1-\cos{\theta})^2 & m=2, j=-2 \\
      -\frac{1}{2}\sin{\theta}(1-\cos{\theta}) & m=2, j=-1\\
      \sqrt{\frac{3}{8}}\sin^2{\theta} & m=2, j=0\\
      -\frac{1}{2}\sin{\theta}(1+\cos{\theta}) & m=2, j=1\\
      \frac{1}{4}(1+\cos{\theta})^2 & m=2, j=2
   \end{cases}
\end{align}
In the matrix notation we can write,
\begin{align}
    d_1^{m,j}(\theta) = \begin{pmatrix}
        \frac{1}{2}\left(1+\cos{\theta}\right) & \frac{1}{\sqrt{2}}\sin{\theta} & \frac{1}{2}\left(1-\cos{\theta}\right) \\
        -\frac{1}{\sqrt{2}}\sin{\theta} & \cos{\theta} & \frac{1}{\sqrt{2}}\sin{\theta}\\
        \frac{1}{2}\left(1-\cos{\theta}\right) & -\frac{1}{\sqrt{2}}\sin{\theta} & \frac{1}{2}\left(1+\cos{\theta}\right)
    \end{pmatrix}
\end{align}
Hence,
\begin{align}
    \mathcal{D}_1^{m,j}(\theta,\phi,\psi) = \sqrt{\frac{3}{8\pi^2}}\begin{pmatrix}
        \frac{1}{2}\left(1+\cos{\theta}\right)e^{i\phi}e^{i\psi} & \frac{1}{\sqrt{2}}\sin{\theta}e^{i\phi} & \frac{1}{2}\left(1-\cos{\theta}\right)e^{i\phi}e^{-i\psi} \\
        -\frac{1}{\sqrt{2}}\sin{\theta}e^{i\psi} & \cos{\theta} & \frac{1}{\sqrt{2}}\sin{\theta}e^{-i\psi}\\
        \frac{1}{2}\left(1-\cos{\theta}\right)e^{-i\phi}e^{i\psi} & -\frac{1}{\sqrt{2}}\sin{\theta}e^{-i\phi} & \frac{1}{2}\left(1+\cos{\theta}\right)e^{-i\phi}e^{-i\psi}
    \end{pmatrix}
\end{align}
The above development lets us write the following equalities,
\begin{align}
    t_3^{(3)}(\Omega) & = \cos{\theta}=\sqrt{\frac{8\pi^2}{3}}\mathcal{D}_1^{0,0}\\
    t_3^{(2)}(\Omega) & = \sin{\theta}\sin{\phi}=\sin{\theta}\left(\frac{e^{i\phi}-e^{-i\phi}}{2i}\right) = -i\sqrt{\frac{4\pi^2}{3}}\left(\mathcal{D}_1^{-1,0}+\mathcal{D}_1^{1,0}\right)\\
    t_3^{(1)}(\Omega) & = \sin{\theta}\cos{\phi}=\sin{\theta}\left(\frac{e^{i\phi}+e^{-i\phi}}{2}\right) = \sqrt{\frac{4\pi^2}{3}}\left(\mathcal{D}_1^{-1,0}-\mathcal{D}_1^{1,0}\right)\\
    t_2^{(3)}(\Omega) & = \sin{\theta}\sin{\psi}=\sin{\theta}\left(\frac{e^{i\psi}-e^{-i\psi}}{2i}\right) = i\sqrt{\frac{4\pi^2}{3}}\left(\mathcal{D}_1^{0,-1}+\mathcal{D}_1^{0,1}\right)\\
    t_1^{(3)}(\Omega) & = -\sin{\theta}\cos{\psi}=-\sin{\theta}\left(\frac{e^{i\psi}+e^{-i\psi}}{2}\right) = \sqrt{\frac{4\pi^2}{3}}\left(\mathcal{D}_1^{0,-1}-\mathcal{D}_1^{0,1}\right)
\end{align}
%% things so far has been thoroughly checked and verified.
The other remaining four terms in the material triads depend on $(\theta,\phi,\psi)$ and can be solved by writing down coupled equations and solving for the triads.
Specifically we find,
\begin{align}
    t_1^{(1)}(\Omega) + t_2^{(2)}(\Omega) = \sqrt{\frac{8\pi^2}{3}}
    \left(\mathcal{D}_1^{1,1}+\mathcal{D}_1^{-1,-1}\right)\\
    t_1^{(1)}(\Omega) - t_2^{(2)}(\Omega) = -\sqrt{\frac{8\pi^2}{3}}
    \left(\mathcal{D}_1^{1,-1}+\mathcal{D}_1^{-1,1}\right)
\end{align}
Hence, the above equations give us,
\begin{align}
     t_1^{(1)}(\Omega) = \frac{1}{2}\sqrt{\frac{8\pi^2}{3}}\left(\mathcal{D}_1^{1,1}+\mathcal{D}_1^{-1,-1}-\mathcal{D}_1^{1,-1}-\mathcal{D}_1^{-1,1}\right)\\
     t_2^{(2)}(\Omega) = \frac{1}{2}\sqrt{\frac{8\pi^2}{3}}\left(\mathcal{D}_1^{1,1}+\mathcal{D}_1^{-1,-1}+\mathcal{D}_1^{1,-1}+\mathcal{D}_1^{-1,1}\right)
\end{align}
The other set of coupled equations give,
\begin{align}
    t_1^{(2)}(\Omega) + t_2^{(1)}(\Omega) = i\sqrt{\frac{8\pi^2}{3}}
    \left(\mathcal{D}_1^{-1,1}-\mathcal{D}_1^{1,-1}\right)\\
    t_1^{(2)}(\Omega) - t_2^{(1)}(\Omega) = i\sqrt{\frac{8\pi^2}{3}}
    \left(\mathcal{D}_1^{1,1}-\mathcal{D}_1^{-1,-1}\right)
\end{align}
Hence,
\begin{align}
     t_1^{(2)}(\Omega) = \frac{i}{2}\sqrt{\frac{8\pi^2}{3}}\left(\mathcal{D}_1^{1,1}-\mathcal{D}_1^{-1,-1}+\mathcal{D}_1^{-1,1}-\mathcal{D}_1^{1,-1}\right)\\
     t_2^{(1)}(\Omega) = \frac{i}{2}\sqrt{\frac{8\pi^2}{3}}\left(-\mathcal{D}_1^{1,1}+\mathcal{D}_1^{-1,-1}+\mathcal{D}_1^{-1,1}-\mathcal{D}_1^{1,-1}\right)
\end{align}
This completes writing down the material triads in terms of Wigner-D functions with the explicit index of $l=1$.
In general we could write the following generic relation,
\begin{equation}
    t_i^{(\alpha)}(\Omega)= \sqrt{\frac{8\pi^2}{3}} \sum_{j,j^\prime=-1}^{1} \mathcal{C}_{i_\alpha}^{j,j^\prime} \mathcal{D}_1^{j,j^\prime}(\Omega)
\end{equation}
where, $\mathcal{C}_{i_\alpha}^{j,j^\prime}$ gives the corresponding coefficients related to the expansion of $t_i^{(\alpha)}(\Omega)$ in terms of Wigner-D functions. 

\newpage
\renewcommand{\theequation}{S4.\arabic{equation}}
\section{Green Function for Twist-Bend Coupled Helical Wormlike Chain or TBcHWLC \label{section:Green}}
\subsection{Path Integrals and Fokker-Planck Equation}
Let us start by defining the conditional probability distribution or the Green function for the polymer as follows~\cite{yamakawa1997helical},
\begin{align}
    \mathcal{G}(\vec{r},\Omega,s=L|\vec{r}_0,\Omega_0,s=0)\equiv \mathcal{G}(\vec{R};\Omega|\Omega_0;L)  & = \text{ Probability that a chain that begins at the origin} \nonumber \\
    & \text{at one end ($s=0$) with fixed orientation $\Omega_0$ will have} \nonumber \\
    & \text{the  other end ($s=L$) at position $\vec{R}$ with fixed} \nonumber \\
    & \text{orientation $\Omega$.}
\end{align}
The conditional probability can be normalized using the following relation,
\begin{align}
    \int \!d\vec{R}\int \!d\Omega \mathcal{G}(\vec{R};\Omega|\Omega_0;L) = 1.
\end{align}
A Fourier transform of the conditional distribution function can be defined as,
\begin{align}
    \hat{\mathcal{G}}(\vec{k};\Omega|\Omega_0;L)
    = \int\! d\vec{R} \exp{(i\vec{k}\cdot\vec{R})} \mathcal{G}(\vec{R};\Omega|\Omega_0;L).
\end{align}
Now, the Green function can be written as a path integral over angular variable paths with Boltzmann weighted elastic energy as,
\begin{align}
    \mathcal{G} 
(\vec{R}, \Omega|\Omega_0; L) 
= \int_{\Omega(s=0)=\Omega_0}^{\Omega(s=L)=\Omega} \mathcal{D}[\Omega(s)] \exp{\left(-\beta\int_0^L ds\mathcal{E}\right)} \delta\left(\vec{R}-\int_0^L ds \vec{t}_3(s)\right)
\end{align}
where, $\beta \mathcal{E}$ is given by Eq.~\ref{eq:elastic_energy}.
The delta function constraint ensures sum over body tangents along the polymer path gives the end-to-end vector of the polymer.
Using the Fourier transform allows us to write the Green function as a path integral over the Lagrangian density as,
\begin{align}
    \hat{\mathcal{G}}(\Vec{k}, \Omega|\Omega_0; L) = \int_{\Omega_0}^{\Omega} \mathcal{D}[\Omega(s)] \exp{\left(-\int_0^L ds \mathcal{L}\right)} 
\end{align}
where, the classical action is the integral over Lagrangian density and $\mathcal{L}$ can be explicitly written as,
\begin{align}
    \mathcal{L} = \beta \mathcal{E} + i\vec{k}\cdot\vec{t}_3.
\end{align}
Here, $i\vec{k}\cdot\vec{t}_3$ originates due to the presence of a constraint and can be thought of as an external potential bit.
From the Langrangian density, we can define the generalized momenta as,
\begin{equation}
    L_i = \frac{\partial \mathcal{L}}{\partial \Omega_i}
\end{equation}
More explicitly, we find,
\begin{align}
    L_1 & = A_1 \Omega_1 \\
    L_2 & = A_2 \Omega_2 + A_4 \Omega_3\\
    L_3 & = A_3 \Omega_3 + A_4 \Omega_2.
\end{align}
Since, the Hamiltonian density ($\mathcal{H}$) and Lagrangian are related by Legendre transforms, we can write,
\begin{align}
    \mathcal{H} & = \sum_{i=1}^{3} L_i\Omega_i - \mathcal{L} \\
    & = A_1\Omega_1^2 + A_2\Omega_2^2 + A_3\Omega_3^2 + A_4(\Omega_2\Omega_3+\Omega_3\Omega_2) - \frac{1}{2}\left(A_1 \Omega_{1}^{2}+A_2 \Omega_{2}^{2}+A_3 \Omega_{3}^{2}+A_4(\Omega_2\Omega_3+\Omega_3\Omega_2)\right) \nonumber \\
    & \hspace{2.9 in} - i\vec{k}\cdot\vec{t}_3 \nonumber \\
    & = \frac{1}{2}\left(A_1 \Omega_{1}^{2}+A_2 \Omega_{2}^{2}+A_3 \Omega_{3}^{2}+A_4(\Omega_2\Omega_3+\Omega_3\Omega_2)\right) -  i\vec{k}\cdot\vec{t}_3 \\
    & \equiv \mathcal{H}_0 - i\vec{k}\cdot\vec{t}_3.
\end{align}
The above Hamiltonian can be manipulated using the following relations,
\begin{align}
    \Omega_1 & = \frac{L_1}{A_1} \\
    \Omega_2 & = \frac{A_4L_3-A_3L_2}{A_4^2-A_2A_3}\\
    \Omega_3 & = \frac{A_4L_2-A_2L_3}{A_4^2-A_2A_3}.
\end{align}
To note, in the limit of $A_4\to 0$, hence no twist-bend coupling, the above relations reduce to, $\displaystyle \Omega_i = \frac{L_i}{A_i}$ for $i=1,2,3$.
The above manipulations gives the final form of Hamiltonian, without the external potential part as,
\begin{align}
    \mathcal{H}_0 = \frac{1}{2A_1}\left[L^2 - \Delta_{13}L_2^2 - \Delta_{12}L_3^2 -\frac{\Delta_{14}}{2}(L_2L_3+L_3L_2) \right]
\end{align}
where, $\displaystyle L^2 = \sum_{i=1}^3 L_i^2$ and
\begin{align}
    \Delta_{13} & = 1+\frac{A_1A_3}{C_m^2}(2C_m+C_p) \label{eq:delta_def1}\\
    \Delta_{12} & = 1+\frac{A_1A_2}{C_m^2}(2C_m+C_p) \label{eq:delta_def2}\\
    \Delta_{14} & = 4\frac{A_1A_4^3}{C_m^2} \label{eq:delta_def3}\\
    C_m & = A_4^2 - A_2 A_3 \label{eq:delta_def4}\\
    C_p & = A_4^2 + A_2 A_3 \label{eq:delta_def5}
\end{align}
In the limit of $A_4\to 0$, the Hamiltonian reduces to,
\begin{equation}
    \mathcal{H}_0 = \frac{1}{2A_1}\left[L^2 - \Delta_{12}^0 L_2^2 - \Delta_{13}^0 L_3^2 \right]
\end{equation}
where,
\begin{align}
    \Delta_{12}^0 = \lim_{A_4\to 0} \Delta_{13} = 1 - \frac{A_1}{A_2} \\
    \Delta_{13}^0 = \lim_{A_4\to 0} \Delta_{12} = 1 - \frac{A_1}{A_3}.
\end{align}
Given the full form of the Hamiltonian a corresponding Schrödinger equation can be written as,
\begin{equation}
    \left(\frac{\partial }{\partial L} - \mathcal{H}_0+i\Vec{k}\cdot \vec{t}_3\right)\hat{\mathcal{G}}(\Vec{k}, \Omega|\Omega_0; L) = 0
\end{equation}
with the initial condition, $\displaystyle \lim_{L\to 0}\hat{\mathcal{G}}(\Vec{k}, \Omega|\Omega_0; L)=\delta(\Omega-\Omega_0)$.
Correspondingly, we can introduce a dimensionless chain length variable as $N = L/(2A_1)$ and make the wave vector $\vec{k}$ dimensionless by multiplying it with the lengthscale $2A_1$ as $\vec{K}= 2A_1\vec{k}$ to rewrite the Schrödinger equation as,
\begin{equation}
    \left(\frac{\partial }{\partial N} - \mathcal{H}_0+i\Vec{K}\cdot \vec{t}_3\right)\hat{\mathcal{G}}(\Vec{K}, \Omega|\Omega_0; N) = 0
\end{equation}
where, the operator $\mathcal{H}_0$ is defined as,
\begin{equation}
    \mathcal{H}_0 = L^2 - \Delta_{13}L_2^2 - \Delta_{12}L_3^2 -\frac{\Delta_{14}}{2}(L_2L_3 + L_3L_2).
\end{equation}
The real space Green function automatically follows as,
\begin{equation}
    \left(\frac{\partial }{\partial N} - \mathcal{H}_0 - \vec{t}_3\cdot \vec{\nabla}_{\vec{R}}\right){\mathcal{G}}(\Vec{R}, \Omega|\Omega_0; N) = 0.
\end{equation}
In real space the initial conditions would satisfy, $\displaystyle \lim_{N\to 0}\hat{\mathcal{G}}(\Vec{R}, {\Omega}|\Omega_0; N)=\delta(\vec{R})\delta(\Omega-\Omega_0)$.

\subsection{Free Particle Green Function and the Solution}
The `free particle' Green function or, the so-called orientation-only Green function can be given as the integral over spatial degrees of freedom of the full Green function as,
\begin{align}
    {\mathcal{G}}_0(\Omega|\Omega_0; L) = \int d\vec{R}{\mathcal{G}}(\Vec{R}, \Omega|\Omega_0; L) \equiv {\mathcal{G}}(\Vec{K}=\vec{0}, \Omega|\Omega_0; L).
\end{align}
This means the orientation only Green function obeys,
\begin{equation}
    \left(\frac{\partial }{\partial L} - \mathcal{H}_0\right){\mathcal{G}}_0(\Omega|\Omega_0; L) = 0
\end{equation}
with the initial condition, $\displaystyle \lim_{L\to 0}\hat{\mathcal{G}}_0( {\Omega}|\Omega_0; L)=\delta(\Omega-\Omega_0)$.
This Green's function can generally be expanded in terms of Wigner's $\mathcal{D}$-functions and can be written as,
\begin{equation}
    \mathcal{G}_0(\Omega|\Omega_{0}; L) = \sum_{lmj} \sum_{l_{0}m_{0}j_{0}}
    \mathcal{D}_{l}^{mj} (\Omega)
    \mathcal{D}_{l_{0}}^{m_{0}j_{0}^{*}} (\Omega_{0})
    g_{l_{0}m_{0}j_{0}}^{lmj}(L)
\end{equation}
The boundary condition reduces the above expansions to terms satisfying, $\delta_{ll_0}\delta_{mm_0}$, hence giving,
\begin{equation}
    \mathcal{G}_0(\Omega|\Omega_{0}; L) = \sum_{l=0}^\infty \sum_{m=-l}^{l}\sum_{j, j_0 = -l}^{l}
    \mathcal{D}_{l}^{mj} (\Omega)
    \mathcal{D}_{l}^{m j_{0}^{*}} (\Omega_{0})
    g_{l}^{jj_0}(L).
\end{equation}
A few operator algebra of the angular momentum operators results gives the following relations for Wigner functions,
\begin{align}
    \hat{L}_1 |\mathcal{D}_{l}^{m,j}\rangle & = \frac{i}{2}c_l^j |\mathcal{D}_{l}^{m,j+1}\rangle + \frac{i}{2}c_l^{-j} |\mathcal{D}_{l}^{m,j-1}\rangle \\
    \hat{L}_2 |\mathcal{D}_{l}^{m,j}\rangle & = -\frac{1}{2}c_l^j |\mathcal{D}_{l}^{m,j+1}\rangle + \frac{1}{2}c_l^{-j} |\mathcal{D}_{l}^{m,j-1}\rangle \\
    \hat{L}_3 |\mathcal{D}_{l}^{m,j}\rangle & = i j  |\mathcal{D}_{l}^{m,j}\rangle 
\end{align}
where, the coefficient is given as,
\begin{equation}
    c_l^j = \left[l(l+1)-j(j+1)\right]^{1/2}
    \label{eq:c_coeff}.
\end{equation}
Using the above relations, we find,
\begin{align}
    \hat{L}_1^2 |\mathcal{D}_{l}^{m,j}\rangle & = -\frac{1}{4}c_l^j c_l^{j+1} |\mathcal{D}_{l}^{m,j+2}\rangle - \frac{1}{4}\left(c_l^{j} c_l^{-(j+1)}+c_l^{-j}c_l^{j-1}\right) |\mathcal{D}_{l}^{m,j}\rangle -\frac{1}{4}c_l^{-j} c_l^{-(j-1)} |\mathcal{D}_{l}^{m,j-2}\rangle \label{eq:p1_squared}\\
    \hat{L}_2^2 |\mathcal{D}_{l}^{m,j}\rangle & = \frac{1}{4}c_l^j c_l^{j+1} |\mathcal{D}_{l}^{m,j+2}\rangle - \frac{1}{4}\left(c_l^{j} c_l^{-(j+1)}+c_l^{-j}c_l^{j-1}\right) |\mathcal{D}_{l}^{m,j}\rangle +\frac{1}{4}c_l^{-j} c_l^{-(j-1)} |\mathcal{D}_{l}^{m,j-2}\rangle \label{eq:p2_squared}\\
    \hat{L}_3^2 |\mathcal{D}_{l}^{m,j}\rangle & = -j^2 |\mathcal{D}_{l}^{m,j}\rangle
    \label{eq:p3_squared}
\end{align}
From the above we can verify that,
\begin{align}
    \hat{L}^2 |\mathcal{D}_{l}^{m,j}\rangle \equiv (\hat{L}_1^2+\hat{L}_2^2+\hat{L}_3^2)|\mathcal{D}_{l}^{m,j}\rangle = -l(l+1) |\mathcal{D}_{l}^{m,j}\rangle.
\end{align}
Hence the $\hat{L}^2$ operator operates the same way for both spherical harmonics and Wigner-D functions (\emph{i.e.} the well-known eigenfunction relation $\hat{L}^2 Y_{l}^{m}=-l(l+1)Y_{l}^{m}$).
Another two relations that we need are,
\begin{align}
    \hat{L}_2 \hat{L}_3 |\mathcal{D}_{l}^{m,j}\rangle & = -\frac{i}{2}j c_l^{j}  |\mathcal{D}_{l}^{m,j+1}\rangle + \frac{i}{2}j c_l^{-j}  |\mathcal{D}_{l}^{m,j-1}\rangle \\
    \hat{L}_3 \hat{L}_2 |\mathcal{D}_{l}^{m,j}\rangle & = -\frac{i}{2}(j+1) c_l^{j}  |\mathcal{D}_{l}^{m,j+1}\rangle + \frac{i}{2}(j-1) c_l^{-j}  |\mathcal{D}_{l}^{m,j-1}\rangle.
\end{align}
Hence,
\begin{align}
    [\hat{L}_2, \hat{L}_3]_{+} |\mathcal{D}_{l}^{m,j}\rangle = -\frac{i}{2}(2j+1) c_l^{j}  |\mathcal{D}_{l}^{m,j+1}\rangle + \frac{i}{2}(2j-1) c_l^{-j}  |\mathcal{D}_{l}^{m,j-1}\rangle.
\end{align}
From the above equations we can note that the angular momentum operators couple $|\mathcal{D}_{l}^{m,j}\rangle$ to $|\mathcal{D}_{l}^{m,j}\rangle$, $|\mathcal{D}_{l}^{m,j\pm 1}\rangle$ and $|\mathcal{D}_{l}^{m,j\pm 2}\rangle$.
In absence of the twist-bend coupling terms, $|\mathcal{D}_{l}^{m,j}\rangle$ does not couple with $|\mathcal{D}_{l}^{m,j\pm 1}\rangle$.
Now we plug in the full expansion of $\mathcal{G}$ into the diffusion equation and integrate over $\displaystyle \int\!\! d\Omega\!\! \int\!\! d\Omega_0$ to write (using the Dirac bra-ket notation in quantum mechanics),
\begin{align}
    \frac{\partial g_{l}^{j,j_0}(L)}{\partial N} & = g_{l}^{j,j_0}(L) \langle \mathcal{D}_{l}^{m,j_0} |\hat{\mathcal{H}_0}|\mathcal{D}_{l}^{m,j}\rangle \nonumber \\
    & = g_{l}^{j,j_0}(L) \langle \mathcal{D}_{l}^{m,j_0} |\hat{L}^2 - \Delta_{13}\hat{L}_2^2 - \Delta_{12}\hat{L}_3^2 -\frac{\Delta_{14}}{2}(\hat{L}_2\hat{L}_3 + \hat{L}_3\hat{L}_2)|\mathcal{D}_{l}^{m,j}\rangle.
\end{align}
The operation $\mathbb{D}_{l}^{j,j_0}=\langle \mathcal{D}_{l}^{m,j_0} |\hat{\mathcal{H}_0}|\mathcal{D}_{l}^{m,j}\rangle$ produces a matrix that can be defined by the following rules,
\begin{align}
    \mathbb{D}_{l}^{j,j_0}=\begin{cases}
        -l(l+1)+\frac{1}{4} \Delta_{13} \left(c_l^{j-1} c_l^{-j}+c_l^{-j-1} c_l^j\right)+\Delta_{12} j^2 & \mathrm{for}\text{ }j=j_0\\
        \displaystyle -\frac{i}{4}\Delta_{14}(2j-1)c_l^{-j} & \mathrm{for}\text{ }j=j_0+1\\
        \displaystyle \frac{i}{4}\Delta_{14}(2j+1)c_l^{j} & \mathrm{for}\text{ }j=j_0-1\\
        \displaystyle\frac{1}{4}\Delta_{12}c_l^{-j}c_l^{-j-1} & \mathrm{for}\text{ }j=j_0+2\\
        \displaystyle\frac{1}{4}\Delta_{12}c_l^{j}c_l^{j+1} & \mathrm{for}\text{ }j=j_0-2
    \end{cases}
    \label{eq:D_expressions}
\end{align}

For $l=0$ we get,
\begin{equation}
    g_0^{0,0}(L)=1
\end{equation}
For $l=1$, we have a matrix equation of the form,
\begin{align}
    \frac{\partial}{\partial L} \begin{pmatrix}
g_1^{-1,-1} & g_1^{-1,0} & g_1^{-1,1}\\
g_1^{0,-1} & g_1^{0,0} & g_1^{0,1} \\
g_1^{1,-1} & g_1^{1,0} & g_1^{1,1} 
\end{pmatrix} = \begin{pmatrix}
D_1^{-1,-1} & D_1^{-1,0} & D_1^{-1,1}\\
D_1^{0,-1} & D_1^{0,0} & D_1^{0,1} \\
D_1^{1,-1} & D_1^{1,0} & D_1^{1,1} 
\end{pmatrix}
\begin{pmatrix}
g_1^{-1,-1} & g_1^{-1,0} & g_1^{-1,1}\\
g_1^{0,-1} & g_1^{0,0} & g_1^{0,1} \\
g_1^{1,-1} & g_1^{1,0} & g_1^{1,1} 
\end{pmatrix}
\end{align}
In the above equation the $g$-matrix is may not be symmetric,
\emph{i.e.}, we can have $g_1^{i,j}\ne g_1^{j,i}$ for $i\ne j$.
In general all $l\ne 0$ equations are matrix equations with matrix dimensions $(2l+1)\times (2l+1)$.
An easier method of solution is Laplace transforming the diffusion equation ($L\to s$) and we obtain (use the notation $\mathbb{D}_{l}^{j,j_0}=\langle \mathcal{D}_{l}^{m,j_0} |\hat{\mathcal{H}_0}|\mathcal{D}_{l}^{m,j}\rangle$),
\begin{equation}
    s\Tilde{g}_l^{j,j_0}(s) - g_l^{j,j_0}(L=0) = \Tilde{g}_l^{j,j_0}(s)\mathbb{D}_{l}^{j,j_0}
\end{equation}
The above equation simplifies to,
\begin{align}
    (s\mathbb{I}_l-\mathbb{D}_{l}^{j,j_0})\Tilde{g}_l^{j,j_0}(s) & = \delta_{j,j_0} \\ 
    \text{or, } \Tilde{g}_l^{j,j_0}(s) & = {[(s\mathbb{I}_l-\mathbb{D}_{l}^{j,j_0})]}^{-1}\delta_{j,j_0}
\end{align}
In Laplace space the above matrix is diagonal.
By taking the inverse Laplace transform, we find,
\begin{align}
    {g}_l^{j,j_0}(L) & = {\mathcal{L}}_{(s\to L)}^{-1}\left[ {[(s \mathbb{I}_l -\mathbb{D}_{l}^{j,j_0})]}^{-1}\delta_{j,j_0} \right]\\
    \text{or, } \mathbb{G}_l & = {\mathcal{L}}_{(s\to L)}^{-1} \left[{(s \mathbb{I}_l -\mathbb{D}_{l})]}^{-1} \mathbb{I}_l \right]
\end{align}
where, $[.]^{-1}$ represents a matrix inverse operation. 
The $\mathbb{D}_1$ matrix can be written as,
\begin{align}
    \mathbb{D}_1 & = 
    \begin{pmatrix}
D_1^{-1,-1} & D_1^{-1,0} & D_1^{-1,1}\\
D_1^{0,-1} & D_1^{0,0} & D_1^{0,1} \\
D_1^{1,-1} & D_1^{1,0} & D_1^{1,1} 
\end{pmatrix} \nonumber \\
& = \begin{pmatrix}
-2+\frac{1}{2}\Delta_{13}+\Delta_{12} & -\frac{i}{2\sqrt{2}}\Delta_{14} & \frac{1}{2}\Delta_{13}\\
\frac{i}{2\sqrt{2}}\Delta_{14} & -2+\Delta_{13} & \frac{i}{2\sqrt{2}}\Delta_{14} \\
\frac{1}{2}\Delta_{13} & -\frac{i}{2\sqrt{2}}\Delta_{14} & -2+\frac{1}{2}\Delta_{13}+\Delta_{12} \label{eq:D_general}
\end{pmatrix}
\end{align}
To note, it is obvious from the $\mathbb{D}_1$ matrix that $\displaystyle \lim_{A_4\to 0}D_{1}^{j,j_0\pm 1} = 0$, hence they both vanish in absence of Twist-Bend coupling. 
Moreover it is to be noted that $j=j_0\pm 1$ elements of the matrix are distributed in an anti-symmetric way and are imaginary.
In this limit we have,
\begin{align}
    \mathbb{D}_1 & = \begin{pmatrix}
-2+\frac{1}{2}\Delta_{12}^0+\Delta_{13}^0 & 0 & \frac{1}{2}\Delta_{12}^0\\
0 & -2+\Delta_{12}^0 & 0 \\
\frac{1}{2}\Delta_{12}^0 & 0 & -2+\frac{1}{2}\Delta_{12}^0+\Delta_{13}^0 
\end{pmatrix}
\end{align}
In this limit, we obtain (using dimensionless chain lengths $N=L/(2A_1)$,
\begin{align}
    \mathbb{G}_1 & = \begin{pmatrix}
\frac{1}{2}e^{(-2+\Delta_{13}^0)N}\left(1+e^{\Delta_{12}^0N}\right) & 0 & \frac{1}{2}e^{(-2+\Delta_{13}^0)N}\left(-1+e^{\Delta_{12}^0N}\right)\\
0 & e^{(-2+\Delta_{12}^0)N} & 0 \\
\frac{1}{2}e^{(-2+\Delta_{13}^0)N}\left(-1+e^{\Delta_{12}^0 N}\right) & 0 & \frac{1}{2}e^{(-2+\Delta_{13}^0 )N}\left(1+e^{\Delta_{12}^0 N}\right) 
\end{pmatrix}
\label{eq:G1_A40}
\end{align}
To note,
\begin{align}
    \displaystyle (-2+\Delta_{12}^0)N & = \left(-2+1-\frac{A_1}{A_2}\right)\frac{L}{2A_1}=-\frac{L}{2}\left(\frac{1}{A_1}+\frac{1}{A_2}\right)\equiv -\frac{L}{2A^{(12)}} \\
    \displaystyle (-2+\Delta_{13}^0)N & = \left(-2+1-\frac{A_1}{A_3}\right)\frac{L}{2A_1}=-\frac{L}{2}\left(\frac{1}{A_1}+\frac{1}{A_3}\right)\equiv -\frac{L}{2A^{(13)}}\\
    (-2+\Delta_{13}^0+ \Delta_{12}^0)N & =\left(-2+1-\frac{A_1}{A_3}+1-\frac{A_1}{A_2}\right)\frac{L}{2A_1}=-\frac{L}{2}\left(\frac{1}{A_2}+\frac{1}{A_3}\right)\equiv -\frac{L}{2A^{(23)}}
\end{align}

\noindent Here we can identify three persistence lengths as, $\displaystyle A^{(ij)}=2 (A_i^{-1}+A_j^{-1})^{-1} \text{, }\forall (i\ne j)$.
The above lets us write,
\begin{align}
    \mathbb{G}_1 & = \begin{pmatrix}
\frac{1}{2}\left(e^{-L/A^{(13)}}+e^{-L/A^{(23)}}\right) & 0 & \frac{1}{2}\left(e^{-L/A^{(23)}}-e^{-L/A^{(13)}}\right)\\
0 & e^{-L/A^{(12)}} & 0 \\
\frac{1}{2}\left(e^{-L/A^{(23)}}-e^{-L/A^{(13)}}\right) & 0 & \frac{1}{2}\left(e^{-L/A^{(13)}}+e^{-L/A^{(23)}}\right) 
\end{pmatrix}
\end{align}
In the limit of $A_1=A_2=A$ and $A_3=A_4=0$, or the case of wormlike chain only $l=l_0=0$ element of $\mathbb{G}_1$ survives and equals $\displaystyle \exp{\left(-{L}/{A}\right)}\equiv \exp{(-2N)}$ [where, $N=L/(2A)$].

\noindent For the general form of $\mathbb{D}_1$ matrix in Eq.~\ref{eq:D_general}, we can write,
\begin{align}
    \mathbb{G}_1 & = 
    \begin{pmatrix}
g_1^{-1,-1} & g_1^{-1,0} & g_1^{-1,1}\\
g_1^{0,-1} & g_1^{0,0} & g_1^{0,1} \\
g_1^{1,-1} & g_1^{1,0} & g_1^{1,1}  
\end{pmatrix} \equiv  
\begin{pmatrix}
    g_1 & -ig_3 & g_2\\
ig_3 & g_0 & ig_3 \\
g_2 & -ig_3 & g_1 
\end{pmatrix}
\end{align}
and identify (again in terms of dimensionless chain length, $N$),
\begin{align}
    g_1 = \frac{1}{4} e^{\frac{1}{2} \left(-\Delta +\Delta _{12}-4\right) N} \left[\frac{\Delta _{12}}{\Delta} e^{\Delta _{13} N} \left(e^{\Delta  N}-1\right)+2 e^{\frac{1}{2}
   \left(\Delta +\Delta _{12}\right) N}+e^{\Delta _{13} N}+e^{\left(\Delta +\Delta _{13}\right) N}\right].
\end{align}
\begin{align}
    g_2=\frac{1}{4} e^{\frac{1}{2} \left(-\Delta +\Delta _{12}-4\right) N} \left[\frac{\Delta _{12}}{\Delta} e^{\Delta _{13} N} \left(e^{\Delta  N}-1\right)-2 e^{\frac{1}{2}
   \left(\Delta +\Delta _{12}\right) N}+e^{\Delta _{13} N}+e^{\left(\Delta +\Delta _{13}\right) N}\right].
\end{align}
\begin{align}
    g_3 = \frac{ \Delta _{14}}{\sqrt{2} \Delta} e^{\frac{1}{2} \left(\Delta _{12}+2 \Delta _{13}-4\right) N} \sinh \left(\frac{\Delta  N}{2}\right)
\end{align}
and,
\begin{align}
    g_0 = {e^{\frac{1}{2} \left(\Delta _{12}+2 \Delta _{13}-4\right) N} \left[\cosh \left(\frac{\Delta  N}{2}\right)-\frac{\Delta _{12}}{\Delta} \sinh \left(\frac{\Delta 
   N}{2}\right)\right]}
\end{align}
where, we define $\displaystyle \Delta=\sqrt{\Delta _{14}^2+\Delta_{12}^2}$ and other coefficients are defined in Eq.~\ref{eq:delta_def1}-Eq.~\ref{eq:delta_def5}.
The limit of $A_4\to 0$ successfully reproduces Eq.~\ref{eq:G1_A40}.
\begin{figure}[h!]
    \centering
    \includegraphics[scale=0.50]{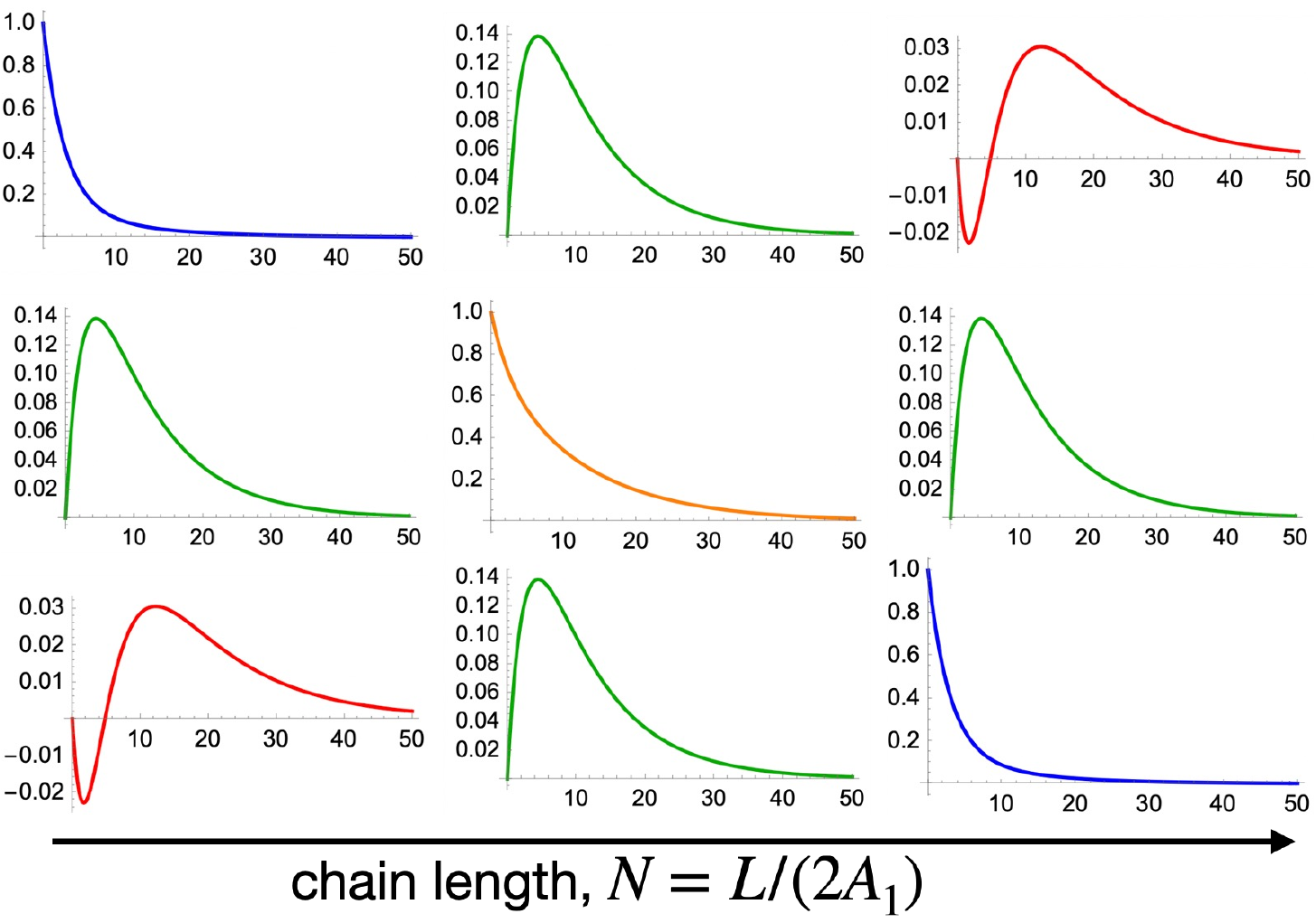}
    \caption{The coefficients $g_1^{m,j}$ of $\mathbb{G}_1$ matrix are plotted for $A_1=10$, $A_2=2$, $A_3=1$ and $A_4=0.7$. }
    \label{fig:suppfig1}
\end{figure}

\noindent The coefficients are plotted as a function of chain length in Fig.~\ref{fig:suppfig1} for the parameter choices of $A_1=10$, $A_2=2$, $A_3=1$ and $A_4=0.7$.
The $g_1^{0,0}$ element (orange) gives the dominant relaxation of the $\mathbb{G}_1$ matrix an correlates with the principal tangent correlation function as discussed later.
As we can see, $g_1^{-1,-1}=g_1^{1,1}$ elements are identical and represent relaxation behavior on a different length scale.
All of the off-diagonal elements start at $0$ for $N=0$ and show a minima or a maxima indicating the lengthscale at which cross-correlations due to coupling becomes important.
To note, only the $g_3$ elements are plotted in green without the proper phase of the correlation. 

\begin{figure}[h!]
    \centering
    \includegraphics[scale=0.50]{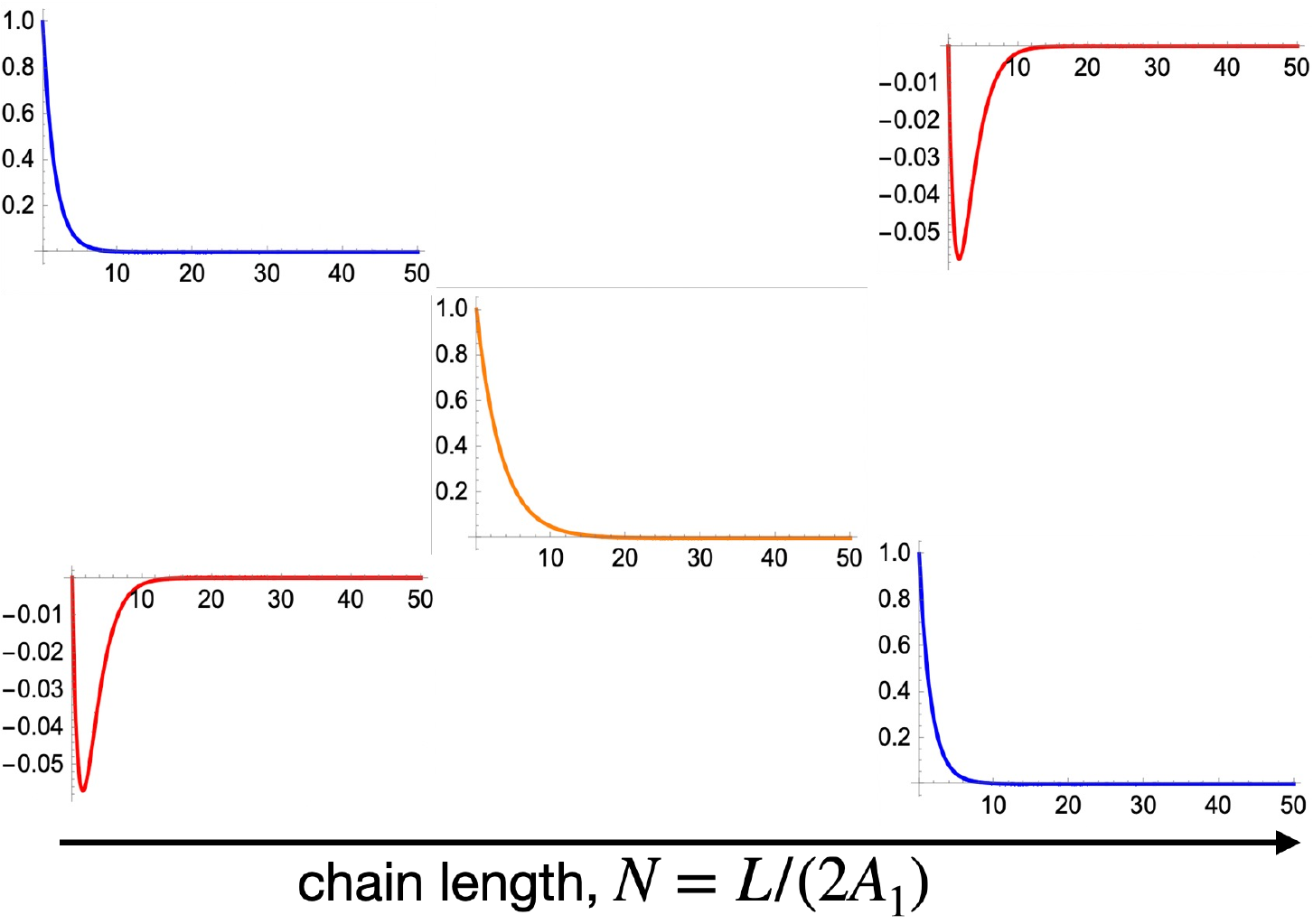}
    \caption{The coefficients $g_1^{m,j}$ of $\mathbb{G}_1$ matrix are plotted for $A_1=10$, $A_2=2$, $A_3=1$ and $A_4=0$. Empty places indicate elements to be zero.}
    \label{fig:suppfig2}
\end{figure}
Figure ~\ref{fig:suppfig2} plots the exact same quantity as a function of chain length in the limit of vanishing twist-bend coupling.
As obvious, $g_1^{j,j\pm 1}$ elements do not exist for this model.
Qualitatively we obtain similar behavior for the other principal decay functions, while quantitatively the off diagonal elements differ much.

\subsection{Utility of Free Particle Solution}
\subsubsection{Tangent Correlations}
\begin{figure}[h!]
    \centering
    \includegraphics[scale=0.50]{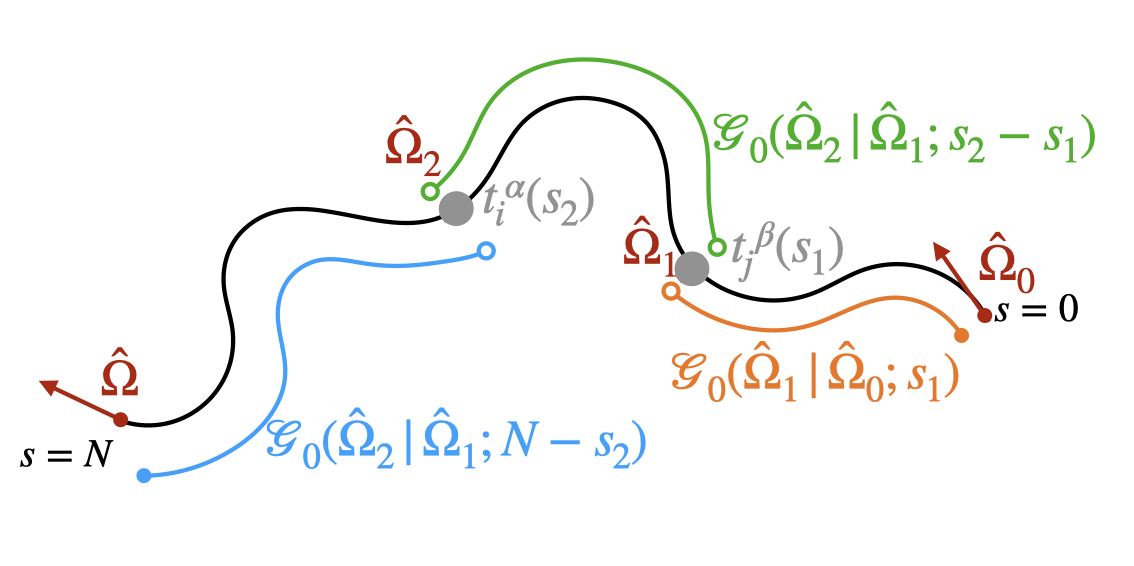}
    \caption{Schematic representation of the calculation of the average $\langle t_{i}^{\alpha}(s_2) t_{j}^{\beta}(s_1) \rangle$. 
    %The mathematical and schematic procedure is exactly same as that represented in Fig.~\ref{fig:wlc_average_schematic}.
    }
    \label{fig:tbchwlc_average_schematic}
\end{figure}

The development above allows to calculate conformation dependent averages. 
A fundamental quantity is the triad triad correlation function at two points along the chain contour.
The calculation of the average is schematically represented in Fig.~\ref{fig:tbchwlc_average_schematic}.

\noindent Mathematically, the above average, which is the correlation between $\alpha^{\mathrm{th}}$-component of $\vec{t}_i$ at arc-length position $s_2$ with the $\beta^{\mathrm{th}}$-component of $\vec{t}_j$ at arc-length position $s_1$ (where in general $i, j$ and $\alpha,\beta$ can be same of different) can be expressed as [without loss of generality we can assume $0\le s_1 \le s_2 \le L$],
\begin{align}
    \langle t_{i}^{\alpha}(s_2) t_{j}^{\beta}(s_1) \rangle  & = \frac{1}{8\pi^2}\int d\Omega \int d\Omega_{0} \int d\Omega_{2} \int d\Omega_{1} \mathcal{G}_0(\Omega|\Omega_{2};L-s_2)t_{i}^{\alpha}(s_2)\mathcal{G}_0(\Omega_{2}|\Omega_{1};s_2-s_1) \nonumber \\
    & \hspace{3.5 in} \times t_{j}^{\beta}(s_1) \mathcal{G}_0(\Omega_{1}|\Omega_{0};s_1)
\end{align}

As an example we can calculate $\langle t_{3}^{(3)}(s_2) t_{3}^{(3)}(s_1) \rangle$ as,

\begin{align}
    \langle t_{3}^{(3)}(s_2) t_{3}^{(3)}(s_1) \rangle & = \frac{1}{8\pi^2}\int d\Omega_{L} \int d\Omega_{0} \int d\Omega_{2} \int d\Omega_{1} \mathcal{G}(\Omega_{L}|\Omega_{2};L-s_2)t_{3}^{(3)}(s_2) \nonumber \\
    & \!\times \mathcal{G}(\Omega_{2}|\Omega_{1};s_2-s_1) t_{3}^{(3)}(s_1) \mathcal{G}(\Omega_{1}|\Omega_{0};s_1) \\
    & = \frac{1}{8\pi^2}\int d\Omega_{L} \int d\Omega_{0} \int d\Omega_{2} \int d\Omega_{1}
    \sum_{l=0}^\infty \sum_{m=-l}^{l}\sum_{j, j^\prime = -l}^{l}
    \mathcal{D}_{l}^{mj} (\Omega_L)
    \mathcal{D}_{l}^{m {j^\prime}^{*}} (\Omega_{2})
    g_{l}^{jj^\prime}(L-s_2)\nonumber \\
    & \times \sqrt{\frac{8\pi^2}{3}} \mathcal{D}_1^{0,0}(\Omega_2) 
    \sum_{l_2=0}^\infty \sum_{m_2=-l_2}^{l_2}\sum_{j_2, j_2^\prime = -l_2}^{l_2}
    \mathcal{D}_{l_2}^{m_2j_2} (\Omega_{2})
    \mathcal{D}_{l_2}^{m_2 {j_2^\prime}^{*}} (\Omega_{1})
    g_{l_2}^{j_2j_2^\prime}(s_2-s_1)\nonumber \\
    & \times \sqrt{\frac{8\pi^2}{3}} \mathcal{D}_1^{0,0}(\Omega_1) \sum_{l_1=0}^\infty \sum_{m_1=-l_1}^{l_1}\sum_{j_1, j_1^\prime = -l_1}^{l_1}
    \mathcal{D}_{l_1}^{m_1j_1} (\Omega_{1})
    \mathcal{D}_{l_1}^{m_1 {j_1^\prime}^{*}} (\Omega_{0})
    g_{l_1}^{j_1j_1^\prime}(s_1) \\ 
    & = {\frac{1}{3}}\int d\Omega_{2} \int d\Omega_{1} \mathcal{D}_{0}^{0,0} (\Omega_{2})\mathcal{D}_1^{0,0}(\Omega_2)\sum_{l=0}^\infty \sum_{m=-l}^{l}\sum_{j, j^\prime = -l}^{l}
    \mathcal{D}_{l}^{mj} (\Omega_{2})
    \mathcal{D}_{l}^{m {j^\prime}^{*}} (\Omega_{1})
    g_{l}^{jj^\prime}(s_2-s_1) \nonumber \\
    & \times 
    \mathcal{D}_1^{0,0}(\Omega_1) \mathcal{D}_{0}^{0,0} (\Omega_{1}) \\
    &  = {\frac{1}{3}}\sum_{l=0}^\infty \sum_{m=-l}^{l}\sum_{j, j^\prime = -l}^{l}
    g_{l}^{jj^\prime}(s_2-s_1) \int d\Omega_{2} \mathcal{D}_{l}^{mj} (\Omega_{2})
    \mathcal{D}_{1}^{0,0} (\Omega_{2}) \int d\Omega_{1} \mathcal{D}_{l}^{m {j^\prime}^{*}} (\Omega_{1})\mathcal{D}_{1}^{0,0} (\Omega_{1}) \\
    & = {\frac{1}{3}} g_{1}^{0,0}(s_2-s_1)
\end{align}
At this point we can argue that, in absence of any external aligning fields (such as dipole perturbations or nematic quadrupole fields), using symmetry arguments we can arrive at,
\begin{align}
    \langle \hat{t}_{i}(s_2) \hat{t}_{j}(s_1) \rangle & = 3\langle t_{i}^{(3)}(s_2) t_{j}^{(3)}(s_1) \rangle
\end{align}
Inspecting the other tangent correlations we can note that all of the tangent correlations can be written in terms of a correlator involving Wigner-D functions.
Thus, it is sufficient to obtain the following correlation, $\langle \mathcal{D}_1^{m,j}(\Omega_2) \mathcal{D}_1^{n,k}(\Omega_1) \rangle$ for $\{m, n, j, k\}\in[-1,0,1]$.

The correlation can be explicitly calculated as,
\begin{align}
    \langle \mathcal{D}_1^{m,j}(\Omega_2) \mathcal{D}_1^{n,k}(\Omega_1) \rangle & = \frac{1}{8\pi^2}\int d\Omega \int d\Omega_{0} \int d\Omega_{2} \int d\Omega_{1} \mathcal{G}_0(\Omega|\Omega_{2};L-s_2)\mathcal{D}_1^{m,j}(\Omega_2) \mathcal{G}_0(\Omega_{2}|\Omega_{1};s_2-s_1)\nonumber \\
    & \hspace{3 in}\times \mathcal{D}_1^{n,k}(\Omega_1)  \mathcal{G}_0(\Omega_{1}|\Omega_{0};s_1) \\
    & \hspace{-1 in}= \frac{1}{8\pi^2}\int d\Omega \int d\Omega_{0} \int d\Omega_{2} \int d\Omega_{1} \sum_{l_2=0}^\infty \sum_{m_2=-l_2}^{l_2}\sum_{j_2, j_2^\prime = -l_2}^{l_2}
    \mathcal{D}_{l_2}^{m_2j_2} (\Omega)
    \mathcal{D}_{l_2}^{m_2 {j_2^\prime}^{*}} (\Omega_{2})
    g_{l_2}^{j_2j_2^\prime}(L-s_2)\nonumber \\
    & \hspace{1 in}\times \mathcal{D}_1^{m,j}(\Omega_2) 
    \sum_{l_1=0}^\infty \sum_{m_1=-l_1}^{l_1}\sum_{j_1, j_1^\prime = -l_1}^{l_1}
    \mathcal{D}_{l_1}^{m_1j_1} (\Omega_{2})
    \mathcal{D}_{l_1}^{m_1 {j_1^\prime}^{*}} (\Omega_{1})
    g_{l_1}^{j_1j_1^\prime}(s_2-s_1)\nonumber \\
    & \hspace{1 in}\times \mathcal{D}_1^{n,k}(\Omega_1) \sum_{l_0=0}^\infty \sum_{m_0=-l_0}^{l_0}\sum_{j_0, j_0^\prime = -l_0}^{l_0}
    \mathcal{D}_{l_0}^{m_0j_0} (\Omega_{1})
    \mathcal{D}_{l_0}^{m_0 {j_0^\prime}^{*}} (\Omega_{0})
    g_{l_0}^{j_0j_0^\prime}(s_1)
\end{align}
Integration over $\displaystyle \int d\Omega$ and  $\displaystyle \int d\Omega_0$ can be performed independently and $\displaystyle \int d\Omega \mathcal{D}_l^{m,j}(\Omega)=\delta_{l,0}\delta_{m,0}\delta_{j,0}$ since, $\mathcal{D}_0^{0,0}(\Omega)=1$. 
This simplifies the above correlation to (also, $g_0^{0,0}(s)=1, \forall s$),
\begin{align}
    \langle \mathcal{D}_1^{m,j}(\Omega_2) \mathcal{D}_1^{n,k}(\Omega_1) \rangle & = \frac{1}{8\pi^2} \sum_{l=0}^\infty \sum_{m_0=-l}^{l}\sum_{j_0, j_0^\prime = -l}^{l} g_{l}^{j_0,{j_0^\prime}}(s_2-s_1) \int d\Omega_{2} \mathcal{D}_1^{m,j}(\Omega_2) \mathcal{D}_{l}^{m_0,j_0} (\Omega_{2}) \nonumber \\
    & \times \int d\Omega_{1} \mathcal{D}_{l}^{m_0,{j_0^\prime}^\star}(\Omega_{1}) \mathcal{D}_1^{n,k}(\Omega_1) \\
    & = \frac{1}{8\pi^2}g_{1}^{j,k}(s_2-s_1)
\end{align}

\noindent Now, explicit calculations of the linear combinations of $\displaystyle \langle \mathcal{D}_1^{m,j}(\Omega_2) \mathcal{D}_1^{n,k}(\Omega_1) \rangle$ lead to the following self correlations,
\begin{align}
    \langle \hat{t}_{3}(s_2) \hat{t}_{3}(s_1) \rangle & = g_1^{0,0}(\Delta s) \text{,    where, }s_2=s_1+\Delta s\\
    \langle \hat{t}_{2}(s_2) \hat{t}_{2}(s_1) \rangle & = \frac{1}{2}\left(g_1^{1,1}(\Delta s)+g_1^{-1,-1}(\Delta s)+g_1^{1,-1}(\Delta s)+g_1^{-1,1}(\Delta s)\right)\\
    \langle \hat{t}_{1}(s_2) \hat{t}_{1}(s_1) \rangle & = \frac{1}{2}\left(g_1^{1,1}(\Delta s)+g_1^{-1,-1}(\Delta s)-g_1^{1,-1}(\Delta s)-g_1^{-1,1}(\Delta s)\right)
\end{align}
Cross-correlations yield the following ($s_2=s_1+\Delta s$),
\begin{align}
    \langle \hat{t}_{3}(s_2) \hat{t}_{1}(s_1) \rangle & =\frac{1}{\sqrt{2}}\left(g_1^{-1,0}(\Delta s)-g_1^{1,0}(\Delta s)\right)\\
    \langle \hat{t}_{1}(s_2) \hat{t}_{3}(s_1) \rangle & =\frac{1}{\sqrt{2}}\left(g_1^{0,-1}(\Delta s)-g_1^{0,1}(\Delta s)\right)\\
    \langle \hat{t}_{3}(s_2) \hat{t}_{2}(s_1) \rangle & =-i\frac{1}{\sqrt{2}}\left(g_1^{-1,0}(\Delta s)+g_1^{1,0}(\Delta s)\right)\\
    \langle \hat{t}_{2}(s_2) \hat{t}_{3}(s_1) \rangle & =i\frac{1}{\sqrt{2}}\left(g_1^{0,-1}(\Delta s)+g_1^{0,1}(\Delta s)\right)
\end{align}
Finally,
\begin{align}
    \langle \hat{t}_{1}(s_2) \hat{t}_{2}(s_1) \rangle & = i\frac{1}{{2}}\left(g_1^{1,1}(\Delta s)-g_1^{-1,-1}(\Delta s)+g_1^{1,-1}(\Delta s)-g_1^{-1,1}(\Delta s)\right)\\
    \langle \hat{t}_{2}(s_2) \hat{t}_{1}(s_1) \rangle & = i\frac{1}{{2}}\left(g_1^{-1,-1}(\Delta s)-g_1^{1,1}(\Delta s)+g_1^{1,-1}(\Delta s)-g_1^{-1,1}(\Delta s)\right)
\end{align}
The explicit form of $\hat{t}_3$ tangent correlations are found to be,
\begin{align}
\langle \hat{t}_{3}(N)\cdot \hat{t}_{3}(0) \rangle &  = {\exp{\left(\frac{1}{2} \left(\Delta _{12}+2 \Delta _{13}-4\right) N\right)} \left[\cosh \left(\frac{\Delta  N}{2}\right)-\frac{\Delta _{12}}{\Delta} \sinh \left(\frac{\Delta 
   N}{2}\right)\right]}
   %  \langle \hat{t}_{3}(L)\cdot \hat{t}_{3}(0) \rangle = & \exp \left(\frac{1}{4} \left(-\frac{\left(A_2+2 A_3\right) \left(A_2 A_3-3 A_4^2\right)}{\left(A_4^2-A_2 A_3\right){}^2}-\frac{1}{A_1}\right) L\right) \nonumber \\
   %  & \Bigg\{ (\cosh
   % \left(\frac{L}{4 A_1}\sqrt{\frac{16 A_1^2 A_4^6}{\left(A_4^2-A_2 A_3\right)^4}+\left(\frac{A_1 A_2 \left(A_2 A_3-3 A_4^2\right)}{\left(A_4^2-A_2 A_3\right){}^2}-1\right)^2}
   % \right) \nonumber \\
   % & -\frac{\left(\left(A_4^2-A_2 A_3\right){}^2+A_1 A_2 \left(3 A_4^2-A_2 A_3\right)\right)}{\left(A_4^2-A_2 A_3\right)^2
   % \sqrt{\frac{16 A_1^2 A_4^6}{\left(A_4^2-A_2 A_3\right){}^4}+\left(\frac{A_1 A_2 \left(A_2 A_3-3 A_4^2\right)}{\left(A_4^2-A_2 A_3\right){}^2}-1\right)^2}} \nonumber \\
   % & \times \sinh \left(\frac{L}{4A_1}\sqrt{\frac{16 A_1^2 A_4^6}{\left(A_4^2-A_2
   % A_3\right){}^4}+\left(\frac{A_1 A_2 \left(A_2 A_3-3 A_4^2\right)}{\left(A_4^2-A_2 A_3\right){}^2}-1\right)^2}\right)
   % \Bigg\}
\end{align}
where, the expressions of $\{\Delta_{1k}\}$ are found in the preceding sections.
In the limit $A_4\to 0$, the above expression in dimensional form simplifies to (using $N=L/(2A_1)$),
\begin{align}
     \langle \hat{t}_{3}(L)\cdot \hat{t}_{3}(0) \rangle = & \exp{\left(-\frac{1}{2} \left(\frac{1}{A_2}+\frac{1}{A_1}\right) L\right)}
\end{align}
which defines the persistence length along the body-tangent as the harmonic mean of $A_1$ and $A_2$, or, $\displaystyle \frac{1}{A}=\frac{1}{2} \left(\frac{1}{A_2}+\frac{1}{A_1}\right)$.
\begin{figure}[h!]
    \centering
    \includegraphics[scale=0.50]{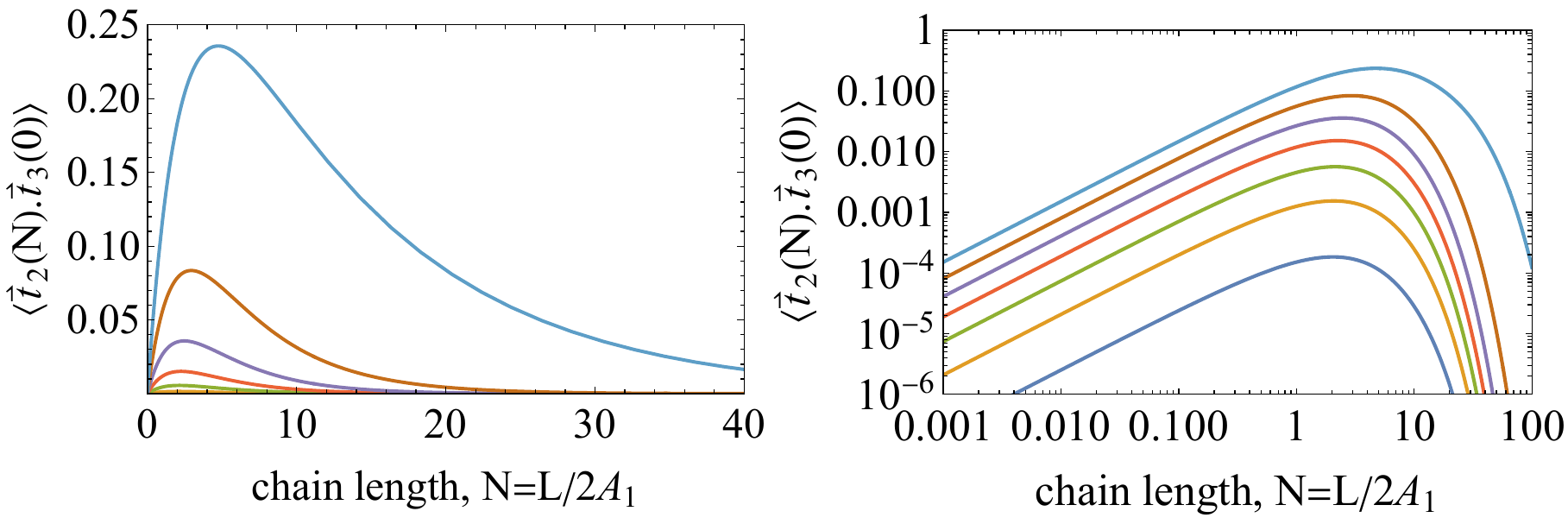}
    \caption{Tangent cross correlation function as a function of chain size in linear-linear (left) and log-log (right) format. 
    The parameter choices are same as Fig.~\ref{fig:suppfig2} for $A_1,A_2,A_3$ and $A_4$ values are chosen as $0.10, 0.20, 0.30, 0.40, 0.50, 0.60$ and $0.70$ (bottom to top).}
    \label{fig:suppfig3}
\end{figure}

\begin{figure}[h!]
    \centering
    \includegraphics[scale=0.50]{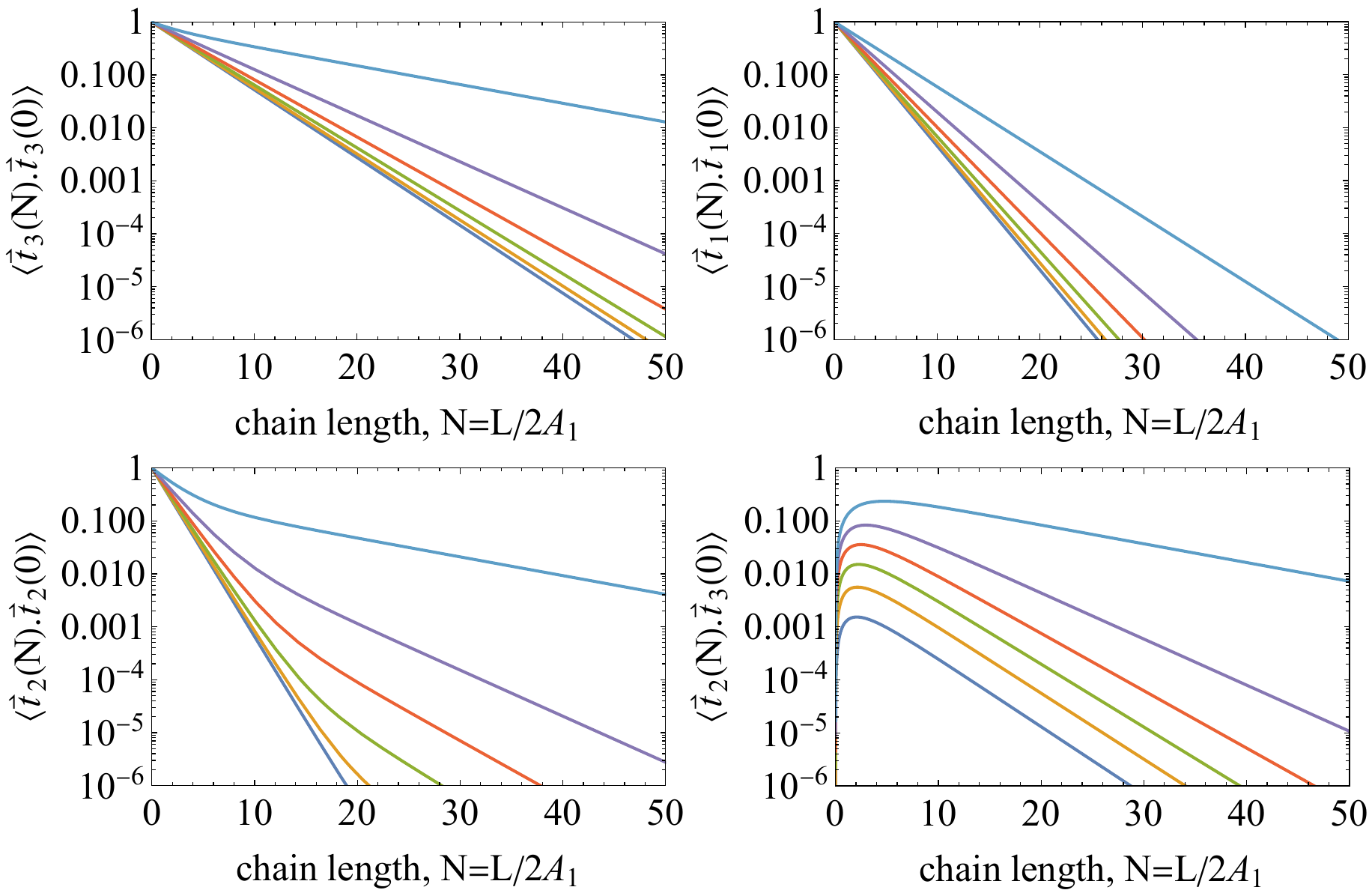}
    \caption{All existing tangent correlation functions are plotted as a function of chain size in log-linear format. 
    The parameter choices are same as Fig.~\ref{fig:suppfig3} for $A_1,A_2,A_3$ and $A_4$.}
    \label{fig:suppfig4}
\end{figure}

\subsubsection{End-to-end distance or Chain size}
The end-to-end vector of a chain can be calculated from the principal tangent only as,
\begin{align}
    \vec{R}(L) = \int_0^L ds \hat{t}_3(s)
\end{align}
Hence the mean squared end-to-end distance of a chain is given as,
\begin{align}
    \langle \vec{R}^2 (L)\rangle & =\int_0^L ds_2 \int_0^L ds_1 \langle \hat{t}_3(s_2) \cdot \hat{t}_3(s_1)\rangle \\
    & = 3 \int_0^L ds_2 \int_0^L ds_1 \langle {t}_3^{(3)}(s_2) {t}_3^{(3)}(s_1)\rangle \text{ (using isotropy)} \\
    & = 6 \int_0^L ds_2 \int_0^{s_2} ds_1 \langle {t}_3^{(3)}(s_2) {t}_3^{(3)}(s_1)\rangle \text{ (using `time' ordering, } s_2>s_1)
\end{align}
The final form of the end to end distance comes to be,

\begin{align}
\langle \vec{R}^2 (L)\rangle & = \frac{2}{4 \left(\Delta _{13}-2\right) \left(\Delta _{12}+\Delta_{13}-2\right)-\Delta _{14}^2} \Bigg[8 L-2 \Delta _{12} L-4 \Delta _{13} L
+\frac{4}{\Delta _{14}^2-4 \left(\Delta_{13}-2\right) \left(\Delta _{12}+\Delta _{13}-2\right)} \nonumber \\
& \hspace{0.5 in}\times\Bigg\{2 \Delta _{12}^2+4 \left(\Delta _{13}-2\right) \Delta _{12}+4 \left(\Delta _{13}-2\right){}^2+\Delta_{14}^2+e^{\frac{L}{2} \left(\Delta _{12}+2 \Delta _{13}-4\right)} \nonumber \\
& \hspace{0.6 in} \times \Bigg(2 \left(\Delta _{12}+2 \Delta _{13}-4\right) \sqrt{\Delta _{12}^2+\Delta _{14}^2} \sinh\left(\frac{L}{2} \sqrt{\Delta _{12}^2+\Delta _{14}^2}\right) \nonumber \\
& \hspace{0.8 in}-\left(2 \Delta _{12}^2+4 \left(\Delta _{13}-2\right) \Delta _{12}+4 \left(\Delta_{13}-2\right){}^2+\Delta _{14}^2\right) \cosh \left(\frac{L}{2} \sqrt{\Delta _{12}^2+\Delta_{14}^2} \right)\Bigg)\Bigg\} \nonumber \\
& \hspace{0.5 in}+\Delta_{12} \Bigg\{-\frac{4 \left(\Delta _{12}+2 \Delta _{13}-4\right)}{\left(4\left(\Delta _{13}-2\right) \left(\Delta_{12}+\Delta_{13}-2\right)-\Delta _{14}^2\right) \sqrt{\Delta _{12}^2+\Delta _{14}^2}} \Bigg(\sqrt{\Delta _{12}^2+\Delta_{14}^2} \nonumber \\
& \hspace{0.8 in} +e^{\frac{L}{2} \left(\Delta _{12}+2 \Delta _{13}-4\right)} \Bigg(\left(\Delta _{12}+2 \Delta _{13}-4\right) \sinh \left(\frac{L}{2} \sqrt{\Delta_{12}^2+\Delta _{14}^2}\right) \nonumber \\
& \hspace{ 1.9 in} -\sqrt{\Delta _{12}^2+\Delta _{14}^2} \cosh \left(\frac{L}{2} \sqrt{\Delta _{12}^2+\Delta _{14}^2} \right)\Bigg)\Bigg) \nonumber \\
& \hspace{0.75 in}+\frac{4}{{4 \left(\Delta_{13}-2\right) \left(\Delta _{12}+\Delta _{13}-2\right)-\Delta _{14}^2}} \Bigg(-\Delta _{12}-2 \Delta_{13}+e^{\frac{L}{2} \left(\Delta _{12}+2 \Delta _{13}-4\right)} \nonumber \\
& \hspace{1.2 in} \times \Bigg(\left(\Delta _{12}+2 \Delta _{13}-4\right) \cosh \left(\frac{L}{2} \sqrt{\Delta _{12}^2+\Delta_{14}^2} \right) \nonumber \\
& \hspace{1.4 in}-\sqrt{\Delta _{12}^2+\Delta _{14}^2} \sinh \left(\frac{L}{2} \sqrt{\Delta _{12}^2+\Delta _{14}^2} \right)\Bigg)+4\Bigg)-2 L\Bigg\}\Bigg]
\end{align}
where, the expressions of $\{\Delta_{1k}\}$ are given previously.
The above equation simplifies for the case of $\Delta_{14}=0$ to be,
\begin{align}
    \langle \vec{R}^2 (L)\rangle & = \displaystyle \frac{2}{\left(\Delta_{12}^{0}-2\right){}^2} \left(-\Delta_{12}^{0} L+e^{\left(\Delta_{12}^{0}-2\right) L}+2 L-1\right)
\end{align}
where, all of the coefficients are defined in Eq.~\ref{eq:delta_def1}-~\ref{eq:delta_def5}.
Not surprisingly, we can note, the above expression is completely independent of twist modulus $A_3$.
Similarly, a radius of gyration of the chain can be given as,
\begin{equation}
    R_g^2(L) = \frac{1}{L^2} \int_0^L ds_2 \int_0^{s_2} ds_1 \langle \vec{R}^2(s_2-s_1)\rangle
\end{equation}
\begin{figure}[h!]
    \centering
    \includegraphics[scale=0.50]{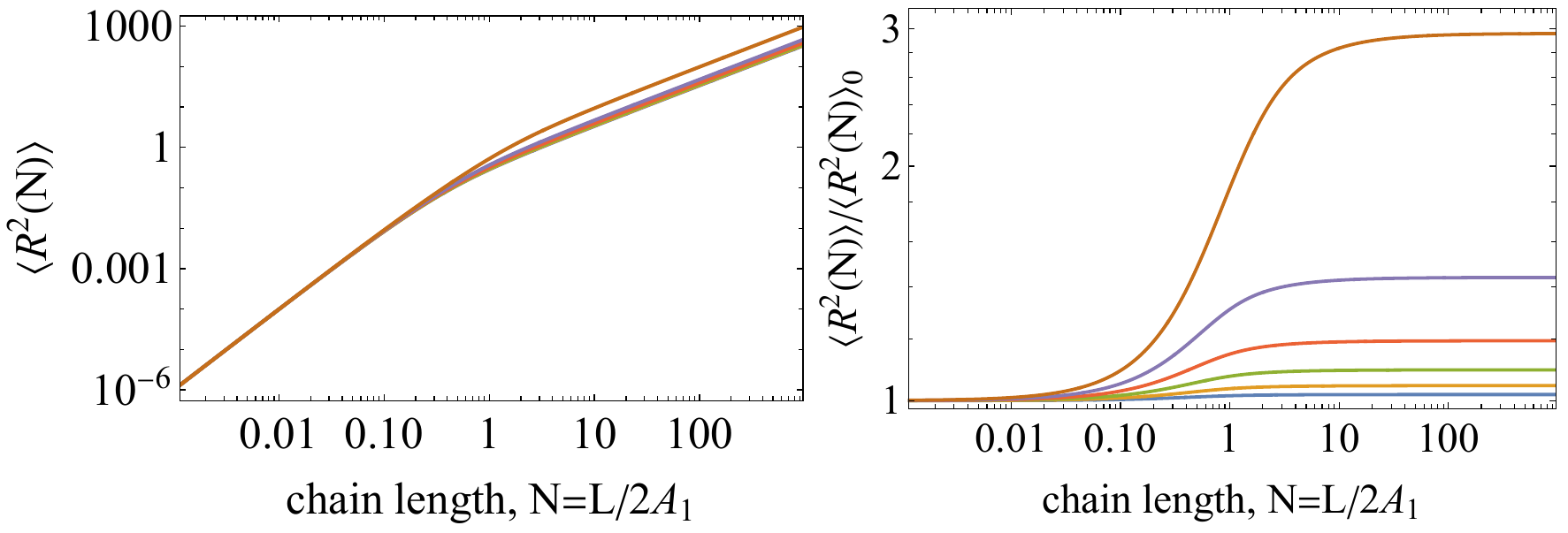}
    \caption{(Left) Mean-squared end to end distance is plotted as a function of chain size in a log-log plot. (Right) The ratio of end-to-end distance for $A_4\ne 0$ and $A_4=0$ are plotted as a function of chain size for the same range of $A_4$ values of $0.10$ to $0.70$ at intervals of $0.10$. In both the plots the other three moduli are kept constant per Fig.~\ref{fig:suppfig4}.}
    \label{fig:suppfig5}
\end{figure}

\subsection{Solution to the Full Green's Function}
Per previous development the effective Hamiltonian for the single chain Green function can be written we,
\begin{align}
    \beta\mathcal{H}=\beta\mathcal{H}_0+\beta \mathcal{H}_{\text{ext}} = \beta\mathcal{H}_0 - i\vec{k}\cdot \int_0^L ds \vec{t}_3(s)
\end{align}
We now turn to the full Green function $\mathcal{G}(\Vec{R}, \Omega|\Omega_0; L)$, which gives the probability that
a chain that begins at the origin with fixed orientation $\Omega_0$ will end at position R with fixed orientation $\Omega$. 
The Green function is given by
\begin{align}
    \mathcal{G}(\Vec{R}, \Omega|\Omega_0; L) = \left\langle \delta\left(\vec{R}-\int_0^L ds \vec{t}_3(s)\right)\right\rangle_{\Omega_0}^{\Omega}
\end{align}
where, $\langle\dots\rangle_{\Omega_0}^{\Omega}$ is an average with respect to the orientation only Green function $\mathcal{G}_0(\Omega|\Omega_0; L)$ with fixed initial and final orientations.
We perform a Fourier transform of the above equation from $\vec{R}\to \vec{k}=k\vec{e}_k$, where $\vec{e}_k(\chi,\Omega)$ gives the direction of $\vec{k}$ defined by the polar and azimuthal angles $\chi$ and $\Omega$, respectively.
Since the elastic deformation energy of the polymer chain is invariant to an arbitrary rotation, we rotate the external coordinate system to align $\vec{e}_k$ with the $z$-axis (\emph{i.e.} $\vec{e}_k\!\parallel\! \hat{\delta}_z$), which rotates the initial orientation to $\Omega_0^\prime$ and the final orientation to $\Omega^\prime$ by the Euler angles $\alpha=-\Omega$, $\beta=-\chi$ and $\gamma=0$.
We now arrive at the expression for the Fourier-space Green function,
\begin{align}
    \mathcal{G}(\Vec{k}, \Omega|\Omega_0; L) = \left\langle \exp\left(ik\int_0^L ds \hat{\delta}_z\cdot\vec{t}_3(s)\right)\right\rangle_{\Omega_0^\prime}^{\hat{\Omega^\prime}}
\end{align}
which reduces our problem to that of a spinner undergoing fluctuations embedded in an imaginary dipole field of strength $k$ along the $z$-axis.

Now, as an aside, let us consider the partition function of the elastic polymer chain subject to a tensile force $\vec{f}$ acting on the end-to-end distance vector $\displaystyle \vec{R}=\int_0^L ds \vec{t}_3(s)$. 
In the equation for total Hamiltonian of the system, the external Hamiltonian now arises from this tensile force and can be written as,
\begin{align}
    \mathcal{H}_{\text{ext}} = -\beta\vec{f}\cdot \int_0^L ds \vec{t}_3(s)
\end{align}
where, without loss of generality, we can assume the force to be aligned along the $z$-axis.
Comparing aforementioned external Hamiltonian with the one written here, we can note the two are related by a simple transformation of the form, $i\vec{k}\leftrightarrow \beta\vec{f}$.
The corresponding Green function will be denoted by, $\mathcal{G}(\Vec{f}, \Omega|\Omega_0; L)$.

The Taylor expansion of the above equation gives an infinite sum of moments of the end-to-end distribution function with fixed end orientations.
The $n^{\text{th}}$ term contains the average,
\begin{align}
    \left\langle \left(\int_0^L ds \cos\theta(s)\right)^n\right\rangle_{\Omega_0^\prime}^{\hat{\Omega^\prime}} = n!  \left\langle \prod_{i=1}^{n} \int_0^{s_{i+1}} ds_{i} \cos\theta(s_{i})\right\rangle_{\Omega_0^\prime}^{\hat{\Omega^\prime}}
\end{align}
where, $\theta$ is the polar angle of the tangent vector $\vec{t}_3$, and $s_{n+1}=L$.
The factor of $n!$ in the right side of the above equation is due to the ``time" ordering \footnote{the naming convention comes from time-ordering of path-integrals context of quantum mechanics.} of the $s$ integrals.
This average is found by placing a chain-orientation Green function $\mathcal{G}_0$ between each successive $\cos \theta$ (hence $n+1$ total propagators for the $n^{\text{th}}$ order average) and integrating over all intermediate triad orientations.
To recapitulate a previous result,
\begin{align}
    \mathcal{G}(\Omega|\Omega_{0}; L) = \sum_{l=0}^\infty \sum_{m=-l}^{l}\sum_{j, j_0 = -l}^{l}
    \mathcal{D}_{l}^{mj} (\Omega)
    \mathcal{D}_{l}^{m j_{0}^{*}} (\Omega_{0})
    g_{l}^{j,j_0}(L)
\end{align}
We define the inner product of Wigner functions as,
\begin{align}
    \langle l, m, j|\mathcal{O}|l^\prime, m^\prime, j^\prime\rangle = \int \!d\Omega  \mathcal{D}_{l}^{mj^\star} (\Omega) \mathcal{O} \mathcal{D}_{l^\prime}^{m^\prime j^\prime} (\Omega)
\end{align}
where $\mathcal{O}$ represents an arbitrary operator.
As an illustration $n=2$ term will be expanded as,
\begin{align}
    \int_0^L\! ds_2  \int_0^{s_2}\! ds_1 \sum_{l_2=0}^\infty \sum_{m_2, j_2=-l_2}^{l_2} g_{l_1}^{m_1,j_1}(s_1) \langle l_1, m_1, j_1|\cos\theta(s_1)|l_2, m_2, j_2\rangle \nonumber \\
    \times
    g_{l_2}^{m_2,j_2}(s_2-s_1) \langle l_2, m_2, j_2|\cos\theta(s_2)|l_3, m_3, j_3\rangle \nonumber \\
    \times g_{l_3}^{m_3,j_3}(L-s_2)
\end{align}
Using the Clebsch-Gordon coefficients~\cite{rose1995elementary,devanathan1999angular}, we note that,
\begin{align}
    \langle l, m, j|\cos\theta|l^\prime, m^\prime, j^\prime\rangle = \left[\alpha_{l}^{mj}\delta_{l,l^\prime+1}+ \beta_{l}^{mj}\delta_{l,l^\prime}+\alpha_{l+1}^{mj}\delta_{l,l^\prime-1}\right]\delta_{m,m^\prime}\delta_{j,j^\prime}
\end{align}
where,
\begin{align}
    \alpha_{l}^{mj} & = \sqrt{\frac{(l-m)(l+m)(l-j)(l+j)}{l^2(4l^2-1)}} \\
    \beta_{l}^{mj} & = \frac{mj}{l(l+1)}
\end{align}
The above rule means that integration over the intermediate triad orientations in the average gives inner products of Wigner functions from each successive propagator; these inner products prescribe a fixed relationship between the eigenvalues of the Wigner functions in successive propagators.
Specifically, the $l$-indices of successive Green functions in the average are either equal or differ by $\pm 1$ leaving $m$ and $j$ indices unchanged.

Below we illustrate $n=2$ order contribution in diagrammatic representation, which are often termed as stone-fence diagrams.
For $n=2$, we will have $n+1=3$ indices to consider as given in an equation in the previous paragraph.

\vspace{0.5cm}
\begin{center}
    \includegraphics[scale=0.40]{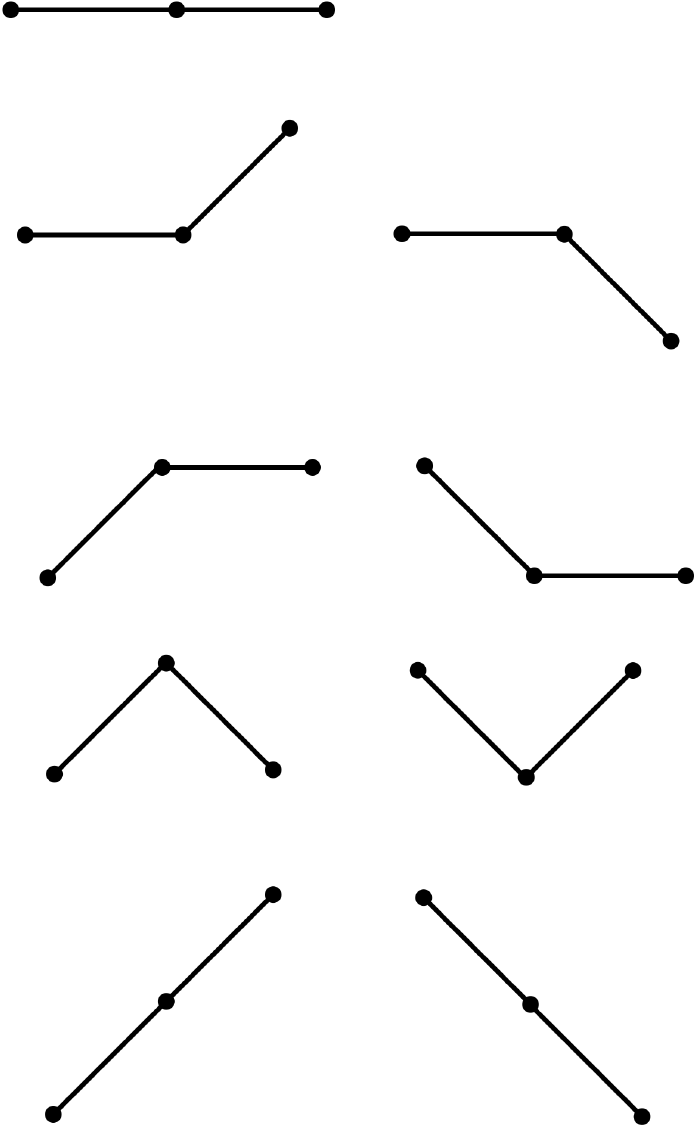}
\end{center}
We note, the above 9 diagrams contribute to the $n=2$ average.
The first line diagram corresponds to $l_1=l_2=l_3$, the second line $l_1=l_2=(l_3\pm 1)$, the third line $l_1=(l_2\pm 1)=l_3$, the forth line $l_1=l_3=(l_2\pm 1)$ and the fifth line is $l_1\ne l_2\ne l_3$ and they can differ only by $\pm 1$.
Accordingly, for the ease of computation, we can formulate the following rules of the diagrams as,
\begin{center}
    \includegraphics[scale=0.40]{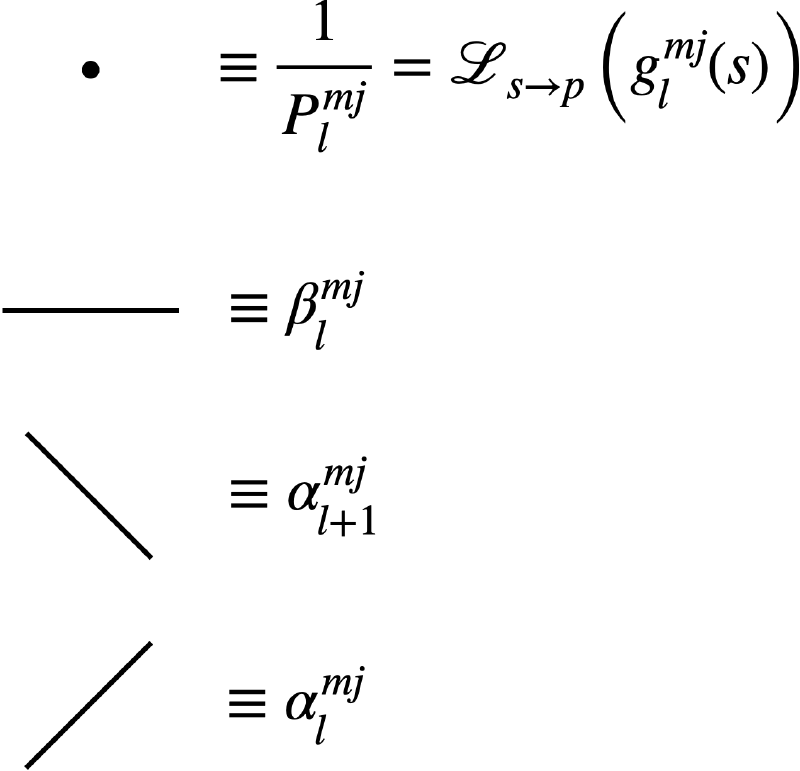}
\end{center}
where, the Laplace transforms have been taken from $s\to p$ space of the coefficients of the Wigner function expansion for the orientation only propagator.
The above rules imply the following algebraic computation,

\vspace{0.50cm}
\begin{center}
    \includegraphics[scale=0.40]{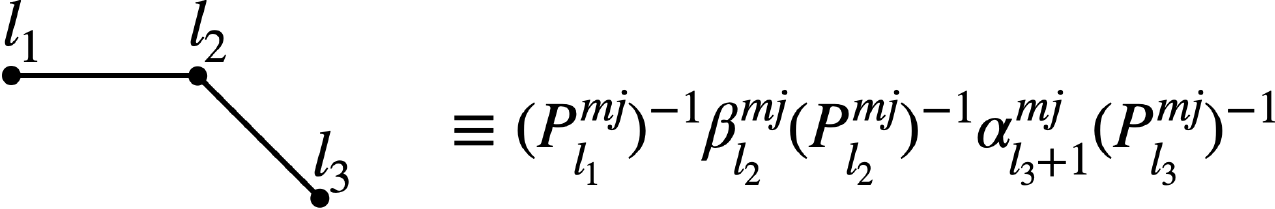}
\end{center}

\noindent This means due to the convolution structure in the evaluation of averages, each valid set of Wigner indices in the Laplace-transformed average contributes a product of terms given by,
\begin{align}
    n^{\text{th}}\text{-order contribution} = (P_{l_1}^{mj})^{-1} \prod_{k=2}^{n+1} ik\left(\alpha_{l_k}^{mj}\delta_{l_k,l_{k-1}+1}+ \beta_{l_k}^{mj}\delta_{l_k,l_{k-1}}+\alpha_{l_k+1}^{mj}\delta_{l_k,l_{k-1}-1}\right)(P_{l_k}^{mj})^{-1}
\end{align}
So, at any order $n$, there are in general $3^{n}$ diagrams that contributes to the averages.
In order to further facilitate computation, let us construct the following diagram,
\begin{center}
    \includegraphics[scale=0.50]{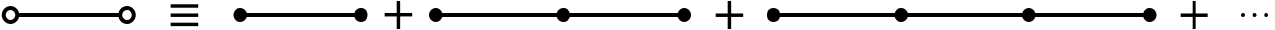}
\end{center}
which contains no factor of the coefficient $\alpha_l^{mj}$, hence it does not act as a ladder operator.
The above diagram (that has the same initial and final $l$ index) can easily be evaluated as the geometric series,
\begin{align}
    \includegraphics[scale=0.50]{Figures/Diagram4.pdf}\nonumber \\
    = \sum_{n=0}^{\infty} \left(\frac{1}{P_{l}^{mj}}\right)^{n+1}(ik\beta_{l}^{mj})^{n} \\
    = \frac{1}{P_{l}^{mj}-ik\beta_{l}^{mj}}
\end{align}
Next we look at those diagrams that start and end with the same $l$ index but never goes below $l$.
We can collect them as,
\begin{center}
    \includegraphics[scale=0.50]{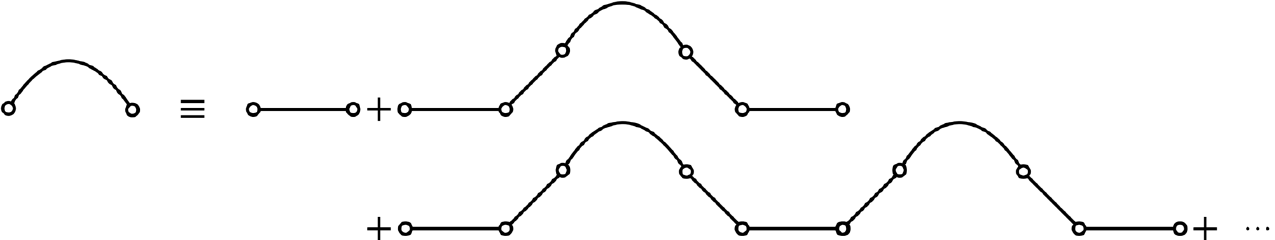} 
\end{center}
Algebraically, 
\begin{align}
    w_l^{(+)m,j} & = \sum_{n=0}^{\infty} \left(\frac{1}{P_{l}^{mj}-ik\beta_{l}^{mj}}\right)^{n+1}\left[\left(ik\alpha_{l+1}^{mj}\right)^2 w_{l+1}^{(+)m,j}\right]^{n} \nonumber \\
    & =\frac{1}{P_{l}^{mj}-ik\beta_{l}^{mj}+\left(k\alpha_{l+1}^{mj}\right)^2 w_{l+1}^{(+)m,j}}
\end{align}
Similarly the diagram that never goes above $l$ for the same starting and ending index is given as,
\vspace{0.5cm}
\begin{center}
    \includegraphics[scale=0.50]{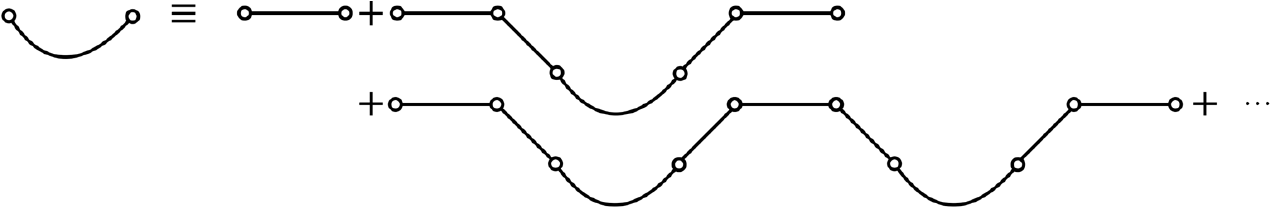} 
\end{center}
Algebraically, 
\begin{align}
    w_l^{(-)m,j} & = \sum_{n=0}^{\infty} \left(\frac{1}{P_{l}^{mj}-ik\beta_{l}^{mj}}\right)^{n+1}\left[\left(ik\alpha_{l}^{mj}\right)^2 w_{l-1}^{(-)m,j}\right]^{n} \nonumber \\
    & =\frac{1}{P_{l}^{mj}-ik\beta_{l}^{mj}+\left(k\alpha_{l}^{mj}\right)^2 w_{l-1}^{(-)m,j}}
\end{align}
In reality for the same starting and ending index of $l$, we can get any possible linear combination of the above diagrams which algebraically evaluates as,
\begin{align}
    W_l^{m,j} & = \sum_{n=0}^{\infty} \left(\frac{1}{P_{l}^{mj}-ik\beta_{l}^{mj}}\right)^{n+1}\left[\left(ik\alpha_{l}^{mj}\right)^2 w_{l-1}^{(-)m,j}+\left(ik\alpha_{l+1}^{mj}\right)^2 w_{l+1}^{(+)m,j}\right]^{n} \nonumber \\
    & =\frac{1}{P_{l}^{mj}-ik\beta_{l}^{mj} +\left(k\alpha_{l+1}^{mj}\right)^2 w_{l+1}^{(+)m,j} +\left(k\alpha_{l}^{mj}\right)^2 w_{l-1}^{(-)m,j}}
\end{align}
The complete expression for $P_{l}^{m,j}$ for the relevant case of $m=j$ is given as,
\begin{align}
    P_l^{m}(p) \equiv P_{l}^{m,j=m}(p)=p+l (l+1)-\frac{1}{4} \Delta_{13} \left(c_l^{j-1} c_l^{-j}+c_l^{-j-1} c_l^j\right)-\Delta_{12} j^2
\end{align}
where, $c_{l}^{j}$ are given by Eq.~\ref{eq:c_coeff}.

The diagrammatic methods outlined in the previous paragraphs provide a simple method of identifying the contributions to a given term in the full Green function.
The diagrams show that the averages are found by determining all of the $l$-index paths connecting the first l-index ($l_0$) to the final l-index ($l_f$).
Hence the full Green function is,
\begin{align}
    \tilde{\mathcal{G}}(\Vec{k}, \Omega|\Omega_0; p) = \frac{1}{8\pi^2}\sum_{l_0, l_f = 0}^{\infty} \sum_{m,j=-\text{min}(l_0,l_f)}^{\text{min}(l_0,l_f)} \mathcal{D}_{l_0}^{mj} (\Omega^\prime)
    \mathcal{D}_{l_f}^{m j^{*}} (\Omega_{0}^\prime) \mathfrak{G}_{l_0l_f}^{mj}(k,p)
    \label{eq:G_k_Omega_p}
\end{align}
The path summation $\mathfrak{G}_{l_0l_f}^{mj}(k,p)$ is constructed using the partial path summations defined in the previous paragraph. 
\begin{align}
    \mathfrak{G}_{l_0l_f}^{mj}(k,p) = 
    \begin{cases} 
      W_l^{mj} & l_0=l_f=l \\
      \displaystyle W_{l_f}^{mj}\prod_{k=l_f+1}^{l_0} (ik\alpha_k^{mj})w_k^{(+)mj} & l_0>l_f \\
      \displaystyle W_{l_0}^{mj}\prod_{k=l_0+1}^{l_f} (ik\alpha_k^{mj})w_k^{(+)mj} & l_0<l_f 
   \end{cases}
\end{align}

\noindent Since, so far we have restricted the initial and end orientations of the propagator, the final orientation independent response can be found to be,
\begin{align}
    \displaystyle \tilde{\mathcal{G}}_P(\Vec{k}; p) & =  \displaystyle \int\! d\Omega \int \! d\Omega_0  \tilde{\mathcal{G}}(\Vec{k}, \Omega|\Omega_0; p) \\
    & = \sum_{l_0, l_f = 0}^{\infty} \sum_{m,j=-\text{min}(l_0,l_f)}^{\text{min}(l_0,l_f)} \int\! d\Omega \int \! d\Omega_0 \mathcal{D}_{l_0}^{mj} (\Omega)
    \mathcal{D}_{l_f}^{m j^{*}} (\Omega_{0}) \mathfrak{G}_{l_0l_f}^{mj}(k,p) \nonumber \\
    & = \sum_{l_0, l_f = 0}^{\infty} \sum_{m,j=-\text{min}(l_0,l_f)}^{\text{min}(l_0,l_f)} \mathfrak{G}_{l_0l_f}^{mj}(k,p) \delta_{l_0,0}\delta_{l_f,0}\delta_{m,0}\delta_{j,0}\nonumber \\
    & = \mathfrak{G}_{00}^{00}(k,p) \\
    & = W_{0}^{00}(k,p)
\end{align}
An explicit expression for $W_{0}^{00}(k,p)$ can be given as (say, up to just $3$ terms in the continued fraction):
\begin{align}
   \displaystyle W_{0}^{00}(k,p) =   \displaystyle\frac{1}{-i k \beta_0+P_0(p)+ \displaystyle\frac{\alpha_1^2 k^2}{-i k \beta_1 +P_1(p)+ \displaystyle\frac{\alpha_2^2 k^2}{-i k \beta_2+P_2(p)+ \displaystyle\frac{\alpha_3^2 k^2}{-i k \beta_3+P_3(p)+ \dots}}}}
\end{align}
where the below $l=0$ diagrams do not contribute due to the condition $l\ge 0$ and $\alpha_l \equiv \alpha_l^{0,0}$, $\beta_l\equiv \beta_l^{0,0}$ and $P_l(p)=P_{l}^{m,j}$.

The Laplace-space inversion or Bromwich integral requires us to determine the following integral,
\begin{align}
    \displaystyle f(t) =\mathcal{L}^{-1}\left\{F(p)\right\}(t) & =\frac{1}{2\pi i} \lim_{T\to \infty}\int_{\lambda-iT}^{\lambda+iT} e^{pt} F(p) dp \\
    & = \sum_{\mu=0}^{\infty} \mathop{\mathrm{Res}}(e^{pt} F(p), a_\mu)
\end{align}
where, $F(p)$ is the Laplace space function and $f(t)$ is the real space value of the same function.
The last equality follows from applying Cauchy's residue theorem and $\mu$ defines the pole\footnote{To note, poles of a function $F(p)$ are zeros of $1/F(p)$, and gives the non-removable singularity of complex function, $F(p)$.\label{footnote:ft1}} index.
\begin{figure}[h!]
    \centering
    \includegraphics[scale=0.50]{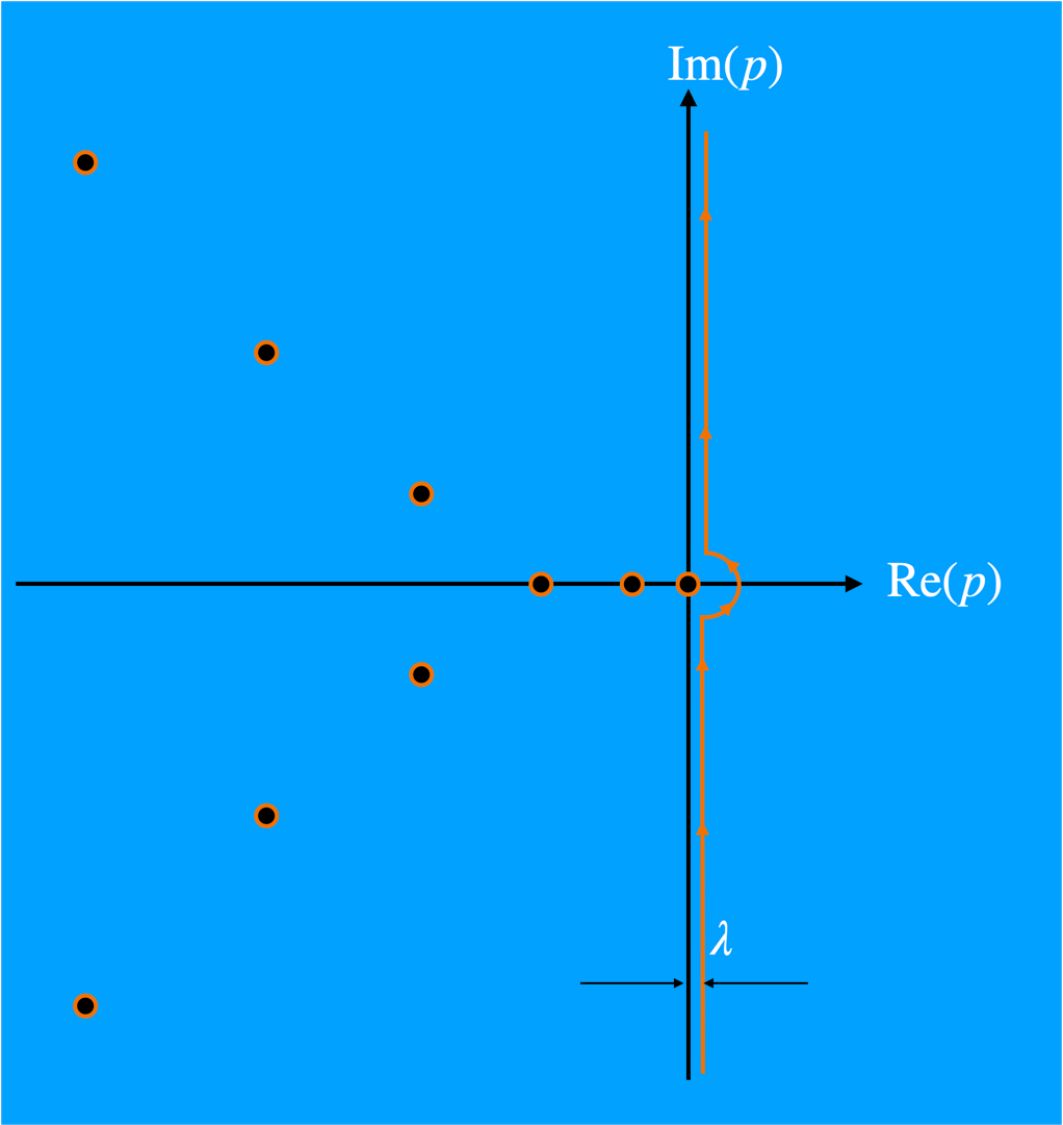}
    \caption{Schematic of the integration path in a typical complex plane of a generic function $F(p)$. Sum over the scaled residues (scaled by the exponential factor) at every pole to the left of the contour (drawn in orange) gives the resultant Bromwich integral. The path direction is given by arrows from $(\lambda-i\infty)$ to $(\lambda+i\infty)$.}
    \label{fig:poles_schematic}
\end{figure}

\begin{figure}[h!]
    \centering
    \includegraphics[scale=0.50]{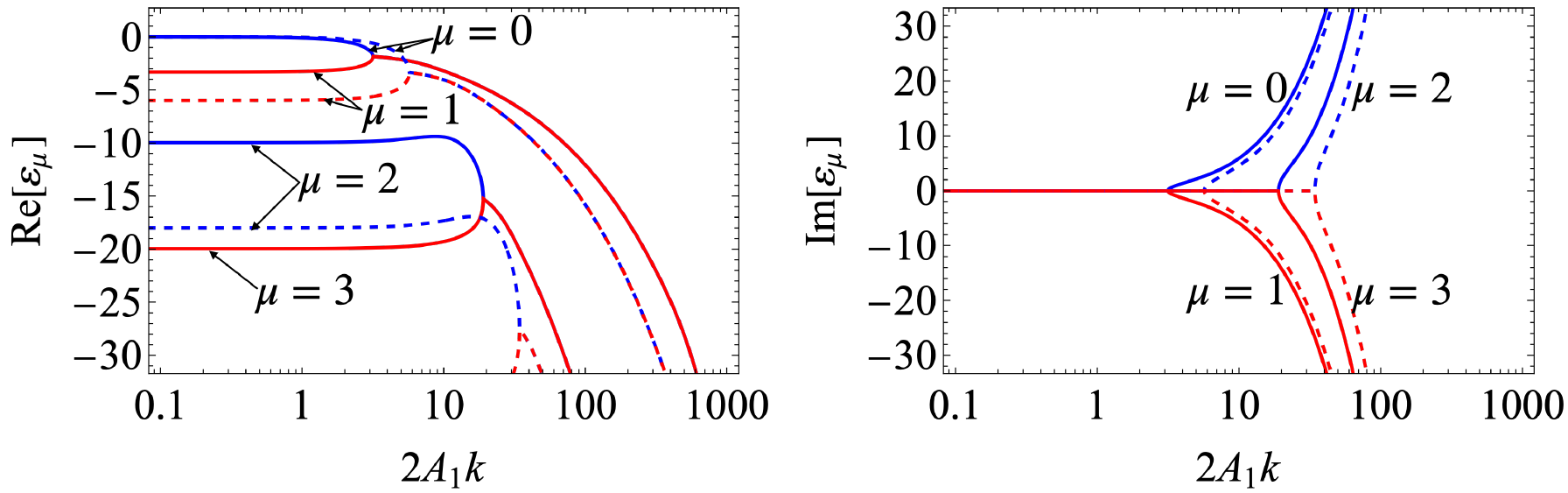}
    \caption{Real (left) and Imaginary (right) parts of the first few poles are plotted as a function of dimensionless wavevector. The $\{A_i\}$ values are chosen same as Fig.~\ref{fig:suppfig5}. Solid lines indicate $A_4\ne 0$ while the dashed lines denote $A_4=0$ case. Values of $\mu$ indicate the order of the pole.}
    \label{fig:suppfig6}
\end{figure}

For our case, let us define the $\mu^{\text{th}}$ pole $\mathcal{E}_\mu(k)$.
Numerical computation of the poles yield a relatively simpler structure of them.
Below, $k<k_c$ poles are all real valued and beyond $k>k_c$ the poles can be written as,
\begin{align}
    \mathcal{E}_\mu=\mathrm{Re}[\varepsilon_\mu]\pm i\mathrm{Im}[\varepsilon_\mu].
\end{align}
We plot the first few poles as a function of dimensionless wavevector in Fig.~\ref{fig:suppfig6}.
The case of $A_4\ne 0$ is plotted in solid lines and $A_4=0$ is given by the dashed line.

Mathematically from the definition of poles we can write$\!\!^{~\ref{footnote:ft1}}$, 
\begin{align}
    j(k;p=\mathcal{E}_\mu)=\frac{1}{\tilde{\mathcal{G}}_P(k;p=\mathcal{E}_\mu)}=0
\end{align}
for $\mu=0,1,2\dots \infty$ gives the infinite set of poles for the infinite continued fraction.
We can write,
\begin{align}
    \mathcal{G}(k;L)& =\sum_{\mu=0}^{\infty} \displaystyle {\text{Res}}_{p=\mathcal{E}_\mu} \left[ \tilde{\mathcal{G}}_P(k;p)e^{pL}\right] = \sum_{\mu=0}^{\infty} \lim_{p\to \mathcal{E}_\mu} \left[(p-\mathcal{E}_\mu)\tilde{\mathcal{G}}_P(k;p)e^{pL}\right] \nonumber \\
    & = \sum_{\mu=0}^{\infty} \frac{1}{\partial_p j(k,p=\mathcal{E}_\mu) } e^{\mathcal{E}_\mu L}
\end{align}
where the final form arises from the fact that all of the poles are simple and we can apply L'Hospital's rule to evaluate the limit.
Finally, Fourier inversion is found by numerically performing the integral,
\begin{align}
    \mathcal{G}(\vec{R};L) = \frac{1}{(2\pi)^3}\int d\vec{k} e^{-i\vec{k}\cdot\vec{R}} \mathcal{G}(k;L)
\end{align}
\begin{figure}[h!]
    \centering
    \includegraphics[scale=0.50]{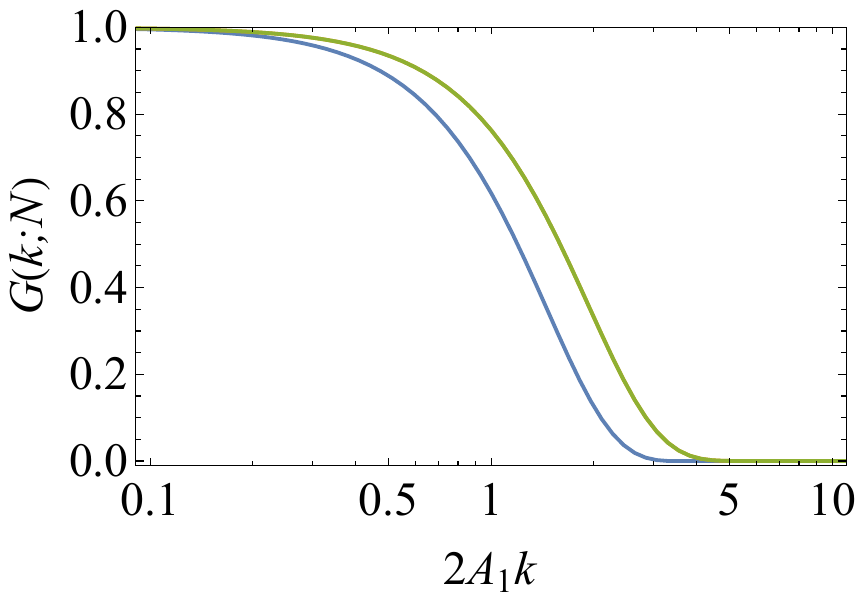}
    \caption{Propagator in Fourier space is plotted as a function of dimensionless wavevector for $N=L/2A_1=5$. The parameter choices $\{A_i\}$ are same as previous figures. Blue line indicates $A_4=0.70$ and green line is for $A_4=0$. The plot for $A_3=A_4=0$ coincides with $A_4=0$.}
    \label{fig:suppfig7}
\end{figure}

\newcommand\xleftrightarrow[1]{
\mathbin{\ooalign{$\,\xrightarrow{#1}$\cr$\xleftarrow{\hphantom{#1}}\,$}}
}

\subsection{Utility of Full Green Function Solution}
\subsubsection{Force Extension Behavior}
As already alluded to in the previous section the full Green function produces the force dependent Green function for free subject to the Wick rotation of the type, $f \xleftrightarrow{\mathrm{Wick \text{ } Rotation}} -ik$ where, $f$ is the magnitude of force and $k$ is the dimensionless wavevector.
In general the derivatives of the partition function or the propagator with respect to force gives the moments of the end-to-end distance along the applied force direction per the definition of the propagator.
As mentioned before, we arbitrarily define the applied force direction to be along the z-direction.
Hence, the moments can be written as,
\begin{align}
   \langle R_{z}^{n}\rangle = \frac{1}{\mathcal{G}(\vec{f},L)} \frac{\partial^{n} {\mathcal{G}(\vec{f};L)}}{\partial f^{n}}
\end{align}
The first order moment yields the average extension or force extension behavior.
The average extension along the applied force direction is given as the first derivative in dimensionless form as 
\begin{align}
    \frac{\langle R_z\rangle}{L} = \frac{2A_1}{L}\frac{\partial \log{\mathcal{G}(\vec{f};L)}}{\partial f}
    \label{eq:extension}
\end{align}
\subsubsection{Static Structure Factor}
%{\color{red} \textbf{Note to self: Show the full derivation.}}

% \begin{figure}[h!]
%     \centering
%     \includegraphics[scale=0.50]{Figures/SuppFig8.pdf}
%     \caption{Scattering function $kS(k)$ is plotted as a function of wavevector for $N=L/(2A_1)=5$ in linear-linear and log-log formats. Blue and Green lines indicate TBcHWLC and HWLC respectively. The scattering function for Gaussian flexible chain and rigid-rod polymer are given by the blacked and blue dashed lines respectively. As expected, the blue dashed line merges with solid blue or green for high wavevectors (small lengthscales), while for small wavevctors (hence, large lengthscales) black dashed lines are equal with the solid blue or solid green.}
%     \label{fig:suppfig8}
% \end{figure}
\noindent The static structure factor of the polymer chain as,
\begin{align}
    S(k) & =\frac{1}{L^2}\int_0^L \!ds_1 \int_0^L \!ds_2 \langle e^{i\vec{k}\cdot[\vec{r}(s_1)-\vec{r}(s_2)]}\rangle \nonumber \\
    & = \frac{2}{L^2}\int_0^L \!ds_1 \int_0^{s_1} \!ds_2 \langle e^{i\vec{k}\cdot[\vec{r}(s_1)-\vec{r}(s_2)]}\rangle \\
    & = \frac{2}{L^2}\int_0^L \!ds_1 \int_0^{s_1} \!ds_2 \mathcal{G}(\vec{k}, s_1-s_2)
\end{align}
where the second line is written using `time-ordering' of integrals.
The third equality uses the definition of $\mathcal{G}(\vec{k},s)$ for an arbitrary separation of two points by a distance $s$ along the contour length.
Laplace transform of $S(k)$ leads to a consolidated form of the Laplace transformed static structure factor.
This lets us write the final form of static structure factor in dimensionless form as ($N=L/(2A_1)$),
\begin{align}
    S(k) & = \frac{2}{N^2}\mathcal{L}^{-1}_{p\to N}\left[\frac{\tilde{\mathcal{G}}_P(\vec{k};p)}{p^2}\right]
\end{align}
where, $\mathcal{L}^{-1}[\cdot]$ indicates the inverse Laplace transform from $p\!\to\! L$.
Also, in the final form $k$ is made dimensionless by multiplying by $2A_1$.
For completeness here we quote the static structure factors for Gaussian chains and rigid rods as~\cite{doi1988theory},
\begin{align}
    S_{\mathrm{Gaussian}}(k) & = \frac{72}{Nk^2}\left(\frac{1}{6}-\frac{1-\exp{\left(-{Nk^2}/{6}\right)}}{Nk^2}\right) \\
    S_{\mathrm{Rigid-rod}}(k) & = \frac{2}{Nk}\left(\int_0^{Nk}\!\frac{\sin{x}}{x} dx - \frac{1-\cos{Nk}}{Nk}\right)
\end{align}

\begin{figure}[h!]
    \centering
    \includegraphics[scale=0.50]{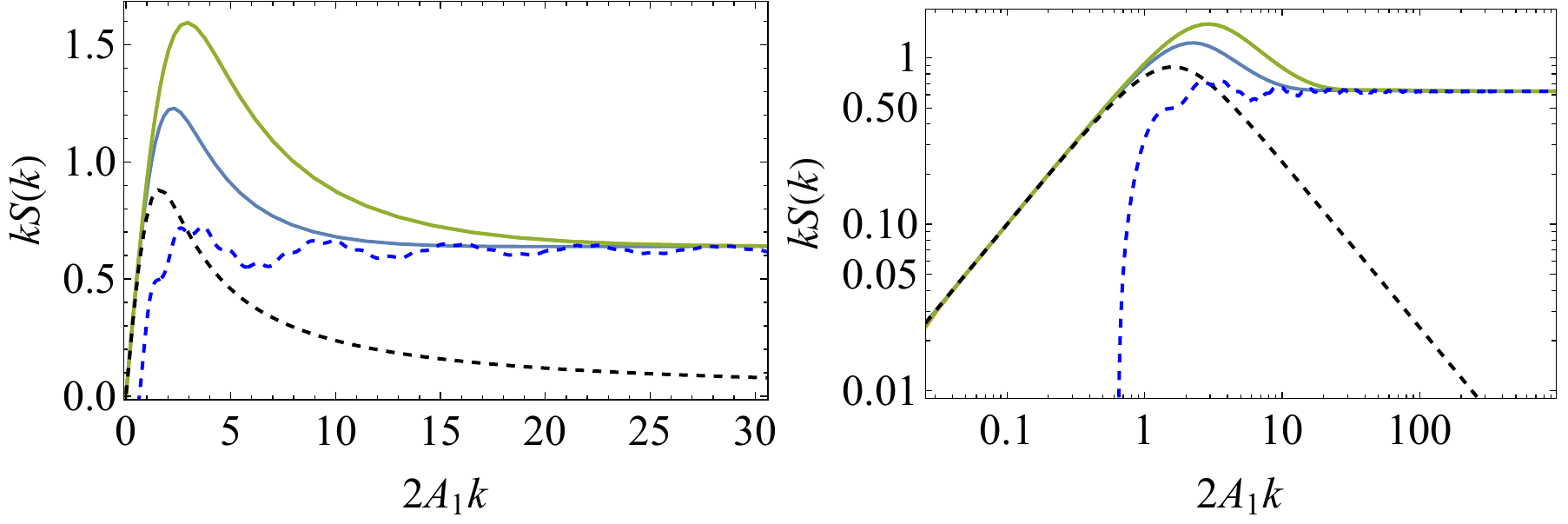}
    \caption{Scattering function $kS(k)$ is plotted as a function of wavevector for $N=L/(2A_1)=5$ in linear-linear and log-log formats. Blue and Green lines indicate TBcHWLC and HWLC respectively. The scattering function for Gaussian flexible chain and rigid-rod polymer are given by the blacked and blue dashed lines respectively. As expected, the blue dashed line merges with solid blue or green for high wavevectors (small lengthscales), while for small wavevctors (hence, large lengthscales) black dashed lines are equal with the solid blue or solid green.}
    \label{fig:suppfig8}
\end{figure}

The scaled static structure factor is plotted in Fig.~\ref{fig:suppfig8} for chain length, $N=5$ to illustrate the difference between various model scaling relations.
At low wavevectors (hence, large lengthscales), we note that both the TBcHWLC (blue solid) and HWLC (green solid) results both coincide with the Gaussian or flexible chain limit (black dashed) indicating random walk behavior of the propagator.
At large wavevector or, small lengthscales both scattering functions merge with the rigid rodlike limit (blue dashed), indicating at short lengthscales the chain behaves exactly same as rigid links.
The log-log plot shows that, while the Gaussian chain decays as $S(k)\!\sim \!k^{-2}$ at long wavenumbers, elastic models (WLC, HWLC or, TBcHWLC) all follow $S(k)\!\sim\! k^{-1}$.

\subsubsection{Ring Closure Probability}
%{\color{red} \textbf{Note to self: Show the full derivation.}}

Before deriving the looping probability, we briefly recap the total Green function and it's meaning.
The full propagator $\mathcal{G}(\Vec{R}, \Omega|\Omega_0; L)$ gives the probability of obtaining a chain configuration with the end-to-end vector distance $\vec{R}$, and chain end orientations of $\Omega(L)=\Omega$ at one end, given the initial orientation is $\Omega(0)=\Omega_0$, where $\Omega(s)=[\theta(s), \phi(s), \psi(s)]$ are the set of Euler angles defined at every arc length parametrization $s\in [0,L]$ and the limits of Euler angles are given as, $\theta\in [0,\pi]$, $\phi\in [0,2\pi]$ and $\psi\in [0,2\pi]$, respectively.
The Jacobson-Stockmayer theory~\cite{jacobson1950intramolecular,flory1969statistical,flory1976macrocyclization,yamakawa1997helical} defines the length dependent looping probability, $J(L)$ as,
\begin{align}
    J(L) & = 8\pi^2\int d\Omega_0\mathcal{G}(\Vec{R}=\vec{0}, \Omega=\Omega_0|\Omega_0; L).
    \label{eq:J_defn}
\end{align}
As demonstrated in Ref.~\cite{shimada1984ring} the cyclization of elastic polymers with end orientations does not necessarily correspond to a homogeneous species in chemical sense, rather contains a `mixture' of topological isomers that ref.~\cite{yamakawa1997helical} calls as \emph{topoisomeric species}, which differ by their knot properties.
More clearly different topoisomers have different linking numbers.
A beautiful demonstration of the physical meaning of this is provided in Fig.7.1 of ref.~\cite{yamakawa1997helical}.
Here, our approach does not require us to define a linking number dependent looping probability, rather the propagator takes care of all possible topoisomers of cyclization.
Here, we rewrite $\mathcal{G}(\Vec{R}, \Omega|\Omega_0; L)$ as the inverse Fourier transform of $\mathcal{G}(\Vec{k}, \Omega|\Omega_0; L)$ as,
\begin{align}
    \mathcal{G}(\Vec{R}, \Omega|\Omega_0; L) & =\frac{1}{(2\pi)^3}\int d\vec{k} e^{-i\vec{k}\cdot\vec{R}} \mathcal{G}(\Vec{k}, \Omega|\Omega_0; L)  \\
    & = \frac{1}{(2\pi)^3}\int d\vec{k} e^{-i\vec{k}\cdot\vec{R}} \mathcal{L}^{-1}_{p\to L}\left[ \tilde{\mathcal{G}}(\Vec{k}, \Omega|\Omega_0; p)\right] 
\end{align}
where, the second line is written as the inverse Fourier, inverse Laplace transform.
Here, we rewrite Eq.~\ref{eq:G_k_Omega_p} for convenience,
\begin{align*}
    \tilde{\mathcal{G}}(\Vec{k}, \Omega|\Omega_0; p) = \frac{1}{8\pi^2}\sum_{l_0, l_f = 0}^{\infty} \sum_{m,j=-\text{min}(l_0,l_f)}^{\text{min}(l_0,l_f)} \mathcal{D}_{l_0}^{mj} (\Omega^\prime)
    \mathcal{D}_{l_f}^{m j^{*}} (\Omega_{0}^\prime) \mathfrak{G}_{l_0l_f}^{mj}(k,p)
\end{align*}
Hence, it immediately follows that,
\begin{align}
    \tilde{\mathcal{G}}(\Vec{k}, \Omega_0|\Omega_0; p) = \frac{1}{8\pi^2}\sum_{l_0, l_f = 0}^{\infty} \sum_{m,j=-\text{min}(l_0,l_f)}^{\text{min}(l_0,l_f)} \mathcal{D}_{l_0}^{mj} (\Omega_0^\prime)
    \mathcal{D}_{l_f}^{m j^{*}} (\Omega_{0}^\prime) \mathfrak{G}_{l_0l_f}^{mj}(k,p)
\end{align}
and thus the orientation independent Green function follows as,
\begin{align}
    \tilde{\mathcal{G}}_P(\vec{k},p)=\underbrace{\int \!d\Omega_0}_{\text{integrate over all possible Euler angles}} \tilde{\mathcal{G}}(\Vec{k}, \Omega_0|\Omega_0; p) = \frac{1}{8\pi^2}\sum_{l= 0}^{\infty} \sum_{m,j=-l}^{l}
    \mathfrak{G}_{ll}^{mj}(k,p).
\end{align}
From the definition of $J$-factor in Eq.~\ref{eq:J_defn}, we obtain,
\begin{align}
    J(L) & = \lim_{\vec{R}\to\vec{0}}\frac{1}{(2\pi)^3}\int d\vec{k} e^{-i\vec{k}\cdot\vec{R}}\mathcal{L}_{p\to L}^{-1}\left[\sum_{l= 0}^{\infty} \sum_{m,j=-l}^{l}
    \mathfrak{G}_{ll}^{mj}(k,p)\right]  \\
    & = \lim_{R\to 0} \frac{1}{2\pi^2} \int_0^\infty dk k^2 j_0(kR) \mathcal{L}_{p\to L}^{-1}\left[\sum_{l= 0}^{\infty} \sum_{m,j=-l}^{l}
    \mathfrak{G}_{ll}^{mj}(k,p)\right]
\end{align}
where, $\displaystyle j_0(x)=\frac{\sin x}{x}$ is the spherical Bessel function of order $0$ and $j_0(0)=1$.

These considerations let us write the final form of ring closure probability from the Jacobson-Stockmayer theory~\cite{jacobson1950intramolecular,yamakawa1997helical}, or the so-called $J$-factor as,
\begin{align}
J(L)=\frac{1}{2\pi^2}\int_0^\infty dk k^2 \left[\sum_{l=0}^\infty\sum_{m,j=-l}^{l}\mathcal{L}^{-1}_{p\to L}\left\{ \mathfrak{G}_{ll}^{mj}(k,p)\right\}\right]
\end{align}
where we can interpret the sum over $l$ originates from topologically distinct species with different Linking numbers that end up in cyclization in accordance with ref.~\cite{shimada1984ring}.
\nopagebreak
\renewcommand{\theequation}{S6.\arabic{equation}}
\begin{center}
\section*{Step-by-step Guide for the Exact Solution\label{sec:step_by_step}}
\end{center}
\begin{tcolorbox}[colback=green!2!white,colframe=green!75!black]
  
$\bullet$ \textbf{Step 1}: Obtain the Hamiltonian from the Legendre transform of Lagrangian density and express in terms of the generalized momentum.

  \begin{tcolorbox}[colback=red!1!white,colframe=red!75!black]
  \begin{align*}
      & \mathrm{Elastic \text{ }Energy}, E \to \mathrm{Lagrangian \text{ } density}, \mathcal{L} \to \mathrm{Hamiltonian \text{ } density}, \mathcal{H} \\
      & \mathcal{L}\xleftrightarrow{\text{Legendre Transform}}\mathcal{H}
  \end{align*}
  \end{tcolorbox}
 
  $\bullet$ \textbf{Step 2}: Write down the path-integral formalism for the Green function or propagator of the system. 
  This allows us to derive the diffusion or Schrödinger equation corresponding to the Hamiltonian of the system.

  \begin{tcolorbox}[colback=red!1!white,colframe=red!75!black]

  \begin{align*}
    \mathcal{G}(\Vec{R}, \Omega|\Omega_0; L) & = \int_{\Omega(s=0)=\Omega_0}^{\Omega(s=L)=\Omega} \mathcal{D}[\Omega(s)] \exp{\left(-\beta\int_0^L ds \mathcal{L}\right)} \\
    & \hspace{1 in}\times \Bigg(\mathrm{constraints \text{ }as \text{ }delta \text{ }functions}\Bigg)
\end{align*}

\begin{equation*}
    \left(\frac{\partial }{\partial L} - \underbrace{\mathcal{H}_0}_{\mathrm{``free-particle" \text { } Hamiltonian}}+\mathrm{``Potential \text{ } term" \text{ }}[ \mathcal{V}]\right)\hat{\mathcal{G}}(\Vec{R}, \Omega|\Omega_0; L) = 0
\end{equation*}
  \end{tcolorbox}

  $\bullet$ \textbf{Step 3}: A free-particle limit of the diffusion equation can be written down. 
  Depending on the problem of interest the free particle solution is often enough.

  \begin{tcolorbox}[colback=red!1!white,colframe=red!75!black]
\begin{align*}
    \mathrm{Set \text{ }} \mathcal{V}=0 \xleftrightarrow{\mathrm{Equivalently}}
    \left(\frac{\partial }{\partial L} - \mathcal{H}_0\right)\underbrace{{\mathcal{G}}_0(\Omega|\Omega_0; L)}_{\mathrm{0} \text{ indicates free-particle solution}} = 0
\end{align*}
  \end{tcolorbox}

  $\bullet$ \textbf{Step 4}: Find the suitable eigenfunctions for the ``time''-independent Schrödinger equation. 
  For the cases considered here, the eigenfunctions are given by the Wigner-D functions.

  \begin{tcolorbox}[colback=red!1!white,colframe=red!75!black]

\begin{equation*}
    \mathcal{G}_0(\Omega|\Omega_{0}; N) = \sum_{lmj} \sum_{l_{0}m_{0}j_{0}}
    \underbrace{\mathcal{D}_{l}^{mj} (\Omega)}_{\text{eigenfunction}}
    \mathcal{D}_{l_{0}}^{m_{0}j_{0}^{*}} (\Omega_{0})
    \underbrace{g_{l_{0}m_{0}j_{0}}^{lmj}(L)}_{\text{coefficient}}
\end{equation*}

  \end{tcolorbox}

\end{tcolorbox}

\begin{tcolorbox}[colback=green!5!white,colframe=green!75!black]
$\bullet$ \textbf{Step 5}: Using the method of separation of variables, a solution to the free particle propagator can be obtained.

  $\bullet$ \textbf{Step 6}: Tangent correlations, mean-squared end-to-end distance (second moment of end-to-end distance), radius of gyration, or, in general any nth-order moment can be obtained using the free-particle solution.

  \begin{tcolorbox}[colback=red!1!white,colframe=red!75!black]
\begin{align*}
    \mathrm{Quantities} = \underbrace{\mathcal{F}\left[\mathcal{G}_0\right]}_{\text{Functionals of } \mathcal{G}_0}
\end{align*}
  \end{tcolorbox}
  
$\bullet$ \textbf{Step 7}: A diagrammatic or algebraic rules depending on the $n$-th order moment can be used to formulate an exact solution for the Green function in arbitrary fields.
  The solution takes the form of a infinite order continued fraction.

  $\bullet$ \textbf{Step 8}: Finally, given the full propagator, statistical quantities such as force-extension of polymer, static structure factor, looping statistics can be formulated.

  \begin{tcolorbox}[colback=red!1!white,colframe=red!75!black]
\begin{align*}
    \mathrm{Quantities} = \underbrace{\mathcal{F}\left[\mathcal{G}\right]}_{\text{Functionals of }\mathcal{G}}
\end{align*}
  \end{tcolorbox}

\end{tcolorbox}
\noindent Using the steps above, we can generate solutions as desired.

In this article, we consider two such cases, a naturally twisted dsDNA with twist-bend coupling and a general elastic wire with all possible coupling constants.

\newpage
\renewcommand{\theequation}{S5.\arabic{equation}}
\section{Natural Twisted ds-DNA\label{sec:naturaA_3wist}}
To consider the case of natural twist for ds-DNA with twist-bend coupling, we consider the following elastic energy as,
\begin{align}
    \beta\mathcal{E} \!= \!\!\int_0^L \!\!ds \left[\frac{A}{2}\left(\Omega_1^2+\Omega_2^2\right) \!+\!\frac{A_3}{2}\left(\Omega_3-\bar{\Omega}\right)^2\!+\!A_4\Omega_2\Omega_3\right]
    \label{eq:dna_energy}
\end{align}
where, $A, A_3 \text{ and } A_4$ indicate the bending, twist and coupling persistence lengths respectively.
To note, here the bending degrees of freedom are isotropic such that, $A_1=A_2=A$.
The natural twist density of dsDNA is considered by implementing $\bar{\Omega}\ne 0$.
Procedurally, we follow the same steps as in Section~\ref{section:Green}.
The Hamiltonian without the external potential, $\mathcal{H}_0$ can be written as,
\begin{align}
    \mathcal{H}_0 & = \frac{1}{2A} \Bigg[L^2 - \left(\frac{A_4^2}{A_4^2-AA_3}\right) L_2^2 -  \left(1+\frac{A^2}{A_4^2-AA_3}\right)L_3^2 +\left(\frac{A_4A}{A_4^2-AA_3}\right)(L_2L_3+L_3L_2) 
    \nonumber \\
    & \hspace{0.65 in}
    +\left(\frac{2AA_3A_4\bar{\Omega}}{A_4^2-AA_3}\right)L_2-\left(\frac{2A^2A_3\bar{\Omega}}{A_4^2-AA_3}\right)L_3 - \frac{AA_4^2A_3 \bar{\Omega}^2}{A_4^2-AA_3}
    \Bigg]
    \\
    & \equiv \frac{1}{2A}\Bigg[L^2+\frac{\phi^2}{\rho+1-\phi^2}L_2^2+\left\{\frac{(\rho+1)^2}{\rho+1-\phi^2}-1\right\} L_3^2-\frac{(\rho+1)\phi}{\rho+1-\phi^2}(L_2L_3+L_3L_2) \nonumber \\
    & \hspace{0.65 in}
    -\frac{\lambda\phi}{\rho+1-\phi^2}L_2 +\frac{\lambda(\rho+1)}{\rho+1-\phi^2}L_3 + \frac{\lambda^2}{4(\rho+1)}\frac{\phi^2}{\rho+1-\phi^2}
    \Bigg]
\end{align}
where $L_i$ is the generalized momenta as defined previously, $\displaystyle \rho=\left(\frac{A}{A_3}-1\right)$, $\lambda = 2A\bar{\Omega}$ and $\displaystyle \phi=\frac{A_4}{A_3}$.
The limit of no twist-bend coupling ($A_4\to 0$ or, equivalently, $\phi\to 0$) gives the form of the Hamiltonian as,
\begin{align}
    \mathcal{H}_0 & = \frac{1}{2A} \Bigg[L^2-\left(1-\frac{A}{A_3}\right)L_3^2+2A\bar{\Omega} L_3\Bigg]\equiv \frac{1}{2A} \left[L^2+\rho L_3^2+\lambda L_3\right].
\end{align}
Given the Hamiltonian, the general steps as written in the step-by-step guide can be taken to solve for both orientation dependent and full Green function solution of the propagator.
In the main article, we specifically look at the (a) tangent tangent correlations and (b) looping factor.

For vanishing twist-bend coupling ($A_4=0$), we find the following simple analytic forms for the tangent correlations ($N=L/2A$),
\begin{align}
    \langle \vec{t}_3(N)\cdot\vec{t}_3(0)\rangle & = \exp{\left(-2N\right)} \\
    \langle \vec{t}_1(N)\cdot\vec{t}_1(0)\rangle & = \langle \vec{t}_2(N)\cdot\vec{t}_2(0)\rangle = \exp{\left(-\{2+\rho\} N\right)}\cos{\left(N\lambda\right)}
\end{align}

\begin{figure}[ht!]
    \centering
    \includegraphics[scale=0.50]{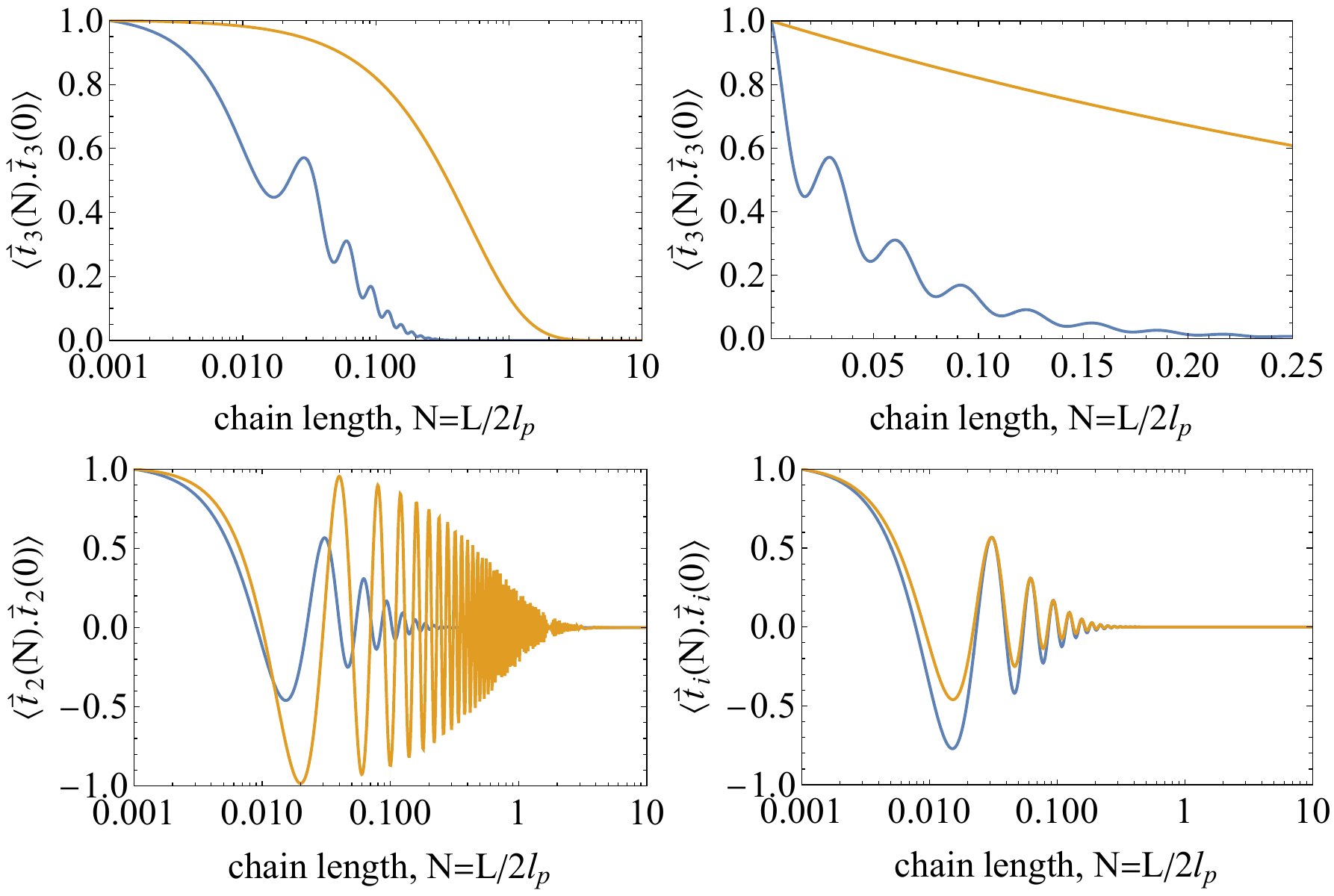}
    \caption{Tangent correlation functions for the coupled dsDNA model, where $l_p=A$. Top panel shows the $\vec{t}_3$ tangent correlation function for non-zero coupling ($A_4=A/2$, blue) and zero coupling (orange). Left bottom panel shows $\vec{t}_2$ tangent correlation function for non-zero coupling ($A_4=A/2$, blue) and zero coupling (orange). For $A_4=0$, $\vec{t}_2$ and $\vec{t}_1$ correlation functions are identical. Bottom right plot gives $\vec{t}_2$ (orange) and $\vec{t}_1$ (blue) correlation functions for $A_4=A/2$.}
    \label{fig:suppfig9}
\end{figure}
Plots for various tangent correlation functions for $\rho=-0.50$, $\phi=0.25$ and $\lambda=\pi/2$ are plotted in Fig.~\ref{fig:suppfig9} as a function of chain size, $N=L/(2A)$.
As described in the main text, $\vec{t}_3$ auto-correlation function shows oscillatory behavior in presence of coupling ($A_4\ne 0$).
The corresponding lengthscale of correlation decay is much smaller in presence of twist-bend coupling as seen from the blue solid lines in the top panel of Fig.~\ref{fig:suppfig9}.
As $A_1=A_2=A$ for the model, it is expected that $\vec{t}_1$ and $\vec{t}_2$ auto-correlation functions are identical when $A_4=0$ .
In presence of coupling this symmetry is broken since only $\vec{t}_2$ couples to $\vec{t}_3$.
This is shown in the right figure in the bottom panel of Fig.~\ref{fig:suppfig9}.

\begin{figure}[htb!]
    \centering
    \includegraphics[scale=0.19]{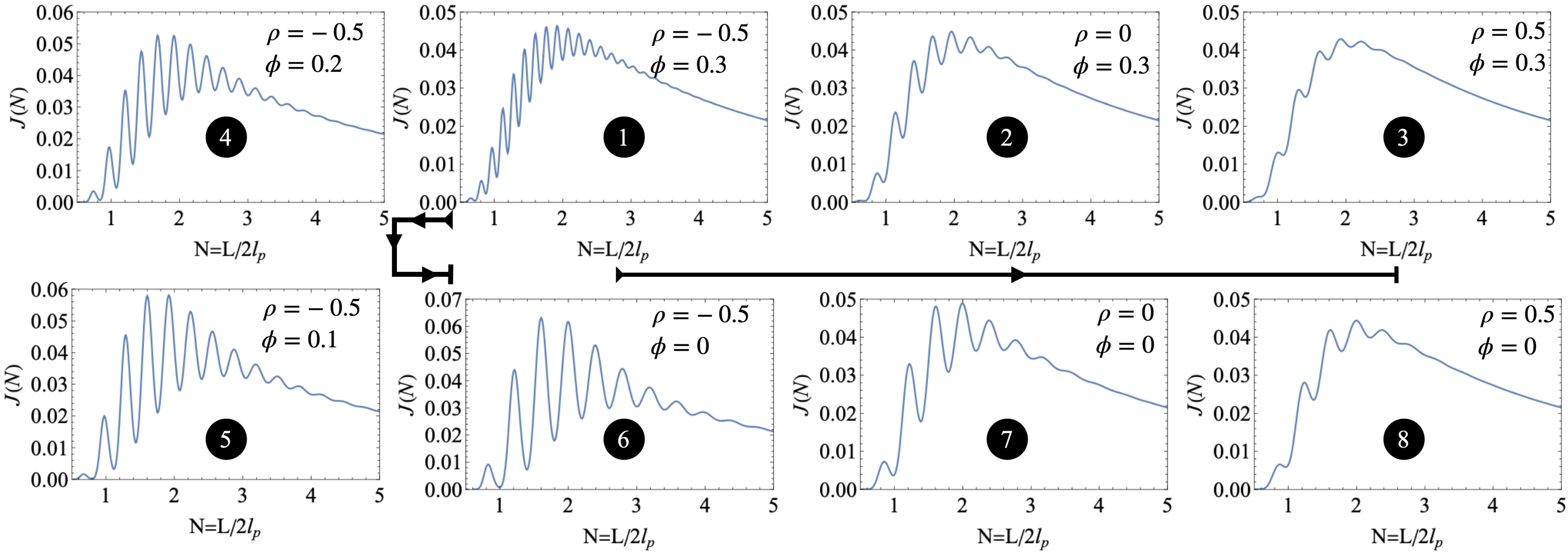}
    \caption{Looping factors are plotted as a function of dimensionless chain length, $N=L/(2A)\equiv L/(2l_p)$. The plots $1\to 2 \to 3$ and $6\to 7 \to 8$ show the effect of $A$ and $A_3$ anisotropy as $\rho=A/A_3-1$ for two values of twist-bend coupling as $\phi=A_4/A_3$. Plots $1\to 4\to 5 \to 6$ show the effect of decreasing twist bend coupling at fixed $\rho$.}
    \label{fig:suppfig10}
\end{figure}

\newpage
\renewcommand{\theequation}{S6.\arabic{equation}}
\section{Generalization to all Possible Elastic Couplings\label{sec:general}}
A complete general case that contains all possible elastic couplings can be written as the following modulus matrix,
\begin{align}
    \mathbb{A} = \begin{pNiceMatrix}
 A_1 & A_5 & A_6  \\
 A_5 & A_2 & A_4  \\
 A_6 & A_4 & A_3
\end{pNiceMatrix}
\end{align}
The corresponding energy of the system is,
\begin{align}
    \beta \mathcal{E} & = \frac{1}{2}\int_0^L ds \left(A_1\Omega_1^2+ A_2\Omega_2^2+A_3\Omega_3^2+A_4(\Omega_2\Omega_3+\Omega_3\Omega_2)+A_5(\Omega_1\Omega_2+\Omega_2\Omega_1)+A_6(\Omega_1\Omega_3+\Omega_3\Omega_1)\right)
    \label{eq:elastic_energy_gen}
\end{align}
which contains two additional coupling constants, $A_5$ and $A_6$.
The modulus $A_5$ interprets as a coupling between bending degrees of freedom, or a coupled bending mode while $A_6$ is another twist-bend coupled mode that has been ignored so far.
The corresponding Hamiltonian can be obtained as,
\begin{align}
    \mathcal{H}_0 & = \frac{1}{2d_{\mathbb{A}}}\Bigg[(A_2A_3-A_4^2)L_1^2 + (A_1A_3-A_6^2)L_2^2 + (A_1A_2-A_5^2)L_3^2 +(A_5A_6-A_1A_4)(L_2L_3+L_3L_2) \nonumber \\
    & \hspace{0.5 in} + (A_4A_6-A_3A_5)(L_1L_2+L_2L_1) +  (A_4A_5-A_2A_6)(L_1L_3+L_3L_1)\Bigg]
\end{align}
where, $\displaystyle d_{\mathbb{A}}$ represents the determinant of the modulus matrix and given as,
\begin{align}
    d_{\mathbb{A}} = \left(A_1A_2A_3+2A_4A_5A_6-A_1A_4^2-A_3A_5^2-A_2A_6^2\right)
\end{align}
It is clear that the limit of no coupling reduces the Hamiltonian to the simpler form given as,
\begin{align}
    \mathcal{H}_0 = \frac{1}{2} \left(\frac{{L_1}^2}{A_1}+\frac{{L_2}^2}{A_2}+\frac{{L_3}^2}{A_3}\right).
\end{align}
For this case, it is perhaps better to directly employ Eq.~\ref{eq:p1_squared}-Eq.~\ref{eq:p3_squared} for the $\{L_i\}$ operators instead of converting the Hamiltonian to have $\displaystyle L^2=\sum_{i=1}^3 L_i^2$ term.
As mentioned in the previous section and the roadmap to derive the final forms of orientation only Green function as well as the full propagator, we can employ the same methodologies for solution.

\newpage
\bibliographystyle{siam}
\bibliography{ref}

\end{sloppypar}